\documentclass[a4paper,12pt]{article}
\usepackage{subcaption}
\usepackage{caption,graphicx}
\usepackage{epsfig}
\usepackage{mathrsfs}
\usepackage{amsmath}
\usepackage{amsfonts}
\usepackage{amssymb}
\allowdisplaybreaks
\usepackage[usenames]{color}
\usepackage[a4paper,left=2.cm,right=2.cm,top=2.5cm,bottom=1.7cm]{geometry}\usepackage[colorlinks=true,linktocpage=true,linkcolor=blue,citecolor=blue]{hyperref}
\usepackage{xcolor}
\usepackage{hyperref}
\usepackage{cite}
\usepackage{caption}
\allowdisplaybreaks
\numberwithin{equation}{section}

\begin{document}

\begin{titlepage}

\thispagestyle{empty}

\begin{center}

\hfill         \phantom{xxx}  EFI-25-4

\vskip 2 cm
\Large{{\bf Stable non-linear evolution in regularised higher derivative effective field theories}}

\vskip 1.25 cm
\small{\bf Pau Figueras\textsuperscript{1,2,3}, \'Aron D. Kov\'acs\textsuperscript{1} and Shunhui Yao\textsuperscript{1}} \vspace{0.3cm}\\
\small{\it \textsuperscript{1}School of Mathematical Sciences, Queen Mary University of London,}\vspace{0.1cm}\\
\small{\it Mile End Road, London E1 4NS, United Kingdom}\vspace{0.3cm}\\
\small{\it \textsuperscript{2}Enrico Fermi Institute \& Kadanoff Center for Theoretical Physics,}\vspace{0.1cm} \\ 
\small{\it \textsuperscript{3}University of Chicago, Chicago, IL 60637, USA} \vspace{0.3cm}\\
\href{mailto:p.figueras@qmul.ac.uk}{\small{\texttt{p.figueras@qmul.ac.uk}}}, 
\href{mailto:a.kovacs@qmul.ac.uk}{\small{\texttt{a.kovacs@qmul.ac.uk}}},
\href{mailto:s.yao@qmul.ac.uk}{\small{\texttt{s.yao@qmul.ac.uk}}}

\end{center}

\date{}

\vskip 1.5 cm

\begin{abstract}
    We study properties of a recently proposed regularisation scheme to formulate the initial value problem for general (relativistic) effective field theories (EFTs) with arbitrary higher order equations of motion. We consider a simple UV theory that describes a massive and a massless scalar degree of freedom. Integrating out the heavy field gives rise to an EFT for the massless scalar. By adding suitable regularising terms to the EFT truncated at the level of dimension-$4$ and dimension-$6$ operators, we show that the resulting regularised theories admit a well-posed initial value problem. The regularised theories are related by a field redefinition to the original truncated EFTs and they propagate massive ghost fields (whose masses can be chosen to be of the order of the UV mass scale), in addition to the light field. We numerically solve the equations of motion of the UV theory and those of the regularised EFTs in $1+1$-dimensional Minkowski space for various choices of initial data and UV mass parameter. When derivatives of the initial data are sufficiently small compared to the UV mass scale, the regularised EFTs exhibit stable evolution in the computational domain and  provide very accurate approximations of the UV theory. On the other hand, when the initial gradients of the light field are comparable to the UV mass scale, the effective field theory description breaks down and the corresponding regularised EFTs exhibit ghost-like/tachyonic instabilities. Finally, we also formulate a conjecture on the global nonlinear stability of the vacuum in the regularised scalar EFTs in $3+1$ dimensions. These results suggest that the regularisation approach  provides a consistent classical description of the UV theory in a regime where effective field theory is applicable. 
\end{abstract}

\end{titlepage}


\newpage
\setcounter{page}{1}

\tableofcontents

\section{Introduction}
\label{sec:intro}
Effective field theories (EFTs) are widely used in physics as they offer a systematic and predictive framework for describing corrections to low-energy (long-distance) physics that may arise due to unknown physics at high energies (see e.g.,\,\cite{Burgess:2020tbq} for a review). 

From a top-down point of view, given a full UV-complete theory containing both heavy and light fields, one can build an EFT for the light fields by integrating out the heavy degrees of freedom. From the bottom-up perspective, EFTs are defined by three basic ingredients: 1) the low energy degrees of freedom, 2) the low energy symmetries that specify the allowed interactions, and 3) a power counting scheme that estimates the size of the possible interactions. Given these three ingredients, the EFT can be constructed by enumerating all possible terms in an effective action consistent with 1) and 2), and organise these terms in a series expansion dictated by the power counting scheme. The simplest example of a power counting scheme is a derivative expansion with a single microscopic length scale $\ell_\text{UV}$ characterising the UV physics.\footnote{Of course, there are EFTs with several characteristic scales and different power counting schemes, see e.g., \cite{Calderon-Infante:2025ldq} and \cite{Cabass:2022avo} for a review on EFTs in cosmology.} In this case, each possible higher derivative interaction in the effective action is multiplied by a dimensionless coefficient (determined by the UV theory) and a suitable power of $\ell_\text{UV}$ (determined by dimensional analysis). 

At any given order in the EFT expansion, there are only a finite number of terms that one can write down. In problems with a large separation of scales, accurate predictions can be made by truncating the series at a low order, allowing one to work with only a handful of interactions. It is then possible to measure or constrain the unknown couplings of the EFT by comparing its predictions against observations, which in turn may reveal valuable information about the underlying principles of the UV theory (see e.g., \cite{Adams:2006sv} or \cite{deRham:2022hpx} for a more recent review).

In this paper we study properties of the classical equations of motion of EFTs. This is motivated by known examples of EFTs where classical higher derivative corrections play an important role, such as in applications of viscous relativistic hydrodynamics in accretion disks around black holes or in the quark/gluon plasma produced in heavy-ion collisions in particle accelerators (see e.g., \cite{Kovtun:2012rj,Romatschke:2009im} for reviews). Another situation where the classical higher derivative corrections might be relevant is in strong gravity: there has been considerable recent discussion in the literature on the possible relevance of EFT extensions of General Relativity in cosmological settings \cite{Cabass:2022avo} and in binary mergers of compact objects \cite{Endlich:2017tqa}.

In order to make predictions about the classical nonlinear dynamics of an EFT, the equations of motion of the theory should admit a well-posed initial value problem. This is a highly nontrivial requirement for the following reason. When using an EFT, we are ignorant about the high frequency behaviour of the theory (beyond a cutoff) since we are only interested in making reliable predictions about the low frequency (low energy) regime of the theory. On the other hand, a well-posed Cauchy problem requires the equations of motion of the EFT to have good behaviour for arbitrarily high frequencies, which generic truncated EFTs usually do not have. Another (but related) obstruction to well-posedness is that the equations of motion of most EFTs are higher than second order in derivatives of the light field(s). Such equations require Cauchy data for higher derivatives of the light field(s), and thus they appear to have more propagating degrees of freedom. It has been argued (see e.g., \cite{Woodard:2015zca}) that generic initial data in higher derivative theories will give rise to ``runaway'' solutions that blow up on a microscopic timescale (i.e., shorter than any other typical timescale in the low energy system). These solutions are usually interpreted as unphysical artifacts of the truncation scheme that need to be eliminated. It seems unlikely, however, that such solutions can be eliminated in general by merely a good choice of initial data.

This problem inspired several works on constructing methods to extract physical solutions of theories with higher than second order equations of motion, see e.g.,\,\cite{Solomon:2017nlh,Flanagan:1996gw} for a survey of some of these methods. One option is to construct solutions perturbatively in a derivative expansion \cite{Simon:1990ic,Simon:1990jn,Okounkova:2019dfo}. In this approach, solutions are constructed iteratively: at every order in the perturbative expansion, the equations to be solved are governed by the zeroth order differential operator (say a wave operator in relativistic theories) and are sourced by the higher derivative terms evaluated on the solution obtained at the previous iteration steps. This method has the advantage of having a well-posed Cauchy problem (provided that the Cauchy problem in the zeroth order theory is well-posed) and the solutions do not exhibit runaway behaviour. However, the error of the iterative scheme exhibits polynomial growth in time (see \cite{Reall:2021ebq} for a rigorous account of this), leading to a secular breakdown of the scheme, even in a regime where the EFT equations should still be valid. Hence, this approach can only provide an accurate approximation to the UV solution for a short period of time (see, however, \cite{GalvezGhersi:2021sxs} for a more refined version of this method). The reason for the secular growth is that solutions of a higher order EFT need not be close to solutions of the zeroth order theory. Even though the relative size of the higher order EFT corrections needs to be small in the equations of motion, these small deviations may accumulate over time and become significant.

An alternative method is the so-called reduction of order \cite{Flanagan:1996gw,Jaen:1986iz,Eliezer:1989cr,Parker:1993dk}. In this case one fixes a slicing of the spacetime and derives a modified version of the equations of motion that has the following two properties: (i) the modified equations are perturbatively equivalent to the original higher derivative equations and, (ii) the modified equations are second order in time derivatives (but, in general, higher order in spatial derivatives). Such a modified equation can be derived iteratively for any higher than second order equation (see e.g., \cite{Flanagan:1996gw} for details). The appeal of this method is that it produces nonlinear equations that only require two pieces of Cauchy data for the light fields, thus eliminating the spurious degrees of freedom mentioned above. On the other hand, the procedure depends on fixing a spacetime slicing which breaks covariance, and hence it is unclear whether it is even possible to make (at least approximately) gauge invariant predictions with this scheme. Furthermore, as mentioned above, the order reduction is generally only possible in derivatives w.r.t. the time coordinate (or any single coordinate) so the resulting modified equations will still contain higher than second order spatial derivatives. Depending on the sign of the coefficients of the higher order spatial derivatives (determined by the EFT couplings) the resulting equations may be either of dispersive type (similar to the Schr\"odinger equation or the Korteweg-De Vries equation) or of the time-reversed diffusive type. In the latter case, the Cauchy problem will be ill-posed (as in the case of the time-reversed heat equation). In the dispersive case, one may be able to make progress (see e.g.,\,\cite{Rubio:2023eva}) but the nature of the Cauchy problem is unclear since very little is known about the mathematical properties of general higher derivative nonlinear dispersive PDEs. 

A third approach to deal with higher order equations of motion emerged originally in the context of relativistic viscous hydrodynamics. The Cauchy problem for equations of relativistic viscous fluids in the usual Landau or Eckart frames is famously ill-posed. Refs. \cite{Muller:1967aa,Israel:1976tn,1979AnPhy.118..341I} (and more recently \cite{Baier:2007ix}) found a well-posed initial value formulation for relativistic viscous fluids, known as the MIS formulation. By adding higher derivative terms to the stress-energy tensor of the fluid beyond the first order viscous terms, the conservation equation becomes third order in time derivatives. Higher derivatives of the thermodynamic fields can be absorbed into the viscous tensor, which is promoted to a new dynamical variable. Prescribing a Maxwell-Cattaneo-type relaxation equation (together with a suitable choice of relaxation timescale) for the viscous tensor yields a first order hyperbolic system of PDEs. In practice, the relaxation timescale in the driver equation for the viscous tensor can be chosen to be much smaller than any other macroscopic timescale in the system so that the particular choice of this parameter does not significantly affect the macroscopic evolution of the fluid. One may object to this seemingly ad-hoc modification of the fluid equations but in the context of hydrodynamics this approach is backed up by both physical and mathematical considerations. On the physics side, the relaxation time introduced in the MIS scheme turns out to be a physically meaningful transport coefficient of the fluid that can be computed from the UV theory (e.g., for holographic fluids, it has been computed from $\mathcal{N}=4$ super Yang-Mills theory \cite{Baier:2007ix}). On the mathematics side, Geroch and Lindblom showed, in a remarkable series of papers \cite{Geroch:1995bx,Lindblom:1995gp}, that a class of ad-hoc modifications of the fluid equations preserves the physical content of the theory, provided that the modified equations satisfy the following conditions: (i) the modified system of equations is strongly/symmetric hyperbolic, (ii) stationary solutions of the modified equations agree with those of the perfect fluid equations, (iii) the stress tensor and the particle-number current are conserved in the modified equations, (iv) the modifying terms in the equations satisfy a dissipation condition. In particular, the MIS theory for relativistic viscous fluid dynamics satisfies these four conditions. Motivated by the success of the MIS approach to hydrodynamics, \cite{Cayuso:2017iqc,Allwright:2018rut} proposed to extend this formulation to higher derivative theories of gravity, and subsequent numerical works have applied it to various gravitational theories \cite{Bezares:2021yek,Corman:2024cdr,Cayuso:2023xbc,Lara:2024rwa,Rubio:2024ryv}. Despite these promising developments, there are still some open questions regarding the validity of an MIS-type approach to gravity, where the corresponding physical and mathematical supporting evidence (analogous to the ones mentioned above in the case of hydrodynamics) is not yet available.

In this paper, we take a fourth approach to solving the equations of motion of EFTs. This is based on the observation that in an EFT there are two types of redundancies at the level of the action. Firstly, total derivatives do not affect the equations of motion, and hence such terms can be removed from the action. Secondly, certain terms can be removed (or added) by performing perturbative field redefinitions. It is well-known that EFTs related by perturbative field redefinitions describe the same low energy physics (see e.g., \cite{Burgess:2020tbq,Grosse-Knetter:1993tae}, and also \cite{Chisholm:1961tha,Kamefuchi:1961sb,Georgi:1991ch,Arzt:1993gz} for related statements on the invariance of the S-matrix). More recently, it was discovered by \cite{Bemfica:2017wps,Kovtun:2019hdm,Bemfica:2019knx,Bemfica:2020zjp} in the context of first order relativistic viscous hydrodynamics and by \cite{Figueras:2024bba} in the context of a general class of gravitational theories that field redefinitions can be exploited to modify the character of the equations of motion and obtain a well-posed initial formulation of EFTs. In particular, field redefinitions that perturbatively shift the light fields by terms that vanish on-shell in the zeroth order theory introduce a special set of higher derivative terms into the equations. These higher derivative terms inherit the good high frequency behaviour of the zeroth order equations of motion.

It is instructive to clarify some of the subtle points of the regularisation scheme of \cite{Figueras:2024bba}. First of all, as pointed out in \cite{Figueras:2024bba}, the regularised higher derivative equations are only weakly hyperbolic, which means that well-posedness is sensitive to some (but not all) of the lower order derivative terms in the equations. However, a careful analysis reveals that the relevant lower derivative terms have a special structure (inherited from the zeroth order equations through the field redefinitions) and the equations of motion of the regularised EFTs admit a (locally) well-posed Cauchy problem. In fact, \cite{Figueras:2024bba} also shows that it is possible to rewrite the regularised higher derivative equations as a system of second order wave equations for a set of auxiliary variables, i.e., in a manifestly symmetric hyperbolic form.

The second remark has to do with the choice of initial data. A feature of the regularisation scheme is that it artificially enlarges the phase space of the theory. However, as mentioned above, only a subset of the phase space is considered to be physically sensible, and solutions arising from arbitrary data are still expected to blow up in finite time. Therefore, if we want to capture the solutions of the regularised theory that are consistent with the EFT regime then we cannot set up initial data for the additional (artificial) degrees of freedom arbitrarily. The data for the extra degrees of freedom should be determined by the IR degrees of freedom. The most straightforward way to achieve this is to use a perturbative order reduction on the initial data slice, see Section \ref{sec:init_dat_EFTs} for more details.

Finally, the regularisation scheme introduces a number of massive ghost fields. One may be concerned that despite our results on local well-posedness and a careful choice of initial data, solutions may still exhibit runaway behaviour. Indeed, due to the nonlinear interactions between the light fields and the ghost fields, the latter will inevitably get excited to some extent during the evolution. Then, energy conservation would allow for a runaway solution, since the ghost fields can carry arbitrarily negative energy while the light fields can carry arbitrarily large positive energy. However, this argument is not a proof of the inevitability of a cascading behaviour for all EFT-compatible initial data, and there may be dynamical obstructions that prevent it. In fact, recent works \cite{Salvio:2019ewf,Deffayet:2023wdg,Held:2023aap,Deffayet:2025lnj} and the results of this paper support this claim. We will return to this issue in the Discussion section (see the subsection `On the global nonlinear stability of the vacuum').

The purpose of this paper is to investigate the regularisation scheme of \cite{Figueras:2024bba} in the context of a simple UV theory (also studied by \cite{Burgess:2014lwa,Allwright:2018rut,Reall:2021ebq}) describing a complex scalar field in a $U(1)$-symmetric potential. This theory exhibits spontaneous symmetry breaking and can be rewritten in terms of a massive real scalar and a massless Goldstone boson (also a real scalar field). Integrating out the heavy field gives rise to an EFT for the massless field. We show how the method of \cite{Figueras:2024bba} applies to various truncations of the resulting EFT. We numerically solve the UV theory and the regularised EFTs and compare the respective solutions for a number of initial data and parameter choices. The rest of the paper is organised as follows. In Section \ref{sec:UV_theory} we introduce the UV theory and the corresponding EFT in more detail. In Section \ref{sec:EFTs} we discuss the regularisation procedure for the EFT truncated at the level of the leading and next-to-leading order corrections to the two-derivative theory in the derivative expansion. In Section \ref{sec:norms} we introduce various norms and conserved charges that we use to monitor the behaviour of the solutions. In Section \ref{sec:numerics} we present the results of our numerical experiments on tests of the various EFTs and their regime of applicability. We summarise and discuss our findings in Section \ref{sec:discussion}. We provide some technical details and consistency checks in the Appendix \ref{sec:consistency_chks}. In Appendix \ref{sec:alt_formulation} we provide an alternative (and equivalent) formulation of one of the EFTs with explicit Klein-Gordon-type equations for the auxiliary massive modes.

\textit{Note added:} As this paper was nearing completion, it has come to our attention that another work \cite{in_prep} studying properties of the regularisation scheme in a different scalar field model is in preparation.

\section{The model}
\label{sec:UV_theory}
In this paper we consider the various EFTs that emerge in a certain low energy limit of  the Abelian-Higgs model. This model is particularly illustrative because the full UV theory is exactly solvable, making it a useful setting for quantitatively assessing how well the various EFTs capture the low energy physics of interest. 

The UV theory on a fixed $d$-dimensional Minkowski background is given by the action \cite{Solomon:2017nlh,Allwright:2018rut,Reall:2021ebq},
\begin{equation}
S=-\int {\rm d}^dx \left[ (\partial^a\phi^\ast)(\partial_a\phi)+V(|\phi|^2)\right]\,,\label{eq:UV_original_action}
\end{equation}
where $\phi$ is a complex scalar field with a potential $V(|\phi|^2)$
\begin{equation}
V(|\phi|^2) = \frac{\lambda}{2}\left(\phi^\ast \phi - \frac{v^2}{2}\right)^2\,.
\end{equation}
The vacuum of the theory is at the minimum of the potential $V$ and it corresponds to $\phi^\ast \phi =\frac{v^2}{2}$. A specific choice of the vacuum state spontaneously breaks the $U(1)$ symmetry $\phi\to e^{\text{i}\alpha}\phi$, leading to a particle spectrum containing a massive field with mass $M^2=\lambda\,v^2$ together with a massless Goldstone boson \cite{Burgess:2014lwa}. To see this, one introduces real (dimensionless) fields $\rho(x)$ and $\theta(x)$ such that
\begin{equation}
\phi(x) = \frac{v^2}{\sqrt{2}}\left[1+\rho(x)\right]\,e^{\text{i}\theta(x)}\,.\label{eq:phi_param}
\end{equation}
In terms of these fields, the action becomes
\begin{align}
\frac{S}{v^2} =& -\int {\rm d}^d x\left[ \frac{1}{2}\,(\partial_a\rho)(\partial^a\rho) + \frac{1}{2}(1+\rho)^2(\partial_a\theta)(\partial^a\theta) + V(\rho)\right]\,,\label{eq:UV_action}\\
V(\rho) =&~\frac{M^2}{2}\left(\rho^2 + \rho^3 + \frac{1}{4}\,\rho^4\right)\,. \label{eq:UV_potential}
\end{align}
In this parametrisation, we see that $\rho$ is a massive field with mass $M$ while $\theta$ is the massless Goldstone. The classical equations of motion can be written as a system of two nonlinear wave equations:
\begin{align}
\Box \rho =&~(1+\rho)(\partial_a\theta)(\partial^a\theta)+ V'(\rho)\,, \label{eq:eq_rho_UV}\\
\Box\theta = &-\frac{2}{1+\rho}\,(\partial_a\rho)(\partial^a\theta)\,. \label{eq:eq_theta_UV}
\end{align}
The Cauchy problem for this system is at least locally well-posed. As pointed out in \cite{Reall:2021ebq}, even though global solutions exist for the theory \eqref{eq:UV_original_action}, the same may not be true for \eqref{eq:eq_rho_UV}-\eqref{eq:eq_theta_UV} since the parametrisation \eqref{eq:phi_param} might break down at a certain point. Nevertheless, we do not observe such a breakdown for the initial data choices discussed later in the paper. 

\subsection{EFT expansion}
\label{sec:EFT_expansion}
Following \cite{Reall:2021ebq}, we will motivate the EFT expansion in an heuristic way, which will also be useful to identify a class of initial conditions for $\rho$ and $\theta$ that will allow us to solve \eqref{eq:eq_rho_UV}--\eqref{eq:eq_theta_UV} while ensuring that these fields are  in the EFT regime (at least for short enough times).  

We start by \textit{assuming} that $\theta$ and $M\,\rho$ and all their derivatives are uniformly bounded as $M\to\infty$.\footnote{Ref. \cite{Reall:2021ebq} proves that, if the initial data for $\rho$ and $\theta$ satisfies this condition, then it continues to hold for the solution at later times. See also \cite{Kadar:2022qkq} for a related rigorous study.} To proceed, we rearrange \eqref{eq:eq_rho_UV} as
\begin{equation}
    \rho = -\frac{1}{M^2}(\partial_a\theta)(\partial^a\theta)-\frac{\rho}{M^2}(\partial_a\theta)(\partial^a\theta)+\frac{1}{M^2}\Box\rho-\rho^2\,W(\rho)\,, \label{eq:eq_rho_UV2}
\end{equation}
with
\begin{equation}
    W(\rho)=\frac{3}{2}+\frac{1}{2}\,\rho\,.
\end{equation}
The assumption of uniform boundedness implies that $\partial_a\theta = O(1)$, $\rho=O(M^{-1})$, $\partial_a\rho = O(M^{-1})$, and $\partial_a\partial_b\rho = O(M^{-1})$. The the first term on the r.h.s. of \eqref{eq:eq_rho_UV2} is $O(M^{-2})$, the second term is $O(M^{-3})$, the third term is also $O(M^{-3})$ and the fourth term is $O(M^{-2})$. Therefore, we conclude that in fact $\rho=O(M^{-2})$, which allows to improve the estimate of the last term to $O(M^{-4})$. Then, the equation of motion for $\rho$, eq. \eqref{eq:eq_rho_UV2}, at $O(M^{-2})$ gives
\begin{equation}
    \rho = \frac{1}{M^2}\,\mathcal{F}_2 + O(M^{-3})\,, 
\end{equation}
with
\begin{equation}
    \mathcal{F}_2 \equiv -(\partial_a\theta)(\partial^a\theta)\,.
    \label{eq:def_F2}
\end{equation}
Now, taking a derivative of \eqref{eq:eq_rho_UV2}:
\begin{equation}
\begin{aligned}
    \partial_b\rho =& -\frac{1}{M^2}\partial_b[(\partial_a\theta)(\partial^a\theta)]-\frac{\rho}{M^2}\partial_b[(\partial_a\theta)(\partial^a\theta)]-\frac{1}{M^2}(\partial_b\rho)(\partial_a\theta)(\partial^a\theta)+\frac{1}{M^2}\partial_b\Box\rho\\
    &-\frac{3}{2}\rho(2+\rho)(\partial_b\rho)\,,
\end{aligned}
\end{equation}
implies that
\begin{equation}
    \partial_a\rho = \frac{1}{M^2}\partial_a\mathcal{F}_2 + O(M^{-3})\,.
\end{equation}
Similarly, for the second derivatives of $\rho$ we get the estimate
\begin{equation}
    \partial_a\partial_b\rho = \frac{1}{M^2}\partial_a\partial_b\mathcal{F}_2 + O(M^{-3})\,,
\end{equation}
and so on for the higher derivatives of $\rho$. Combining these estimates for the derivatives of $\rho$, we can improve the approximation for $\rho$ itself as follows. Note that,
\begin{equation}
    \rho^2W(\rho) = \frac{3}{2}\rho^2+\frac{1}{2}\rho^3 = \frac{3}{2\,M^4}\mathcal{F}_2 + O(M^{-5})\,.
\end{equation}
Then, substituting the estimates at this order back into \eqref{eq:eq_rho_UV2} gives,
\begin{equation}
    \rho = \frac{1}{M^2}\mathcal{F}_2 +\frac{1}{M^4}\,\mathcal{F}_2^2 + \frac{1}{M^4}\Box\mathcal{F}_2 -\frac{3}{2\,M^4}\,\mathcal{F}_2^2 + O(M^{-5})\,,
\end{equation}
from which it follows that 
\begin{equation}
    \rho = \frac{1}{M^2}\mathcal{F}_2 +\frac{1}{M^4}\,\mathcal{F}_4 + O(M^{-5})\,, \label{eq:rho_expansion}
\end{equation}
with
\begin{equation}
    \mathcal{F}_4\equiv\Box\mathcal{F}_2-\frac{1}{2}\mathcal{F}_2^2\,.
    \label{eq:def_F4}
\end{equation} 

Proceeding in the same way, at the next order one has,
\begin{equation}
\rho = \frac{1}{M^2}\mathcal{F}_2 +\frac{1}{M^4}\,\mathcal{F}_4 + \frac{1}{M^6}\,\mathcal{F}_6+O(M^{-7})\,, \label{eq:rho_expansion2}
\end{equation}
with
\begin{equation}
    \mathcal{F}_6\equiv\Box\mathcal{F}_4-2\,\mathcal{F}_2\,\mathcal{F}_4-\frac{1}{2}\mathcal{F}_2^3\,.
    \label{eq:def_F6}
\end{equation} 
One can iterate this process to find a full asymptotic expansion for $\rho$.  

With the uniform boundedness assumption for $\partial_a\theta$, we can similarly obtain an asymptotic expansion for $\Box\theta$, which results in a higher derivative equation of motion for $\theta$. To see this, Taylor-expanding the r.h.s. of \eqref{eq:eq_theta_UV} in $\rho$ and substituting \eqref{eq:rho_expansion} gives
\begin{equation}
\begin{aligned}
    \Box\theta\approx& -2\left(1-\rho+\rho^2+\ldots\right)(\partial_a\rho)(\partial^a\theta) \\
    \approx&-2\left[1-\left(\frac{1}{M^2}\mathcal{F}_2 +\frac{1}{M^4}\,\mathcal{F}_4 + \ldots\right)+\left(\frac{1}{M^2}\mathcal{F}_2 + \ldots\right)^2\right]\partial_a\left(\frac{1}{M^2}\mathcal{F}_2 +\frac{1}{M^4}\,\mathcal{F}_4 + \ldots\right)(\partial^a\theta)\\
    =&-\frac{2}{M^2}(\partial_a\mathcal{F}_2)(\partial^a\theta)\\
    &-\frac{2}{M^4}\partial_a\left(\mathcal{F}_4-\tfrac{1}{2}\mathcal{F}_2^2\right)(\partial^a\theta)\\
&-\frac{2}{M^6}\partial_a\left(\mathcal{F}_6-\mathcal{F}_2\,\mathcal{F}_4+\tfrac{1}{3}\mathcal{F}_2^3\right)(\partial^a\theta) +O(M^{-7}) \,.
\end{aligned}
\label{eq:boxtheta_expansion}
\end{equation}
Note that the uniform boundedness assumption allows us to take derivatives of \eqref{eq:boxtheta_expansion} and thereby obtain an expansion in powers of $1/M$ for the derivatives of $\Box\theta$. From \eqref{eq:boxtheta_expansion} it follows that $\partial^k\Box\theta\sim O(M^{-2})$. The equations of motion for $\theta$ obtained in this way do not derive from a Lagrangian, which implies that they will not give rise to a conserved energy (or other conserved charges).

We can also ``integrate out''  $\rho$ at the level of the action by plugging \eqref{eq:rho_expansion} into the action of the UV theory, \eqref{eq:UV_action}, and expanding it in powers of $1/M$ to obtain the low energy effective action for light field $\theta$. Being able to derive the EFT from an action gives extra structure to the EFT; more specifically, it guarantees the existence of a conserved energy (and other conserved charges). Up to $O(M^{-6})$, we get,
\begin{equation}
\begin{aligned}
\frac{S}{v^2}\approx -\int {\rm d}^dx\bigg\{&-\frac{1}{2}\,\mathcal{F}_2 - \frac{1}{2M^2}\,\mathcal{F}_2^2+\frac{1}{2M^4}(\partial_a\mathcal{F}_2)(\partial^a\mathcal{F}_2) \\
&+ \frac{1}{M^6}\Big[(\partial^a\mathcal{F}_2)(\partial_a\mathcal{F}_4)+\tfrac{1}{8}(2\,\mathcal{F}_4+\mathcal{F}_2^2)^2\Big] + O(M^{-7})\bigg\}\,.
\end{aligned}
\label{eq:low_energy_action}
\end{equation}

The equation of motion for $\theta$ can now be obtained by varying this action at the desired order in the $1/M$ expansion, and it should agree, in the EFT sense, with \eqref{eq:boxtheta_expansion}. At $O(M^{-2})$ we get:
\begin{equation}
\begin{aligned}
\Box\theta =&-\frac{2}{M^2}(\partial_a\mathcal{F}_2)(\partial^a\theta)\\
&-\frac{2}{M^2}\,\mathcal{F}_2(\Box\theta)\,.
\end{aligned}
\label{eq:eoms_M2_gen_action}
\end{equation}
Note that the term in the first line of the r.h.s. coincides with the first term on the r.h.s. of \eqref{eq:boxtheta_expansion}, which we obtained from directly integrating out $\rho$ at the level of the equations of motion. However, under the uniform boundedness assumptions, the term in the second line of the r.h.s. above is $O(M^{-4})$ because $\Box\theta\sim O(M^{-2})$. Therefore, we see that the equations of motion that we obtained from varying the low energy effective action at $O(M^{-2})$ agree with \eqref{eq:boxtheta_expansion} at $O(M^{-2})$ but they include an additional sub-leading $O(M^{-4})$ term. At $O(M^{-4})$ we get:
\begin{equation}
\begin{aligned}
\Box\theta =&-\frac{2}{M^2}(\partial_a\mathcal{F}_2)(\partial^a\theta)\\
&-\frac{2}{M^4}\partial_a\left(\mathcal{F}_4-\tfrac{1}{2}\mathcal{F}_2^2\right)(\partial^a\theta)\\
&
-\frac{2}{M^2}\,\mathcal{F}_2\Big[\Box\theta + \frac{2}{M^2}(\partial_a\mathcal{F}_2)(\partial^a\theta)\Big]\\
&-\frac{2}{M^4}\left(\mathcal{F}_4+\tfrac{1}{2}\,\mathcal{F}_2^2\right)(\Box\theta)+ O(M^{-6})\,,
\end{aligned}
\end{equation}
which again, contains the same terms as \eqref{eq:boxtheta_expansion} up to $O(M^{-4})$, plus two additional terms on the third and fourth lines; these terms are proportional to the lower order equations of motion and hence they are $O(M^{-6})$. At $O(M^{-6})$ we observe the same pattern:
\begin{equation}
\begin{aligned}
\Box\theta =&-\frac{2}{M^2}(\partial_a\mathcal{F}_2)(\partial^a\theta)\\
&-\frac{2}{M^4}\partial_a\left(\mathcal{F}_4-\tfrac{1}{2}\mathcal{F}_2^2\right)(\partial^a\theta)\\
&-\frac{2}{M^6}\partial_a\left(\mathcal{F}_6-\mathcal{F}_2\,\mathcal{F}_4+\tfrac{1}{3}\mathcal{F}_2^3\right)(\partial^a\theta)\\
&-\frac{2}{M^2}\,\mathcal{F}_2\Big[\Box\theta + \frac{2}{M^2}(\partial_a\mathcal{F}_2)(\partial^a\theta)+\frac{2}{M^4}\partial_a\left(\mathcal{F}_4-\tfrac{1}{2}\mathcal{F}_2^2\right)(\partial^a\theta)\Big] \\
&-\frac{2}{M^4}\left(\mathcal{F}_4+\tfrac{1}{2}\,\mathcal{F}_2^2\right)\Big[ \Box\theta + \frac{2}{M^2}(\partial_a\mathcal{F}_2)(\partial^a\theta)\Big] \\
&-\frac{2}{M^6}(\mathcal{F}_6+\mathcal{F}_2\,\mathcal{F}_4)(\Box\theta) + O(M^{-8})\,,
\end{aligned}
\end{equation}
and so on to all orders in perturbation theory.

\subsection{Initial data}
Initial data for \eqref{eq:eq_rho_UV}--\eqref{eq:eq_theta_UV} is simply given by $\rho|_{t=0}$, $\partial_t\rho|_{t=0}$, $\theta|_{t=0}$, $\partial_t\theta|_{t=0}$, all of which can be freely specified.\footnote{Here we are not concerned with identifying classes of initial conditions that give rise to global (in time) solutions of \eqref{eq:eq_rho_UV}--\eqref{eq:eq_theta_UV}.} Since in this article we will be interested in comparing various EFTs with the UV theory, we need to identify initial data for the UV theory that is in the EFT regime. Following \cite{Reall:2021ebq}, we require $\theta_0\equiv\theta|_{t=0}$, and $\theta_1\equiv\partial_t\theta|_{t=0}$ to be $O(1)$. Then,  we can compute $\rho|_{t=0}$, $\partial_t\rho|_{t=0}$ from the EFT expansion using $\theta_0$ and $\theta_1$ and their derivatives. For instance,  $\rho|_{t=0}$ and  $\partial_t\rho|_{t=0}$ up to $O(M^{-4})$ are obtained from  \eqref{eq:rho_expansion} evaluated at $t=0$:
\begin{align}
\rho|_{t=0}=&~\left(\frac{1}{M^2}\mathcal{F}_2 +\frac{1}{M^4}\,\mathcal{F}_4\right)\bigg|_{t=0} + O(M^{-5})\,, \label{eq:rho_0}\\
\partial_t\rho|_{t=0}=&~\left(\frac{1}{M^2}\partial_t\mathcal{F}_2 +\frac{1}{M^4}\,\partial_t\mathcal{F}_4\right)\bigg|_{t=0} + O(M^{-5}) \label{eq:rho_1}\,.
\end{align}
This procedure to construct consistent initial data for the UV theory requires second and higher time derivatives of $\theta$ at $t=0$. Those can be computed to the required order in $1/M$ from equation \eqref{eq:boxtheta_expansion} and its time derivatives. For instance, in 1+1 dimensions and up to $O(M^{-4})$ we get
\begin{align}
    &\rho|_{t=0}=\frac{1}{M^2}\left[(\theta_1)^2-(\partial_x\theta_0)^2\right]
        +\frac{1}{M^4}\left[4\big((\partial_x\theta_1)^2-(\partial_x^2\theta_0)^2\big)-\tfrac{1}{2}\big((\theta_1)^2-(\partial_x\theta_0)^2\big)^2\right]\,,\\
        &\partial_t\rho|_{t=0}=\frac{2}{M^2}\left[\theta_1\partial_x^2\theta_0-\partial_x\theta_1\,\partial_x\theta_0 \right]\nonumber\\
        &\hspace{1.25cm}+\frac{2}{M^4}\left[\partial_x\theta_1\left(4\,\partial_x^3\theta_0-(\partial_x\theta_0)^3+9\,\theta_1^2\,\partial_x\theta_0\right)-\partial_x^2\theta_0\left(4\,\partial_x^2\theta_1+5\,\theta_1^3+3\,\theta_1(\partial_x\theta_0)^2\right)\right]\,.
\end{align}

\section{Effective field theories}
\label{sec:EFTs}
In this section we consider different low energy EFTs. As it is often the case in practical applications, one truncates the EFT expansion at some finite order and then constructs solutions of the truncated equations. For practical purposes we shall consider EFTs that are valid up to $O(M^{-4})$ in the expansion. In this case, the equations of motion are of fourth order in derivatives of the light field $\theta$. However, we emphasise that the regularisation scheme applies to EFTs with equations of motion of arbitrary high order.

\subsection{Linearisation}
\label{sec:linearisation}

Ref. \cite{Reall:2021ebq} studied solutions of the truncated EFT equations, \eqref{eq:boxtheta_expansion}, that can be constructed as a series expansion in inverse powers of $M$ up to some order, say, $M^{-2m}$:
\begin{equation}
    \theta = \sum_{k=0}^m \frac{\theta^{(2k)}}{M^{2k}}\,. \label{eq:theta_expansion}
\end{equation}
Note that in this approach there are two expansions: one is the EFT expansion of the equations of motion and the second one is the expansion of the light field $\theta$, \eqref{eq:theta_expansion}. In practice, the latter has to go up to sufficiently high order in $1/M$ to ensure that the error in the solution is dominated by the error of the truncation of the equations of motion.  

In this scheme, by plugging the expansion of the low energy field $\theta$, eq. \eqref{eq:boxtheta_expansion}, into the truncated equations of motion and collecting the powers of $1/M$, one finds that each term $\theta^{(2k)}$ in the expansion satisfies a linear wave equation with a source that depends on the lower order fields. Therefore, this scheme amounts to linearise \eqref{eq:boxtheta_expansion} order by order in perturbation theory and the equations that one has to solve at each order trivially admit a well-posed initial value problem. Henceforth, we shall refer to this EFT as EFT$_0$. Note that in a more general setting, e.g., in an EFT of gravity, the corresponding version of this scheme would be to construct solutions in a perturbative expansion by solving the non-linear Einstein equations sourced by the higher derivative terms evaluated on the lower order solutions (see e.g., \cite{Okounkova:2019dfo} for an example of this). 

For example, if we are interested in constructing a solution that is accurate up to $O(M^{-4})$, we need to truncate the equations of motion at this order while expanding $\theta$ up to at least $O(M^{-6})$. In this case, the evolution equations that we have to solve for the terms in \eqref{eq:theta_expansion} up to $O(M^{-6})$ are:
\begin{align}
    \Box\theta^{(0)} =&~ 0\,, \label{eq:theta0_EFT0}\\
    \Box\theta^{(2)} =&-2\,(\partial_a\mathcal{F}_2^{(0)})(\partial^a\theta^{(0)})\,, \label{eq:theta2_EFT0}\\
    \Box\theta^{(4)} =&-2\big[(\partial_a\mathcal{F}_2^{(0)})(\partial^a\theta^{(2)})+(\partial_a\mathcal{F}_2^{(2)})(\partial^a\theta^{(0)})\big]-2\,\partial_a\left(\mathcal{F}_4^{(0)}-\tfrac{1}{2}(\mathcal{F}_2^{(0)})^2\right)(\partial^a\theta^{(0)})\,,\\
    \Box\theta^{(6)} =&-2\big[(\partial_a\mathcal{F}_2^{(0)})(\partial^a\theta^{(4)})+(\partial_a\mathcal{F}_2^{(4)})(\partial^a\theta^{(0)})+(\partial_a\mathcal{F}_2^{(2)})(\partial^a\theta^{(2)})\big]\nonumber\\
    &-2\,\partial_a\left(\mathcal{F}_4^{(0)}-\tfrac{1}{2}(\mathcal{F}_2^{(0)})^2\right)(\partial^a\theta^{(2)}) - 2\,\partial_a\left(\mathcal{F}_4^{(2)}-\mathcal{F}_2^{(0)}\,\mathcal{F}_2^{(2)}\right)(\partial^a\theta^{(0)})\,,
    \label{eq:theta4_EFT0}
\end{align}
where
\begin{equation}
    \begin{aligned}
        \mathcal{F}_2^{(0)} =& -(\partial_a\theta^{(0)})(\partial^a\theta^{(0)})\,,\\
        \mathcal{F}_2^{(2)} =& -2\,(\partial_a\theta^{(2)})(\partial^a\theta^{(0)})\,,\\
        \mathcal{F}_2^{(4)} =& -2\,(\partial_a\theta^{(4)})(\partial^a\theta^{(0)})-(\partial_a\theta^{(2)})(\partial^a\theta^{(2)})\,,\\
        \mathcal{F}_4^{(0)} =& ~\Box\mathcal{F}_2^{(0)}-\tfrac{1}{2}(\mathcal{F}_2^{(0)})^2\,,\\
        \mathcal{F}_4^{(2)} =&~ \Box\mathcal{F}_2^{(2)}-\mathcal{F}_2^{(0)}\,\mathcal{F}_2^{(2)}\,.
    \end{aligned}
\end{equation}

The solutions to the EFT equations \eqref{eq:boxtheta_expansion} constructed with this approach suffer from secular error growth, and hence the validity of the expansion \eqref{eq:theta_expansion} is limited in time, see \cite{Reall:2021ebq}. Specifically, for solutions that are accurate up to $O(M^{-l})$, the error at time $t$ is given by \cite{Reall:2021ebq}:
\begin{equation}
    ||\theta_{\rm UV} - \theta_{\text{EFT}_0}||_{C_t^0 L^2_x} \leq \frac{C\,t^{\lfloor \frac{l}{2}\rfloor+2}}{M^{l+1}}\,,
    \label{eq:EFT0_secular_growth}
\end{equation}
where $C$ is a constant,  $t\leq \tilde{C}\,M^{\lambda}$ for some other constant $\tilde{C}$, and $0\leq \lambda<2$. The precise definition of the $C_t^0 L^2_x$ norm can found in Section \ref{sec:norms}. In Section \ref{sec:numerics} we perform some numerical experiments to verify the expected accuracy \eqref{eq:EFT0_secular_growth} of the EFT$_0$ solutions. 

\subsubsection{Initial data} 
Consistency with the EFT expansion in this case simply requires that $\theta^{(2k)}|_{t=0}$ and $\partial_t\theta^{(2k)}|_{t=0}$ are $O(1)$ $\forall k$. Then, for each equation in the expansion, e.g., \eqref{eq:theta0_EFT0}--\eqref{eq:theta4_EFT0} above, one can specify two free pieces of data. In the numerical experiments of Section \ref{sec:numerics}, where we compare this EFT with the UV theory, we will use the same initial conditions for $\theta^{(0)}$ and the UV $\theta$-field, and we set $\theta^{(2k)}|_{t=0}$ and $\partial_t\theta^{(2k)}|_{t=0}=0$ $\forall k>0$.

\subsection{Second order EFTs}
In this subsection we discuss EFTs that are accurate up to $O(M^{-2})$. The key difference between the solutions constructed with the linearisation scheme of Section \ref{sec:linearisation} and those in this Section is that here we provide schemes to construct fully non-linear solutions of the truncated equations of motion. As we shall see in Section \ref{sec:numerics}, this has important consequences for the scaling (in time) of the error terms. For the purposes of our presentation, it will be convenient to work at the level of the action and derive the equations of motion by varying the action.

The low energy action for the light field $\theta$ accurate up to $O(M^{-2})$ is \eqref{eq:low_energy_action}:
\begin{equation}
\frac{S}{v^2}  \approx -\int {\rm d}^d x\left[ \frac{1}{2}(\partial_a\theta)(\partial^a\theta) - \frac{1}{2M^2}\big((\partial_a\theta)(\partial^a\theta) \big)^2 + O(M^{-4})\right]\,. \label{eq:action_M2}
\end{equation}
The equations of motion that follow from varying this action with respect to $\theta$ are given by \eqref{eq:eoms_M2_gen_action}:
\begin{equation}
\Box\theta = \frac{2}{M^2}\left[ (\partial_a\theta)(\partial^a\theta)\Box\theta + 2(\partial^a\theta)(\partial^b\theta)\partial_a\partial_b\theta\right]
\label{eq:eoms_M2}
\end{equation}
We shall refer to the EFT defined by \eqref{eq:action_M2} and \eqref{eq:eoms_M2} as the EFT$_1$.

As we discussed in Section \ref{sec:EFT_expansion}, under the assumption of uniform boundedness of $\theta$ and its derivatives, $\Box\theta = O(M^{-2})$ and hence this term on the r.h.s. of \eqref{eq:eoms_M2} is subleading. Therefore, solutions to \eqref{eq:eoms_M2} should agree with the solutions of effective equations of motion for $\theta$ obtained by integrating out the heavy field $\rho$, eq. \eqref{eq:boxtheta_expansion}, up to $O(M^{-2})$:
\begin{equation}
    \Box\theta = \frac{4}{M^2}(\partial^a\theta)(\partial^b\theta)\partial_a\partial_b\theta\,.
\label{eq:eoms_M2_tmp}
\end{equation}

Equation \eqref{eq:eoms_M2} is a second order quasilinear equation for $\theta$ so the usual PDE theory on local existence and uniqueness of solutions of these equations applies. The evolution of $\theta$ is governed by the effective metric:
\begin{equation}
G^{ab} = \left(1-\frac{2}{M^2}(\partial\theta)^2\right)\eta^{ab} - \frac{4}{M^2}(\partial^a\theta)(\partial^b\theta)\,.
\label{eq:eff_metric_EFT1}
\end{equation}
This metric has Lorentzian signature as long as $(\partial\theta)^2<M^2/6$. Therefore, as long as this condition holds, \eqref{eq:eoms_M2} is strongly hyperbolic and we can ``straightforwardly'' solve the initial value problem. Note that to solve this EFT we can specify the same initial data for the $\theta$ field as in the UV theory; in particular, the EFT$_1$ only propagates one degree of freedom, namely the massless field $\theta$. The expectation is that the $O(M^{-2})$ terms in the equations encode the effects of the heavy field $\rho$ on the low energy dynamics of the light field $\theta$.

We can obtain another EFT that should provide an equivalent description of the dynamics of $\theta$ up to $O(M^{-2})$, following the prescription of \cite{Figueras:2024bba}. By making a perturbative field redefinition to $O(M^{-2})$,\footnote{In the regime of validity of EFT, these field redefinitions are clearly invertible.}
\begin{equation}
\theta \to \tilde\theta\equiv \theta + \frac{\alpha}{M^2}\,\Box\theta\,, \label{eq:field_redef_M2}
\end{equation}
where $\alpha$ is a dimensionless constant, the new action  to $O(M^{-2})$ becomes
\begin{equation}
\frac{S_\text{reg}}{v^2}  \approx -\int {\rm d}^d x\left[ \frac{1}{2}(\partial_a\theta)(\partial^a\theta) - \frac{\alpha}{M^2}\,\theta \Box^2\theta - \frac{1}{2M^2}\big((\partial_a\theta)(\partial^a\theta) \big)^2 + O(M^{-4})\right]\,.
\label{eq:reg_action_M2}
\end{equation}
The new equations of motion that follow from varying this action with respect to $\theta$ are:
\begin{equation}
\Box\theta = \frac{2}{M^2}\left[ (\partial_a\theta)(\partial^a\theta)\Box\theta + 2(\partial^a\theta)(\partial^b\theta)\partial_a\partial_b\theta\right] - \frac{2\,\alpha}{M^2}\,\Box^2\theta\,, \label{eq:eom_reg}
\end{equation}
Defining new variables,
\begin{equation}
    \theta^{(1,0)}_a \equiv \partial_a\theta\,,\quad \theta^{(0,1)} \equiv  \Box\theta\,, \label{eq:def_vars_M2}
\end{equation}
we can rewrite \eqref{eq:eom_reg} as:
\begin{align}
\Box \theta = &~\theta^{(0,1)}\,, \label{eq:eq_theta}\\
\Box\theta^{(1,0)}_a = &~\partial_a\theta^{(0,1)} \,, \label{eq:eq_theta10_M2}\\
\left(\Box +\frac{M^2}{2\,\alpha} \right)\theta^{(0,1)} =&~  \frac{1}{\alpha} \left( \theta^{(1,0)}_a\,\theta^{(1,0)a}\,\theta^{(0,1)} + 2\,\theta^{(1,0)a}\,\theta^{(1,0)b}\,\partial_a\theta^{(1,0)}_b\right) 
\label{eq:eq_box_theta}
\end{align}
We see that equations \eqref{eq:eq_theta}--\eqref{eq:eq_box_theta} constitute a diagonal system of wave equations for $\theta$, $\theta^{(1,0)}_a$ and $\theta^{(0,1)}$, and hence they have a locally well-posed initial value problem, regardless of the size of $\partial\theta$ compared to $M$. We shall refer to the EFT defined by the action \eqref{eq:reg_action_M2} and the equations of motion \eqref{eq:eq_theta}--\eqref{eq:eq_box_theta} as the EFT$_2$. 

Whilst the EFT$_1$ only propagates the massless field $\theta$, the EFT$_2$ propagates, in addition to $\theta$, another massive degree of freedom (corresponding to $\theta^{(0,1)}$ at the linearised level, see below). On the other hand, $\theta^{(1,0)}_a$ does not propagate any new degrees of freedom. To see this, note that in \eqref{eq:eq_theta10_M2} there are $d$ equations, one for each component of $\theta^{(1,0)}_a$ but the definition of this variable in \eqref{eq:def_vars_M2} imposes $d$ constraints. In fact, the equation of motion \eqref{eq:eq_theta10_M2} can be written as a sourceless wave equation for this constraint, so the constraints are trivially propagated, see Appendix \ref{sec:consistency_chks}.  

To identify the degrees of freedom that the EFT$_2$ propagates, it is useful to linearise \eqref{eq:eom_reg} around the trivial solution $\theta=0$ to obtain,
\begin{equation}
\Box\delta\theta+\frac{2\alpha}{M^2}\,\Box^2\delta\theta = 0\,.
\end{equation}
Considering plane wave solutions $\delta\theta=\Theta\, e^{\text{i}k\cdot x}$ with $\Theta=\text{const.}$, we obtain the dispersion relation
\begin{equation}
    k^2\left(1-\frac{2\alpha}{M^2}\,k^2\right)=0\,.
\end{equation}
This shows that the theory propagates a massless mode $\theta$ with $k^2=0$, and a new massive mode $\theta^{(0,1)}$ with 
\begin{equation}
    k^2=\frac{M^2}{2\alpha} \quad \Rightarrow \quad m^2=-\frac{M^2}{2\alpha}
\end{equation}
In order to prevent $\theta^{(0,1)}$ from being tachyonic (and hence to give rise to exponentially growing solutions) we need to impose $\alpha<0$. This condition on $\alpha$ must be satisfied in order to construct consistent solutions to \eqref{eq:eq_theta}--\eqref{eq:eq_box_theta} that remain bounded for sufficiently long times; the numerical experiments in Section \ref{sec:numerics} suggest that this is indeed the case. 

Note that the frequency of oscillation of $\theta^{(0,1)}$ is
\begin{equation}
\omega = \pm \sqrt{|\vec k|^2-\frac{M^2}{2\,\alpha}}\,,
\end{equation}
where $\vec k$ is the wavenumber. This implies that even for small wavenumbers, i.e., long wavelengths, this massive field will be rapidly oscillating with frequency $\omega\sim M$, as one would expect. Note, however, that to obtain solutions compatible with the EFT regime, the amplitudes of these oscillations must be sufficiently small, i.e., consistent initial data for this field must be at most $\theta^{(0,1)}\sim O(M^{-2})$ initially. As we shall see in Section \ref{sec:numerics}, for large enough $M$, and hence large enough mass $m$ of this field, solutions to \eqref{eq:eq_box_theta} remain $O(M^{-2})$ for long enough times. Therefore, in the regime of validity of the EFT expansion, effects of the heavy field $\theta^{(0,1)}$ on the dynamics of the light field $\theta$ through \eqref{eq:eq_theta} remain at the expected $O(M^{-2})$ at all times, which is consistent with the intuition that in this regime the heavy field does not get significantly excited.

\subsection{Regularised fourth order EFT}

Let us now consider the EFT truncated at one order higher. In this case, the Lagrangian is \cite{Burgess:2014lwa}
\begin{equation}
\frac{S}{v^2}  \approx -\int {\rm d}^d x\left[ \frac{1}{2}(\partial_a\theta)(\partial^a\theta) - \frac{1}{2M^2}\big((\partial_a\theta)(\partial^a\theta) \big)^2 + \frac{2}{M^4}\,(\partial_a\partial_b\theta)(\partial^a\partial^c\theta)(\partial^b\theta)(\partial_c\theta)  + O(M^{-6})\right]\,,
\label{eq:action_M4}
 \end{equation}
 and the equations of motion to $O(M^{-4})$ are \cite{Allwright:2018rut}
 \begin{equation}
 \begin{aligned}
 \Box\theta =&~\frac{2}{M^2}\partial_a\big[(\partial\theta)^2\partial^a\theta \big]\\ 
 &~+\frac{4}{M^4}\big[(\partial^a\partial^b\Box\theta)(\partial_a\theta)(\partial_b\theta) + (\Box\theta)(\partial^a\theta)(\partial_a\Box\theta)+(\partial^a\theta)(\partial_a\partial_b\theta)(\partial^b\Box\theta)\\
 &\hspace{1.5cm}+(\Box\theta)(\partial^a\partial^b\theta)(\partial_a\partial_b\theta)+2(\partial^a\theta)(\partial^b\partial^c\theta)(\partial_a\partial_b\partial_c\theta)\big]
 \label{eq:eoms_M4}
 \end{aligned}
 \end{equation}

These equations of motion contain up to fourth order derivatives of $\theta$ and it is not clear how to solve them, fully non-linearly, with standard techniques. The prescription of \cite{Figueras:2024bba} provides a way to construct solutions to \eqref{eq:eoms_M4} that are consistent with the EFT expansion. We make a perturbative field redefinition up to order $O(M^{-4})$:
\begin{equation}
    \theta \to \tilde\theta\equiv\theta + \frac{\alpha_1'}{M^2}\Big[ \Box\theta + \frac{2}{M^2}\partial_a\big((\partial\theta)^2\partial^a\theta \big) \Big] + \frac{\alpha_2'}{M^4}\,\Box^2\theta + O(M^{-5})\,,
    \label{eq:field_redef_M4}
\end{equation}
and substitute it into the original action up to $O(M^{-4})$, eq. \eqref{eq:action_M4}, which leads to the new action
\begin{equation}
\begin{aligned}
    \frac{S}{v^2}\approx-\int {\rm d}^d x\Big[&~ \frac{1}{2}(\partial_a\theta)(\partial^a\theta) - \frac{1}{2M^2}\big((\partial_a\theta)(\partial^a\theta) \big)^2 + \frac{2}{M^4}\,(\partial_a\partial_b\theta)(\partial^a\partial^c\theta)(\partial^b\theta)(\partial_c\theta) \\
&-\frac{\alpha_1}{M^2}\,\theta\Box^2\theta 
-\frac{\alpha_2}{M^4}\,\theta\Box^3\theta + O(M^{-6})\Big]\,,
\end{aligned} 
\label{eq:reg_action_M4}
\end{equation}
where $\alpha_1=\alpha_1'$ and $\alpha_2=\tfrac{1}{2}\alpha_1'^2+\alpha_2'$. Notice that in \eqref{eq:field_redef_M4} we introduced an $O(M^{-4})$ term proportional to $\alpha_1'$ that is not of the form $\Box^n\theta$ for some $n$; this term in the field redefinition is needed in order to ensure that the terms in the action \eqref{eq:reg_action_M4} that are proportional to the free parameters in the field redefintion are all of the form $\theta\,\Box^n\theta$ for some integer $n$ that depends on the order of the truncation ($n=1,2$ for the case at hand).

The  equations of motion up to $O(M^{-4})$ that are derived from the new action \eqref{eq:reg_action_M4} are
\begin{equation}
\begin{aligned}
\Box\theta =&~ \frac{2}{M^2}\partial_a\big[(\partial\theta)^2\partial^a\theta \big]\\
&-\frac{2\,\alpha_1}{M^2}\,\Box^2\theta\\
&~+\frac{4}{M^4}\big[(\partial^a\partial^b\Box\theta)(\partial_a\theta)(\partial_b\theta) + (\Box\theta)(\partial^a\theta)(\partial_a\Box\theta)+(\partial^a\theta)(\partial_a\partial_b\theta)(\partial^b\Box\theta)\\
&\hspace{1.5cm}+(\Box\theta)(\partial^a\partial^b\theta)(\partial_a\partial_b\theta)+2(\partial^a\theta)(\partial^b\partial^c\theta)(\partial_a\partial_b\partial_c\theta)\big]\\
&-\frac{2\,\alpha_2}{M^4}\,\Box^3\theta\,.
\end{aligned}
\label{eq:eoms_M4_reg}
\end{equation}

We will now show that equations \eqref{eq:eoms_M4} can be written as a diagonal system of wave equations. We proceed as before, and introduce new variables that absorb the derivatives of $\theta$:
\begin{align}
    \Box\theta =&~ \theta^{(0,1)}\,,\\
    \Box\theta^{(0,1)} =&~\theta^{(0,2)}\,,
\end{align}
and
\begin{equation}
    \theta^{(1,0)}_a\equiv \partial_a\theta \,,\quad \theta^{(2,0)}_{ab}\equiv \partial_a\partial_b\theta\,,\quad \theta^{(1,1)}_a\equiv \partial_a\Box\theta\,, \label{eq:defs_M4}
\end{equation}
Then, the equations of motion in terms of the new variables become
\begin{align}
    \Box\theta =&~\theta^{(0,1)}\,, \label{eq:eq_theta_M4}\\
    \Box\theta^{(1,0)}_a=&~\partial_a\theta^{(0,1)}\,,\label{eq:eq_theta10_M4}\\
    \Box\theta^{(0,1)} = &~\theta^{(0,2)}\,, \label{eq:eq_theta01_M4}\\
    \Box\theta^{(2,0)}_{ab} =&~\partial_{a}\theta^{(1,1)}_{b}\,, \label{eq:eq_theta20_M4}\\
    \Box\theta^{(1,1)}_a =&~\partial_a\theta^{(0,2)}\,, \label{eq:eq_theta11_M4}
\end{align}
together with equation \eqref{eq:eoms_M4_reg}, which can be written as a massive (non-linear) wave equation for $\theta^{(0,2)}$:
\begin{equation}
\begin{aligned}
    \left(\Box+\frac{\alpha_1}{\alpha_2}\,M^2\right)\theta^{(0,2)} =&-\frac{M^4}{2\,\alpha_2}\,\theta^{(0,1)}\\
    &+\frac{M^2}{\alpha_2}\big[ \theta^{(0,1)}\eta^{ab} + 2\,\theta^{(2,0)ab}\big]\theta^{(1,0)}_a\theta^{(1,0)}_b \\
    &+\frac{2}{\alpha_2}\,\big[~\theta^{(1,0)a}\theta^{(1,0)b}\partial_a\theta^{(1,1)}_b+\theta^{(0,1)}\theta^{(1,0)a}\theta^{(1,1)}_a \\
&\hspace{1.1cm}+\theta^{(1,0)a}\theta^{(1,1)b}\theta^{(2,0)}_{ab}+\theta^{(0,1)}\theta^{(2,0)ab}\theta^{(2,0)}_{ab} +2\,\theta^{(1,0)a}\theta^{(2,0)bc}\partial_{(a}\theta^{(2,0)}_{bc)}\big]\,.
\end{aligned}
\label{eq:eq_theta02_M4}
\end{equation}

The equation of motion for $\theta^{(0,2)}$ has a source with terms that are multiplied by powers of $M$. In the regime of validity of EFT, $M$ has to be suitably large, which implies that $\theta^{(0,2)}$ has a very large source. One may worry that this may destabilize the evolution of this variable, but it turns out that, for suitable initial data, having a large and positive mass is enough for $\theta^{(0,2)}$ to remain bounded at all times. In the Discussion we show that the system \eqref{eq:eq_theta_M4}--\eqref{eq:eq_theta02_M4} can be equivalently written in terms of dimensionless variables and, in this way, the highest power of the UV mass scale that appears is $M^2$. Finally, we note that just as in the EFT$_2$, in this case the equations of motion for the constraint variables \eqref{eq:defs_M4}, are nothing but the wave equations (without sources) for the actual constraints, so the latter are trivially propagated. 

It is interesting to comment on the spectrum of the regularised theory \eqref{eq:eoms_M4_reg}. Consider linear perturbations around the trivial solution $\theta\equiv 0$ on a Minkowski background. The linearised equation of motion for the fluctuations $\delta \theta$ is given by
\begin{equation}
   \left( \Box+\frac{2\alpha_1}{M^2}\Box^2+\frac{2\alpha_2}{M^4}\Box^3\right)\delta\theta=0.
\end{equation}
To find the spectrum, we consider plane wave solutions of the form $\delta\theta=\Theta\,e^{\text{i}kx}$, which gives
\begin{equation}
    k^2\left[1-2\alpha_1\left(\frac{k}{M}\right)^2+ 2\alpha_2\left(\frac{k}{M}\right)^4\right]=0.
\end{equation}
Therefore, the spectrum  consists of the usual massless Goldstone boson with dispersion relation $k^2=0$ and two massive scalars with dispersion relation,
\begin{equation}
    k^2 = M^2\left(\frac{\alpha_1\pm \sqrt{\alpha_1^2-2\,\alpha_2}}{2\,\alpha_2}\right) 
\end{equation}
and corresponding masses given by
\begin{equation}
    \left(\frac{m_\pm}{M}\right)^2=-\frac{\alpha_1\pm \sqrt{\alpha_1^2-2\,\alpha_2}}{2\,\alpha_2}\,.
\end{equation}
To ensure that the masses are real and positive we can choose $\alpha_1<0$ and $0<\alpha_2\leq\frac{1}{2}\alpha_1^2$. These conditions must be satisfied in order to be able to construct EFT solutions of the equation of motion \eqref{eq:eoms_M4_reg}. We note that when linearising around a more general background configuration, the dispersion relations and masses receive background-dependent corrections, hence changing the particle masses. However, within the EFT regime these corrections are expected to be small, in which case the effective masses should remain real and positive. In Appendix \ref{sec:alt_formulation} we present an alternative (and equivalent) formulation of \eqref{eq:eq_theta_M4}--\eqref{eq:eq_theta02_M4} with explicit Klein-Gordon-type-of equations for the massive modes.

\subsection{Initial data}
\label{sec:init_dat_EFTs}

Since the equations of motion for the EFT$_1$ are second order, we only need to specify $\theta_0\equiv \theta|_{t=0}$ and $\theta_1\equiv \partial_t\theta|_{t=0}$. To make the comparisons with the UV we will use the same values for $\theta_0$ and $\theta_1$ for both theories. 

To specify initial data for the EFT$_2$ we need the second and  third time derivatives of $\theta$ at $t=0$ up to $O(M^{-2})$. We can obtain them consistently with the EFT expansion using the reduction of order procedure   \cite{Flanagan:1996gw}. By imposing the equations of motion at $t=0$ we can compute the required higher time derivatives at the initial time.  In the particular case of the EFT$_2$, we can impose either \eqref{eq:eom_reg} or \eqref{eq:eoms_M2} at $t=0$ to compute $\partial_t^2\theta|_{t=0}$, since the contributions from $\Box\theta$ are sub-leading in the EFT regime.\footnote{This is equivalent to imposing \eqref{eq:eoms_M2} to find $\Box\theta|_{t=0}$ (and hence $\theta^{(0,1)}|_{t=0}$), and similarly for its time derivative. } We get:
\begin{equation}
    (\partial_t^2\theta)\Big|_{t=0} = \frac{(\partial_x^2\theta_0)\left[M^2-6(\partial_x\theta_0)^2+2\,\theta_1^2\right]+8\,\theta_1(\partial_x\theta_0)(\partial_x\theta_1)}{M^2-2(\partial_x\theta_0)^2+6\,\theta_1^2}\,.
    \label{eq:d2theta0_reg_M2}
\end{equation}
Taking another time derivative of this expression, we can thus obtain $(\partial_t^3\theta)|_{t=0}$ from previously known data:
\begin{equation}
    \begin{aligned}
        (\partial_t^3\theta)\Big|_{t=0} =&\,\frac{1}{[M^2-2(\partial_x\theta_0)^2+6\,\theta_1^2]^3}\Big[
        16\,\partial_x\theta_1\,\partial_x\theta_0\,\partial_x^2\theta_0\Big(-6\,\theta_1^4 +6\,\big(\partial_x\theta_0)^4-3\,M^2\,(\partial_x\theta_0)^2\\
        &\hspace{5cm}+\theta_1^2(96\,(\partial_x\theta_0)^2-13\,M^2)\big)\Big) \\
        &+\partial_x^2\theta_1\left(6\,\theta_1^2 - 2\,(\partial_x\theta_0)^2+M^2\right)\big( 
        12\,\theta_1^4 + 12\,(\partial_x\theta_0)^4-8\,M^2\,(\partial_x\theta_0)^2+M^4\\
        &\hspace{6cm}+8\,\theta_1^2(3\,(\partial_x\theta_0)^2+M^2)\big)\\
        &-8\,\theta_1\,(\partial_x^2\theta_0)^2\big( 48(\partial_x\theta_0)^2-6\,M^2\,(\partial_x\theta_0)^2+M^4 + 2\,\theta_1^2(24(\partial_x\theta_0)^2+M^2) \big)\\
        &+8\,\theta_1\big(
        \partial_x\theta_0\,\partial_x^3\theta_0\left( 2\,\theta_1^2-6\,(\partial_x\theta_0)^2+M^2\right)\left( 6\,\theta_1^2-2\,(\partial_x\theta_0)^2+M^2\right) \\
        &\hspace{1.5cm}+(\partial_x\theta_1)^2\left(-36(\partial_x\theta_0)^4+16(\partial_x\theta_0)^2(-6\,\theta_1^2+M^2)+(6\,\theta_1^2+M^2)^2\right)
        \big)
        \Big]\,.
    \end{aligned}
    \label{eq:d3theta0_reg_M2}
\end{equation}
Equations \eqref{eq:d2theta0_reg_M2} and \eqref{eq:d3theta0_reg_M2} can be further expanded in $1/M$ keeping only the terms up to $O(M^{-2})$. Therefore, initial data for \eqref{eq:eq_theta}--\eqref{eq:eq_box_theta} is given by $\theta_0$, $\theta_1$ and
\begin{equation}
    \begin{aligned}
        &\theta^{(1,0)}_t\big|_{t=0} = \theta_1\,,\quad \partial_t\theta^{(1,0)}_t\big|_{t=0}=(\partial_t^2\theta)\big|_{t=0}\,,\\
        &\theta^{(1,0)}_x\big|_{t=0} = \partial_x\theta_0\,,\quad \partial_t\theta^{(1,0)}_x\big|_{t=0} = \partial_x\theta_1\,,\\
        &\theta^{(0,1)}\big|_{t=0} = -(\partial_t^2\theta)\big|_{t=0} + \partial_x^2\theta_0\,,\quad \partial_t\theta^{(0,1)}\big|_{t=0} = -(\partial_t^3\theta)\big|_{t=0} + \partial_x^2\theta_1\,,
    \end{aligned}
\end{equation}
with $(\partial_t^2\theta)\big|_{t=0}$ and $(\partial_t^3\theta)\big|_{t=0}$ given by eqs. \eqref{eq:d2theta0_reg_M2} and \eqref{eq:d3theta0_reg_M2}
respectively. 

In the case of the the EFT$_4$, we need up to the $5^{\rm th}$ time derivative of $\theta$ at $t=0$, and  we impose the equations of motion \eqref{eq:eoms_M4_reg} at $t=0$ to the required order in $1/M$ to compute the time derivatives of $\theta$ that we need. Note that in the case of EFT$_4$, it is important to consider the field redefinition to construct consistent initial data.

\section{Norms and conserved charges}
\label{sec:norms}

\subsection{Notations and definitions of norms}

We start by defining some norms that we will use to monitor the solutions to the individual EFTs and their deviations from the UV solutions. For functions $u:[-1,1]\mapsto \mathbb{R}$ and non-negative integers $s$ we can define the Sobolev norms
\begin{equation}
    ||u||_{H^s}=\left(\sum\limits_{k\leq s}\int {\rm d}x\, |\partial_x^k u|^2\right)^{1/2}.
\end{equation}
For $s=0$ this coincides with the $L^2$ norm. For functions $u:[0,T]\times [-1,1]\mapsto \mathbb{R}$ we will also make use of
\begin{equation}
    || u||_{C^0_tH^s_x}=\sup_{t\in [0,T]}||u(t,\cdot) ||_{H^s}\,,
\end{equation}
which can be thought of as the upper envelope of the function $||u||_{H^s_x}(t)$.

\subsection{Conserved charges}

The theories under consideration admit conserved charges. It is instructive to compute these quantities to describe certain aspects of the dynamics and also to check the consistency of our numerical simulations.

Consider first the UV theory \eqref{eq:UV_action} and let us compute the conserved energy in this theory. To this end, we compute the stress tensor by covariantizing the action \eqref{eq:UV_action},
\begin{equation}
\frac{S}{v^2} = -\int {\rm d}^d x\sqrt{-g}\left[ \frac{1}{2}\,(\nabla_a\rho)(\nabla^a\rho) + \frac{1}{2}(1+\rho)^2(\nabla_a\theta)(\nabla^a\theta) + V(\rho)\right]\,,
\end{equation}
from which we obtain
\begin{equation}
\begin{aligned}
    T_{ab}^{\rm{UV}}=&-\frac{2}{v^2}\frac{\delta S}{\delta g^{ab}}\\
    =&~(\nabla_a\rho)(\nabla_b\rho)+(1+\rho)^2(\nabla_a\theta)(\nabla_b\theta)
    -g_{ab}\left[ \tfrac{1}{2}\,(\nabla\rho)^2+\tfrac{1}{2}(1+\rho)^2(\nabla\theta)^2 + V(\rho)\right]\,. 
\end{aligned}
\label{eq:UV_stress_tensor}
\end{equation}
The conserved energy associated with \eqref{eq:UV_stress_tensor} on a slice of constant time $t$ in $d$-dimensional Minkowski space is given by
\begin{equation}
    \mathcal{E}_{\rm UV}[\theta,\rho](t)=\int {\rm d}^{d-1}x~\left\{\frac12 (\partial_t\rho)^2+\frac12 (\partial_i\rho)(\partial^i\rho)+\frac12(1+\rho)^2\left[(\partial_t\theta)^2+(\partial_i\theta)(\partial^i\theta)\right]+V(\rho)\right\}\,.
\end{equation}
with $V(\rho)$ given in \eqref{eq:UV_potential}. This energy is clearly non-negative, as expected from a classically ``healthy'' UV theory. 

The action \eqref{eq:UV_action} also admits a shift symmetry in the $\theta$ field: it is invariant under $\theta\to \theta+{\rm const.}$ This symmetry implies that the equation of motion for $\theta$ is the divergence of a current
\begin{equation}
    \frac{1}{v^2}\frac{\delta S}{\delta \theta}=-\partial_a[(1+\rho)^2\partial^a\theta]=-\partial_a J^a\,,
\end{equation}
and therefore $J^a$ is conserved on-shell. On a $d$-dimensional Minkowski background, this symmetry gives rise to the conserved charge
\begin{equation}
    \mathcal{Q}_{\rm UV}=\int{\rm d}^{d-1}x\,(1+\rho)^2\,\partial_t\theta\,.
\end{equation}

Similarly, each of the EFTs that we are considering can be derived from an action, and therefore they also admit a corresponding conserved energy $\mathcal E$ and conserved charge $\mathcal{Q}$ associated to the shift symmetry. We shall derive these charges for the regularised fourth order EFT, then one can straightforwardly read off the corresponding quantities for the lower order EFTs as well. On the other hand, the EFTs obtained from integrating out the heavy field $\rho$ at the level of the equations of motion, e.g., \eqref{eq:eoms_M2_tmp} at $O(M^{-2})$, do not admit a conserved energy nor a conserved charge. Therefore, it is more useful to work with the EFTs arising from an action.

To obtain the conserved energy for \eqref{eq:action_M4}, we vary the action
\begin{align}
    \frac{S}{v^2}\approx-\int {\rm d}^d x\sqrt{-g}\bigg[&~ \frac{1}{2}(\nabla_a\theta)(\nabla^a\theta) - \frac{1}{2M^2}\big((\nabla_a\theta)(\nabla^a\theta) \big)^2 + \frac{2}{M^4}\,(\nabla_a\nabla_b\theta)(\nabla^a\nabla^c\theta)(\nabla^b\theta)(\nabla_c\theta) \nonumber \\
&-\frac{\alpha_1}{M^2}\,\theta\Box^2\theta 
-\frac{\alpha_2}{M^4}\,\theta\Box^3\theta \bigg]\, \label{eq:reg_action_M4_curved}
\end{align} 
w.r.t. the metric to obtain the energy-momentum tensor. In flat space, it takes the form
\begin{align}
    T_{ab}^{{\rm EFT}_4}&=\left(-\frac12 (\partial\theta)^2+\frac{1}{2M^2}\big((\partial_a\theta)(\partial^a\theta)\big)^2\right)\eta_{ab}+\left(1-\frac{2}{M^2}(\partial\theta)^2\right)(\partial_a\theta)(\partial_b \theta) \nonumber \\
    &+\frac{2\alpha_1}{M^2}\biggl[(\partial_a \Box \theta)(\partial_b\theta)+(\partial_b \Box\theta)(\partial_a\theta)-\eta_{ab}\,(\partial^c\theta)( \partial_c \Box \theta)-\frac12 \eta_{ab} (\Box\theta)^2\biggr] \nonumber \\
    &+\frac{4}{M^4}\biggl[(\partial_a\partial_c\theta)(\partial_b\partial_d\theta)(\partial^c\theta)(\partial^d\theta)-(\partial_a\theta)(\partial_b\theta )(\partial_c\partial_d \theta)(\partial^c\partial^d \theta) \nonumber\\
    &\hspace{1.5cm}-(\partial_a\theta)(\partial_b\theta)(\partial^c\theta)(\partial_c\Box\theta)-\frac12 \eta_{ab}(\partial^c\theta)(\partial^d\theta)(\partial_e\partial_c \theta)(\partial^e\partial_d \theta)\biggr] \nonumber \\
    &+\frac{2\alpha_2}{M^4}\biggl[(\partial_a\Box^2\theta)(\partial_b\theta) +(\partial_b\Box^2\theta)(\partial_a\theta)+(\partial_a\Box\theta)(\partial_b\Box\theta)\nonumber\\
    &\hspace{1.5cm}-\eta_{ab}(\Box\theta)(\Box^2\theta)-\eta_{ab}(\partial_c\Box^2\theta)(\partial^c\theta)-\frac12\eta_{ab}(\partial_c\Box\theta)(\partial^c\Box\theta)\biggr]\,. \label{eq:stress_tensor_EFT4}
\end{align}
Then the conserved energy is simply given by
\begin{equation}
    \mathcal{E}_{{\rm EFT}_4}[\theta](t)=\int {\rm d}^{d-1}x~T_{tt}^{{\rm EFT}_4}\,. \label{eq:conserved_E}
\end{equation}
From \eqref{eq:stress_tensor_EFT4} and \eqref{eq:conserved_E} we can easily obtain the formulae for the conserved energies of the other EFTs considered in this paper.\footnote{For both the  EFT$_1$ and the EFT$_2$ we ignore the terms proportional to $1/M^4$ in the expressions below. For the EFT$_1$ we further set $\alpha_1=0$, and for the EFT$_2$ we set $\alpha=\alpha_1$. } In $1+1$ dimensions we have:
\begin{align}
    \mathcal{E}_{\text{EFT}_4}[\theta](t)&=\int {\rm d}x \Biggl[\frac12 (\theta_t^{(1,0)})^2+\frac12 (\theta_x^{(1,0)})^2  -\frac{1}{2M^2}\left(-(\theta_t^{(1,0)})^2+(\theta_x^{(1,0)})^2\right) \left(3\,(\theta_t^{(1,0)})^2+(\theta_x^{(1,0)})^2\right) \nonumber \\
    &\qquad \qquad +\frac{2\alpha_1}{M^2}\left((\partial_t\theta)\,\partial_t\theta^{(0,1)}+(\partial_x\theta)\,\partial_x\theta^{(0,1)}+\frac12 (\theta^{(0,1)})^2\right) \nonumber \\\
    &\qquad\qquad +\frac{2}{M^4}\biggl((\theta_x^{(1,0)})^2(\theta_{tx}^{(2,0)})^2+(\theta_x^{(1,0)})^2(\theta_{xx}^{(2,0)})^2 \nonumber \\
    & \qquad\qquad\qquad \qquad-(\theta_t^{(1,0)})^2(\theta_{tt}^{(2,0)})^2-2(\theta_t^{(1,0)})^2(\theta_{xx}^{(2,0)})^2 \nonumber \\
    & \qquad\qquad \qquad \qquad+5(\theta_t^{(1,0)})^2(\theta_{tx}^{(2,0)})^2+2(\theta_t^{(1,0)})^3(\theta_{t}^{(1,1)})-2(\theta_t^{(1,0)})^2(\theta_x^{(1,0)})(\theta_{x}^{(1,1)}) \nonumber \\
    &\qquad \qquad \qquad \qquad -2(\theta_t^{(1,0)})(\theta_x^{(1,0)})(\theta_{tx}^{(2,0)})(\theta_{xx}^{(2,0)})-2(\theta_t^{(1,0)})(\theta_x^{(1,0)})(\theta_{tt}^{(2,0)})(\theta_{tx}^{(2,0)})\biggr) \nonumber \\
    &\qquad\qquad+\frac{2\alpha_2}{M^4}\biggl((\theta_t^{(1,0)})(\partial_t\theta^{(0,2)})+(\theta_x^{(1,0)})(\partial_x\theta^{(0,2)})\nonumber \\
    &\qquad\qquad \qquad\qquad+(\theta^{(0,1)})(\theta^{(0,2)})+\frac12(\theta_{t}^{(1,1)})^2+\frac12(\theta_{x}^{(1,1)})^2\biggr)\Biggr]\,,
    \label{eq:energy_EFT4}
\end{align}
Note that the contributions of the higher derivative terms to the energy, i.e., the terms in \eqref{eq:energy_EFT4} proportional to $1/M^2$ and $1/M^4$, are not positive definite. In particular, the EFT$_1$ is a Horndeski-type theory that only propagates a ``healthy'' massless degree of freedom, $\theta$, and yet the energy is not manifestly positive. Therefore, we can see that it is possible to choose initial conditions in these higher derivative theories such that the conserved energy is negative. This requires, however, $\partial^k\theta\sim O(M^k)$, i.e., initial data that lies outside the regime of validity of EFT. Even if the total energy is positive, for some solutions of the EFT$_2$ and EFT$_4$, the contributions of the ghost fields to the energy (i.e., the terms proportional to the free parameters $\alpha_i$ in the field redefinitions) may have arbitrarily large negative values while the terms in the first line of \eqref{eq:energy_EFT4} (the energy carried by the massless degree of freedom) may have large positive values. Nevertheless, one may still be able to derive bounds on certain positive definite quantities such as the $H^s$ norm of the solution, ensuring the existence of long-lived or possibly even global solutions (at least for some open set of initial data). We shall expand on this point in the Discussion section.

Considering the variation of the action \eqref{eq:reg_action_M4} with respect to $\theta$, one can straightforwardly deduce that for the fourth order regularised EFT, the conserved current associated with the shift symmetry is
\begin{equation}
    J^a_{{\rm EFT}_4}=\Bigl[1+\frac{2}{M^2}(\partial\theta)^2-\frac{2}{M^4}(\Box \mathcal{F}_2)\Bigr](\partial^a\theta)-\frac{\alpha_1}{M^2}\,(\partial^a\Box\theta)-\frac{\alpha_2}{M^4}\,(\partial^a\Box^2\theta)\,,
\end{equation}
and the associated conserved charge is
\begin{equation}
    \mathcal{Q}_{{\rm EFT}_4}=\int {\rm d}^{d-1}x~J_{t}^{{\rm EFT}_4}\,.
\end{equation}
In $1+1$ dimensions this reads as
\begin{align}
    \mathcal{Q}_{{\rm EFT}_4}&=\int {\rm d}x~\Bigl\{\Bigl[1+\frac{2}{M^2}(\partial\theta)^2-\frac{2}{M^4}(\Box \mathcal{F}_2)\Bigr](\partial_t\theta) -\frac{\,\alpha_1}{M^2}\,(\partial_t\Box\theta)-\frac{\alpha_2}{M^4}\,(\partial_t\Box^2\theta)\Bigr\}\,,
\end{align}
From this expression, we can deduce the conserved charges for the other EFTs that we consider.

\section{Numerical experiments}
\label{sec:numerics}

In this Section we carry out numerical experiments to explore the dynamics of the UV theory and the various EFTs for different classes of initial conditions and values of the mass scale $M$. For simplicity, we carry out the simulations in (1+1)-dimensional Minkowski space with a compact spatial domain $x\in [-1,\,1]$ and periodic boundary conditions. Therefore, the length $L$ of the spatial domain is $L=2$. These simulations are simple enough so that Mathematica's \texttt{NDSolve} function suffices; we use fourth order finite differences and a fourth order Runge-Kutta time integrator unless otherwise stated.

\begin{figure}[t]
    \centering
    \includegraphics[width=0.49\linewidth]{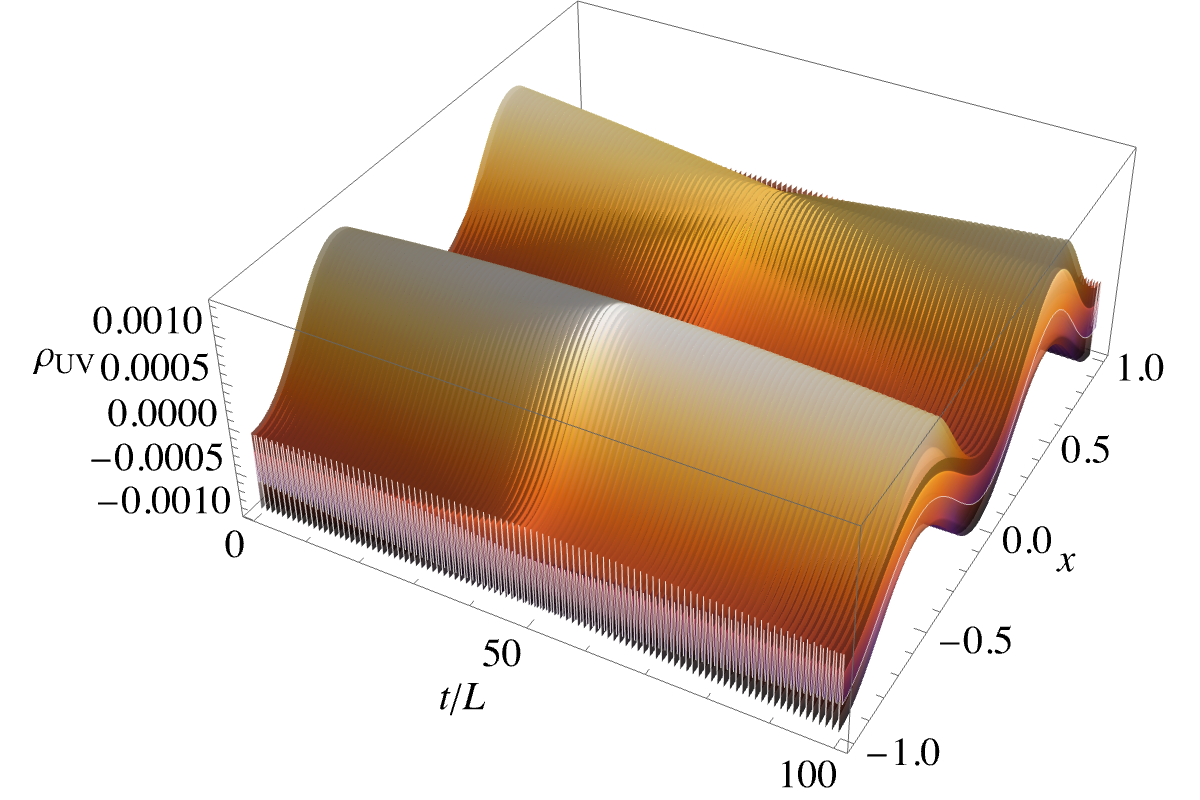}
    \includegraphics[width=0.49\linewidth]{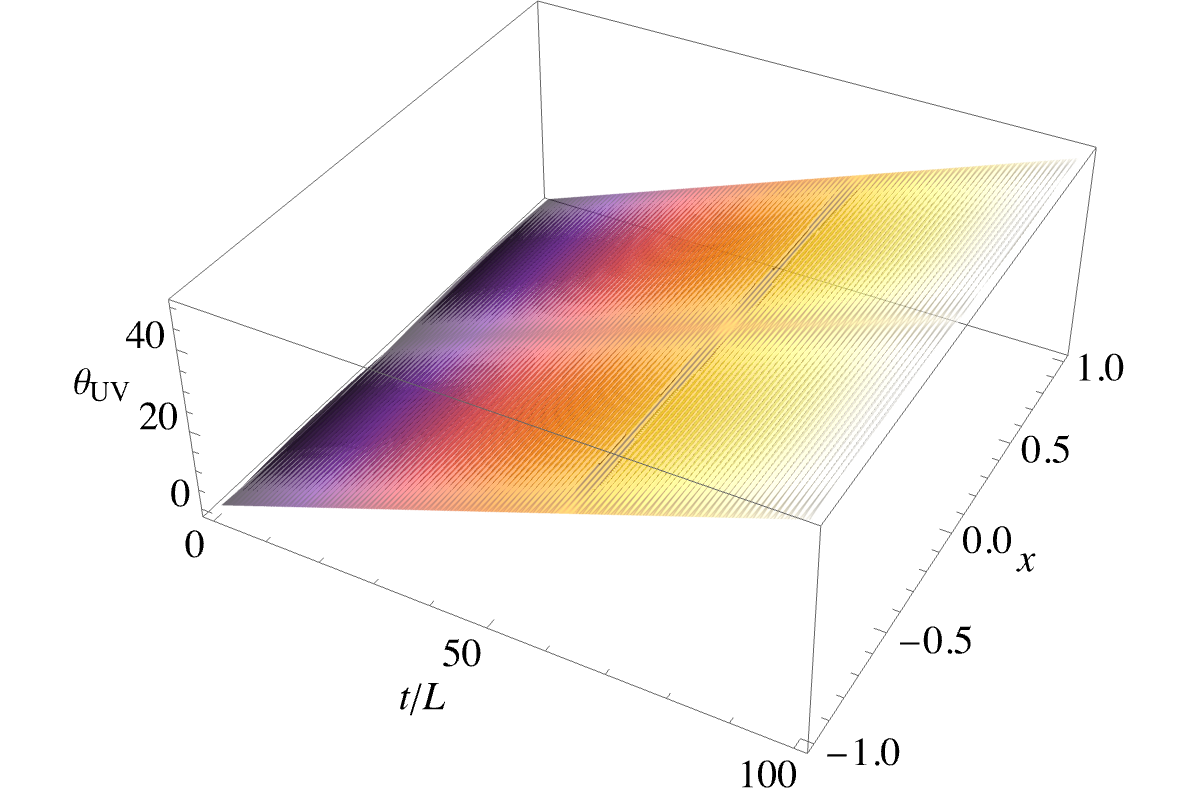}
    \includegraphics[width=0.49\linewidth]{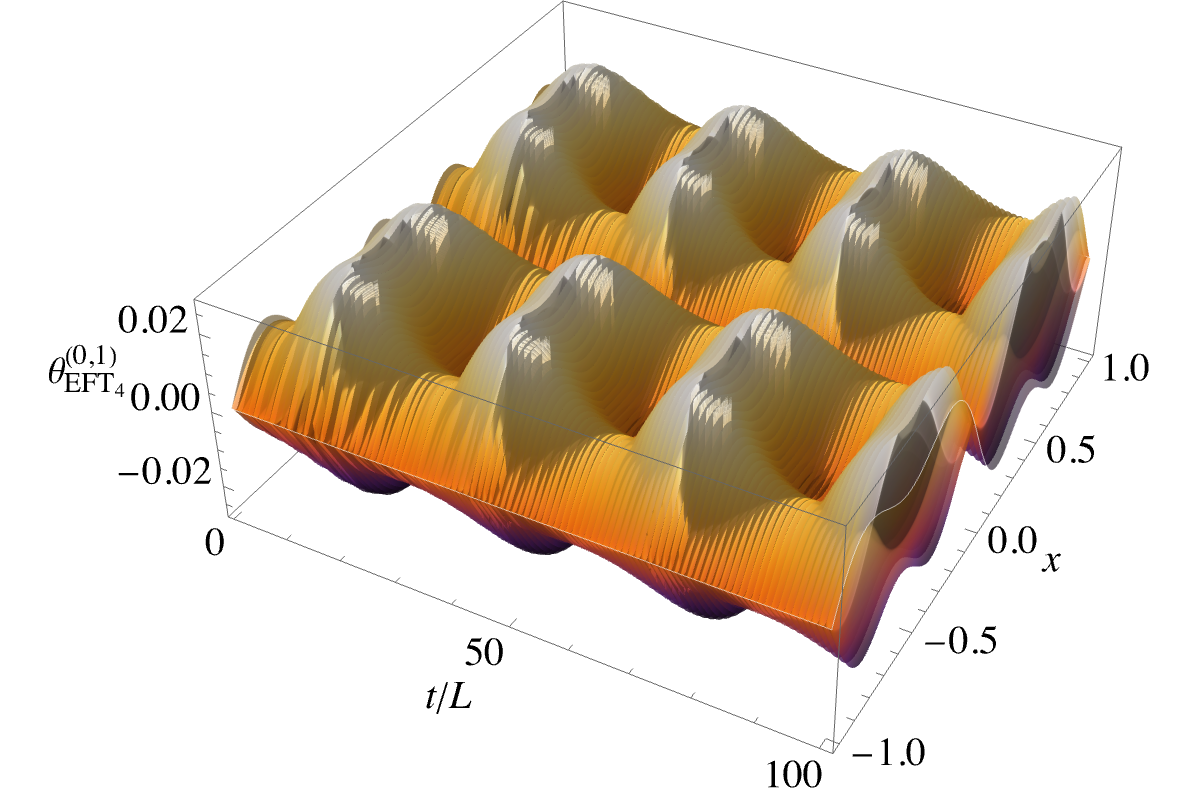}
    \includegraphics[width=0.49\linewidth]{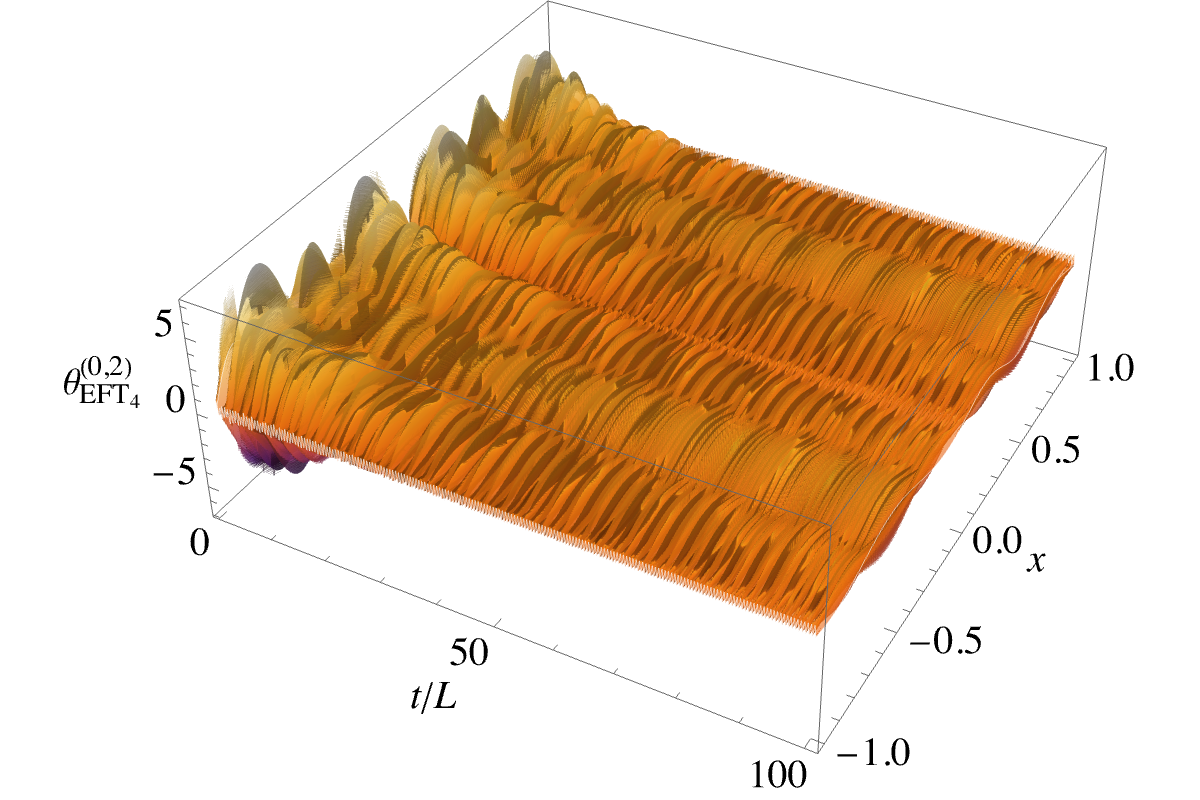}
    \captionsetup{width=.9\linewidth}
    \caption{Representitive plots for the evolution of $\rho$ (top left), $\theta$ (top right), $\theta^{(0,1)}$ (bottom left) and $\theta^{(0,2)}$ (bottom right). $\rho$ and $\theta$ are the solutions of the UV theory while $\theta^{(0,1)}$ and $\theta^{(0,2)}$ are solutions of the EFT$_4$. We use \eqref{eq:init_dat_1} as initial data and set $M=100$ and $\alpha_1=-0.8$, $\alpha_2=0.3$ to evolve the EFT$_4$.}
    \label{fig:all_plots}
\end{figure}

Note that in this setup the vacuum in the UV theory is not stable. For initial data with a non-zero charge $\mathcal{Q}_{\text{UV}}$, the solution $\theta_\text{UV}$ exhibits linear growth and $\rho_\text{UV}$ does not decay (see Fig. \ref{fig:all_plots}). Due to the absence of a decay mechanism in $1+1$ dimensions, we do not expect global solutions to exist in the effective theories. Nevertheless, we will see that for large enough masses and choices of initial data compatible with the EFT assumptions, we obtain long-lived solutions. In $(3+1)$-dimensional Minkowski spacetime, where the vacuum is stable (see e.g., \cite{Dong:2019ttn}), we may expect global solutions to exist in EFTs, see more on this in the Discussion section.

\noindent
We first consider smooth initial data for $\theta$ given by
\begin{equation}
   \theta|_{t=0}\equiv \theta_0=\sin\left(\pi\,x\right)\,,\quad \partial_t\theta|_{t=0}\equiv \theta_1=\tfrac12\,\cos^2(\pi\,x)\,. \label{eq:init_dat_1}
\end{equation}
This initial data satisfies the uniform boundedness assumption (being independent of $M$) and $\sup_{x\in[-1,1]}|\partial\theta|\sim \pi$ on the initial data slice. Hence, we expect the EFT solutions to provide a good approximation to the UV solution when $M$ is significantly larger than $\pi$. Furthermore, the amplitudes of the waves in \eqref{eq:init_dat_1} are $O(1)$, and hence the nonlinearities in the equation of motion play an important role in the dynamics even at early times. Finally, note that the conserved charge $\mathcal{Q}$ for this initial data is non-zero.

Given $\theta_0$ and $\theta_1$ as above, in Section \ref{sec:EFT_expansion} we explain how we specify the initial data for the ``heavy'' field $\rho$ such that it is in the EFT regime. Likewise, in Section \ref{sec:init_dat_EFTs} we give details on how to obtain consistent initial data for the remaining fields in the EFT$_2$ and EFT$_4$. 

In Fig. \ref{fig:all_plots} (top left panel) we show the evolution of the heavy field $\rho$, together with $\theta$ (top right panel), $\theta^{(0,1)}\equiv\Box\theta$ (bottom left panel) and $\theta^{(0,2)}\equiv\Box^2\theta$ (bottom right panel) obtained by evolving the initial data \eqref{eq:init_dat_1} in the UV theory and in EFT$_4$. On the scale of this plot one cannot tell the difference between $\theta_{\rm UV}$ and $\theta_{\text{EFT}_4}$. As this plot shows, throughout the evolution the amplitude of $\rho$ is small and it exhibits high frequency oscillations, as one would expect from the ``heavy'' field. With the choice of initial data \eqref{eq:init_dat_1}, $\theta$ grows linearly with time; superposed to this linear growth there are some small amplitude oscillations (with $\omega\sim M)$, which are the effects of interaction with the field $\rho$. In the EFTs, these effects are induced by the higher derivative terms in the equations. The evolution of $\Box\theta$ and $\Box^2\theta$ shows that these fields remain bounded at all times. Furthermore, these fields also exhibit high frequency oscillations, with $\omega\sim M$ since the masses of these fields are of this order.

\begin{figure}[t]
    \centering
    \includegraphics[width=0.49\linewidth]{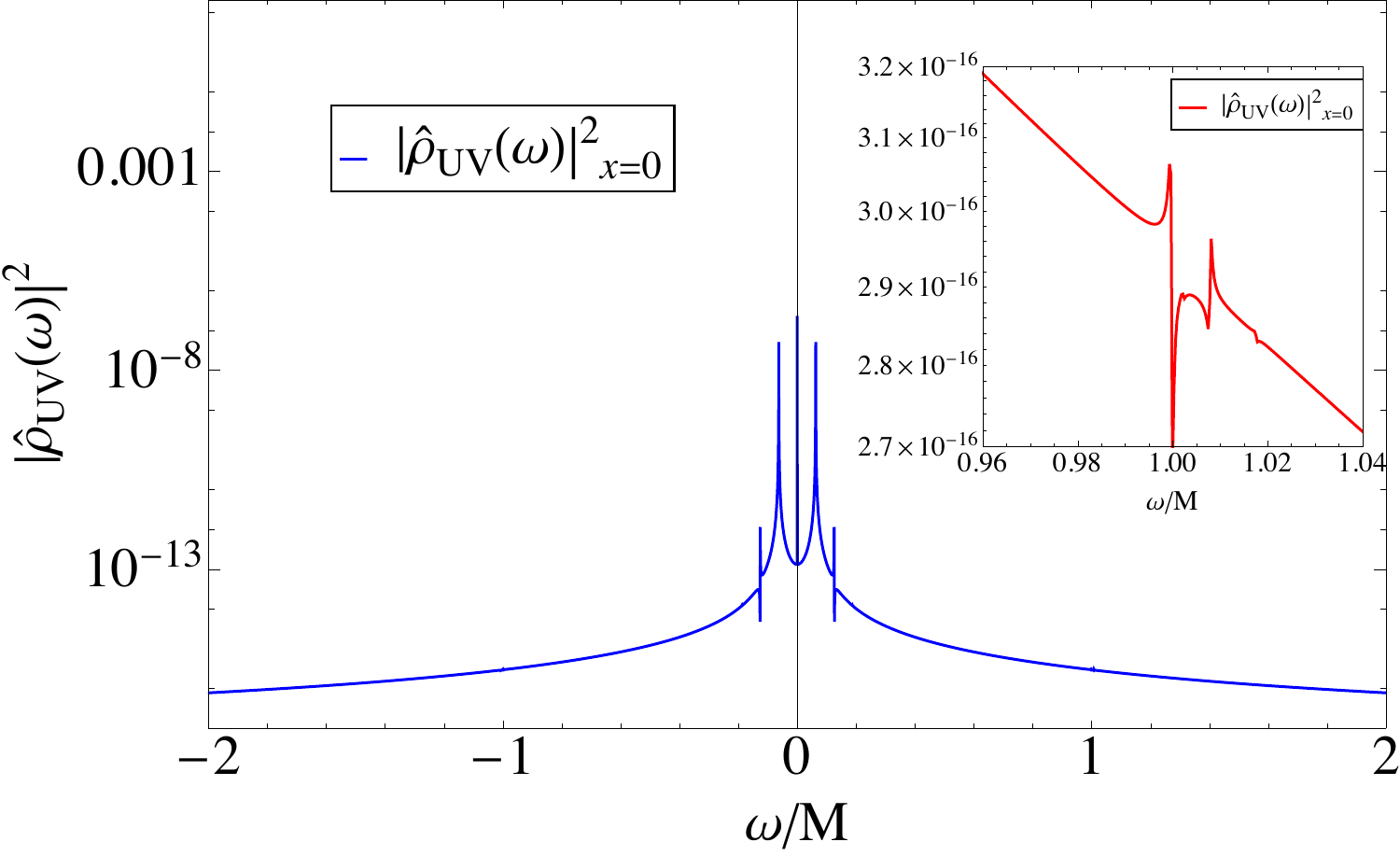}
    \includegraphics[width=0.49\linewidth]{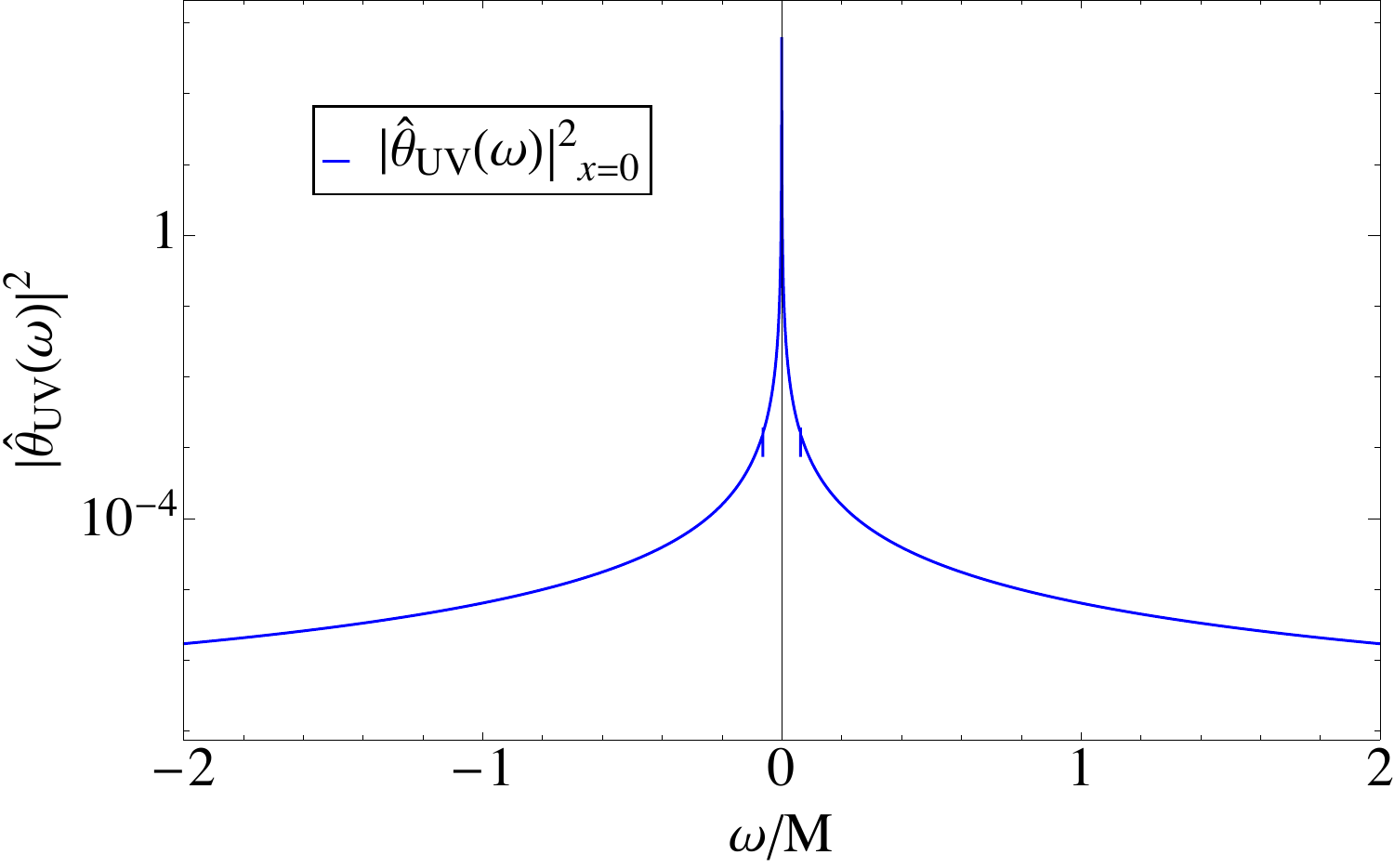}
    \includegraphics[width=0.49\linewidth]{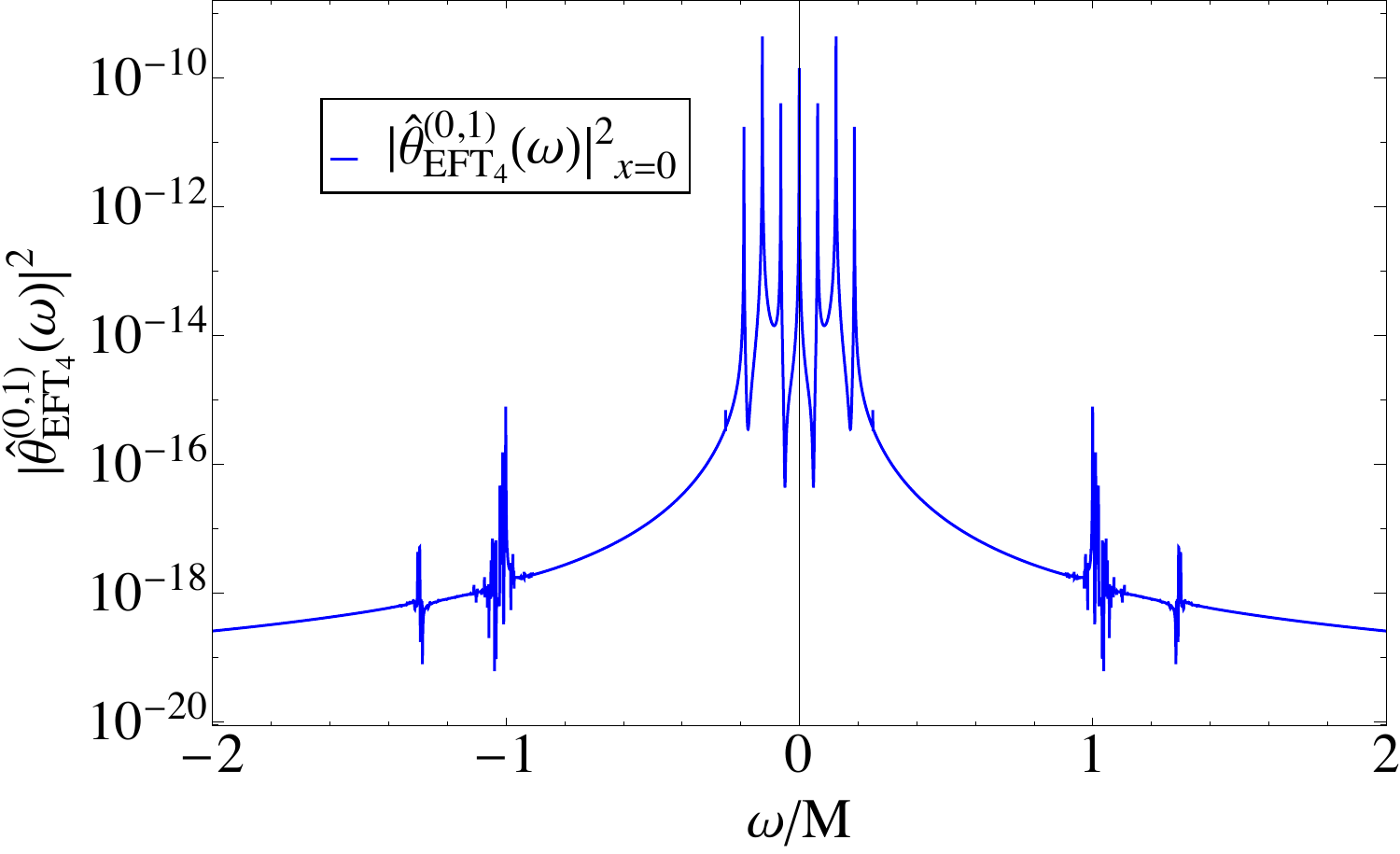}
    \includegraphics[width=0.49\linewidth]{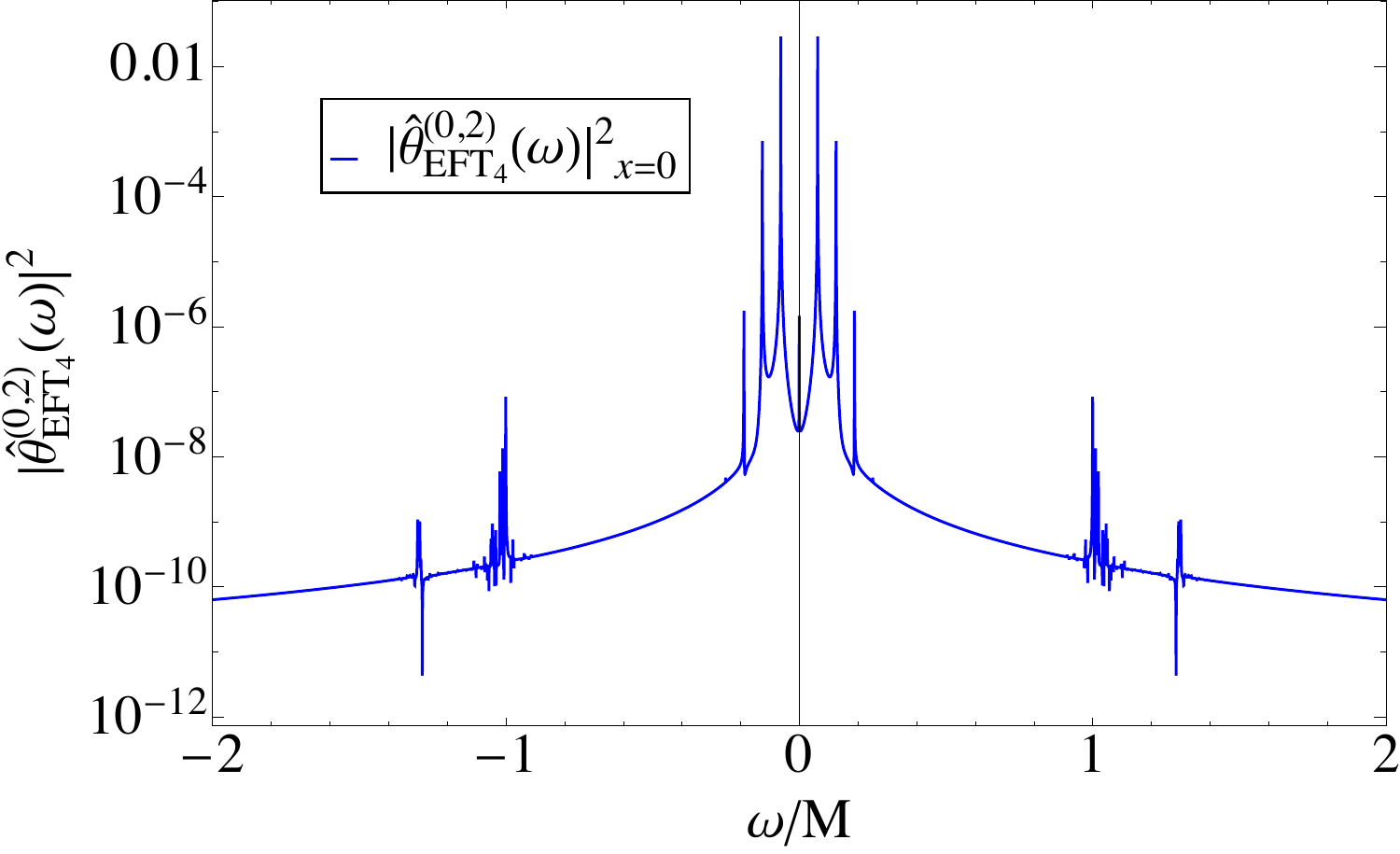}
    \captionsetup{width=.9\linewidth}
    \caption{The spectral density of the fields plotted in Fig. \ref{fig:all_plots}: $\rho$ (top left), $\theta$ (top right), $\theta^{(0,1)}$ (bottom left) and $\theta^{(0,2)}$ (bottom right).}
    \label{fig:spec_all_plots}
\end{figure}

To obtain further information about the Fourier spectrum of the solution, we numerically computed the Fourier transform of the solutions for $\theta_{\rm UV}$, $\rho_{UV}$ in the UV theory and $\theta^{(0,1)}$, $\theta^{(0,2)}$ in EFT$_4$. We show the spectral density of these fields in Fig. \ref{fig:spec_all_plots}. It is clear that $\theta_{\rm UV}$ and $\rho_{\rm UV}$ are dominated by low frequency modes: the amplitudes of modes with frequencies $\gtrsim M$ are highly suppressed (although there is a small peak in $|\hat\rho_{\rm UV}|^2$ at around $\omega\sim M$ as shown on the inset of the top left panel). This behaviour of the UV solution is also captured well by the solution to EFT$_4$. The spectral density for $\theta_{\text{EFT}_4}$ is indistinguishable from that of $\theta_{\rm UV}$ on the scale of the plots, hence we only include one of the in Fig. \ref{fig:spec_all_plots} in order to avoid repetition. On the other hand, the regularising terms have a small but noticeable effect on derivatives of $\theta$: in the spectral densities of $\theta^{(0,1)}$, $\theta^{(0,2)}$ in EFT$_4$ (displayed in the bottom left and right panels of Fig. \ref{fig:spec_all_plots}) have small peaks around frequencies equal to the masses of the auxiliary massive fields. Nevertheless, the amplitudes of these high frequency modes are several orders of magnitude smaller than the amplitudes of the relevant low frequency modes and these deviations lie beyond the expected accuracy of EFT$_4$. These results suggest that the regularising terms do not affect the solution below the cutoff frequency.

\subsection{Evolution of norms}

In this subsection we quantify the deviation of EFT solutions from the UV solution. To this end, we evolve the UV theory and the EFTs for different values of the mass $M$, starting from the initial data \eqref{eq:init_dat_1}. We compute various norms of the difference between the UV field $\theta_\text{UV}$ and the solutions of the different EFTs. These norms allow us to quantitatively assess whether the EFTs remain in their regime of validity. In the case of the EFT$_2$ and EFT$_4$ we also take into account the (perturbative) field redefinitions, eqs. \eqref{eq:field_redef_M2} and \eqref{eq:field_redef_M4} respectively, to make comparisons with the UV theory. At the end of this section we comment on the effects of field redefinitions on the solutions. 

\begin{figure}[t!]
    \centering
    \includegraphics[width=0.8\linewidth]{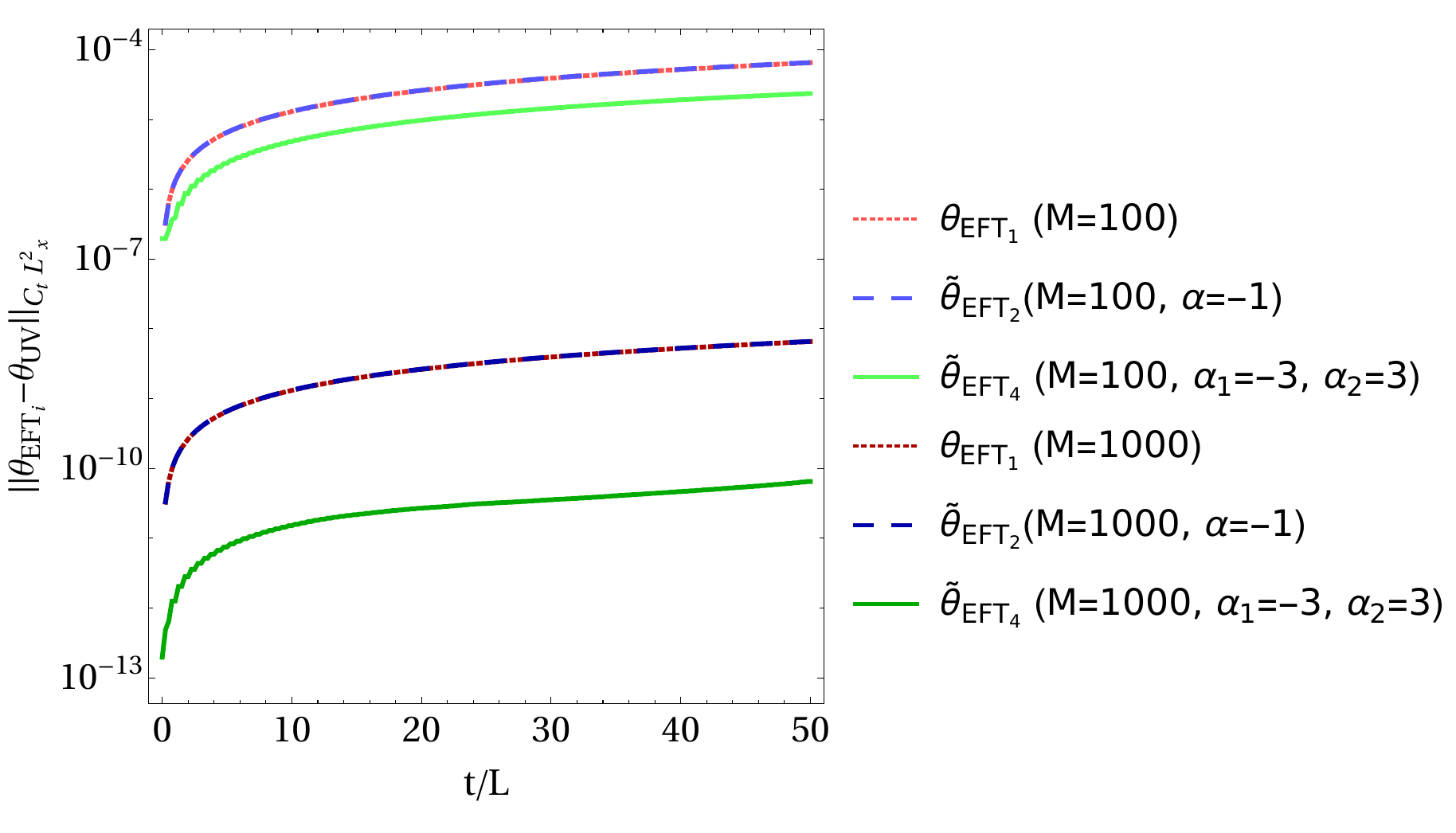}
    \captionsetup{width=.9\linewidth}
    \caption{$C_t\,L^2_x$ norm of the difference between the UV solution and the various EFTs, for $M=100$ and $M=1000$. Note that the EFT$_1$ and EFT$_2$ provide equally accurate solutions for both choices of UV mass scale, and their respective curves on this plot are on top of each other.}
    \label{fig:sup_norm}
\end{figure}

In Fig. \ref{fig:sup_norm} we examine the $C_t\,L^2_x$ norm of the difference between the UV field, $\theta_{\rm UV}$, and the various EFT solutions, $\theta_{\rm{EFT}_i}$, for different choices of the UV mass scale $M$. In this setting EFT$_1$ and EFT$_2$ provide equally accurate solutions for both choices of $M$, that is, for fixed $M$ the norms associated with EFT$_1$ and EFT$_2$ are indistinguishable on the scale of the plot. For $M=100$, EFT$_4$ provides a marginally more accurate solution than the two other EFTs, indicating that this choice of $M$ lies near the border of applicability of EFT. On the other hand, for $M=1000$ there is a clear hierarchy between the EFTs that are accurate up to $O(M^{-2})$, and the EFT$_4$, which is accurate up to $O(M^{-4})$, with the error associated to the latter being much smaller. To make the comparisons for the EFT$_2$ and the EFT$_4$ we took into account the perturbative field redefinitions, namely equations \eqref{eq:field_redef_M2} and \eqref{eq:field_redef_M4} respectively. This implies that for these two EFTs, the $C_t\,L^2_x$ norm also measures the size of certain derivatives of the light field $\theta$ through the auxiliary massive fields that the regularisation scheme introduces. Therefore, the fact that for these EFTs the $C_t\,L^2_x$ norm is bounded and well-behaved strongly supports that their respective solutions provide very accurate approximations of the UV solution for sufficiently long times.  

\begin{figure}[t!]
    \centering
    \includegraphics[width=0.8\linewidth]{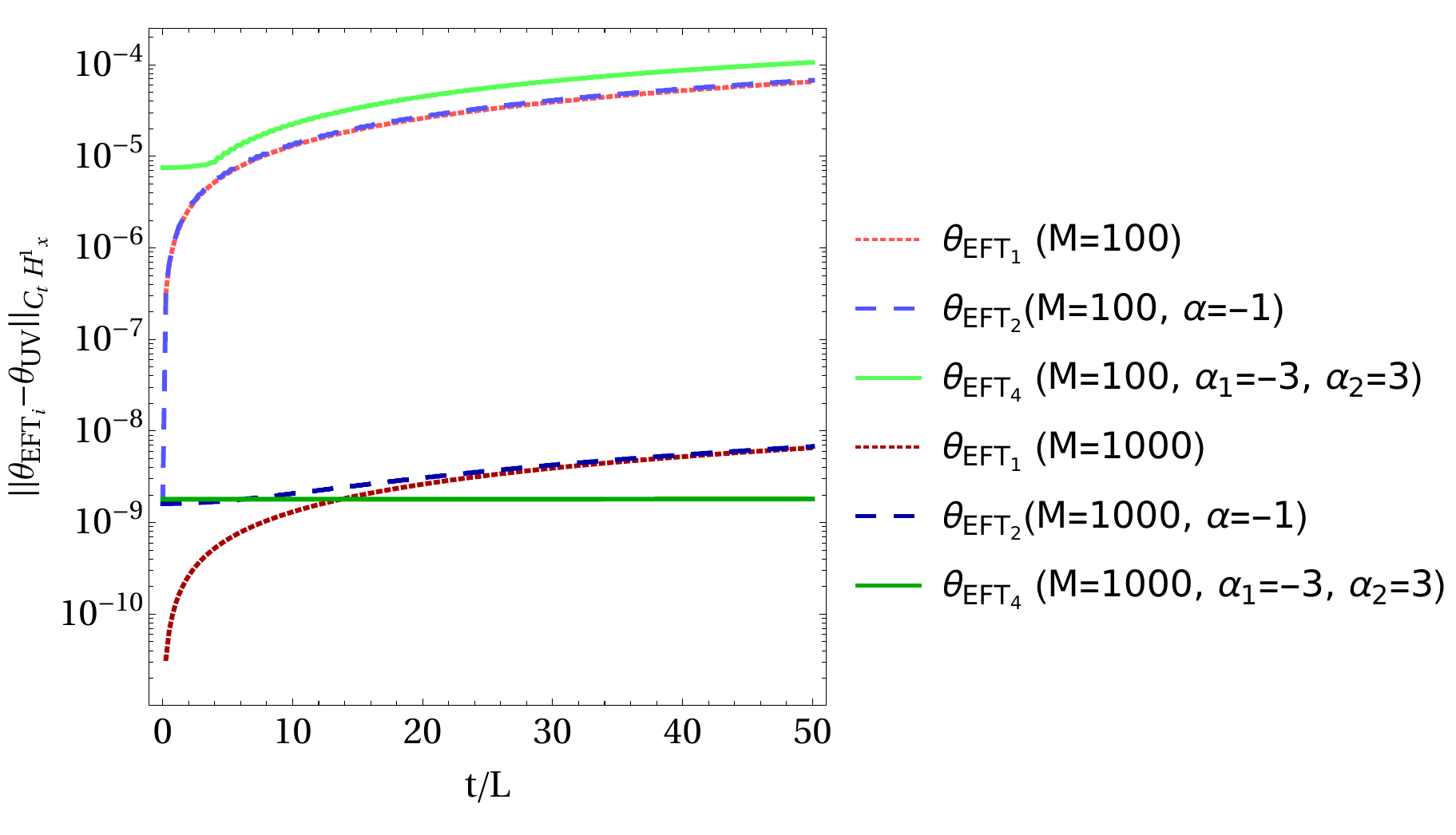}
    \captionsetup{width=.9\linewidth}
    \caption{$C_t\,H^1_x$ norm of the difference between the EFT solutions and the UV solution  remain small and bounded throughout the evolution.}
    \label{fig:sup_H1_norm}
\end{figure}

The behaviour of the $C_t\,H^1_x$ norm of the difference between the EFT solutions and the UV solution is similar to the previous norm, see Fig. \ref{fig:sup_H1_norm}. In this case we do not take into account the field redefinition, and hence the larger initial error in the EFT$_4$. This plot shows that the $C_t\,H^1_x$ norm also remains bounded, which further supports that the EFT solutions and their derivatives are under control. 

\begin{figure}[t!]
    \centering
    \includegraphics[width=0.8\linewidth]{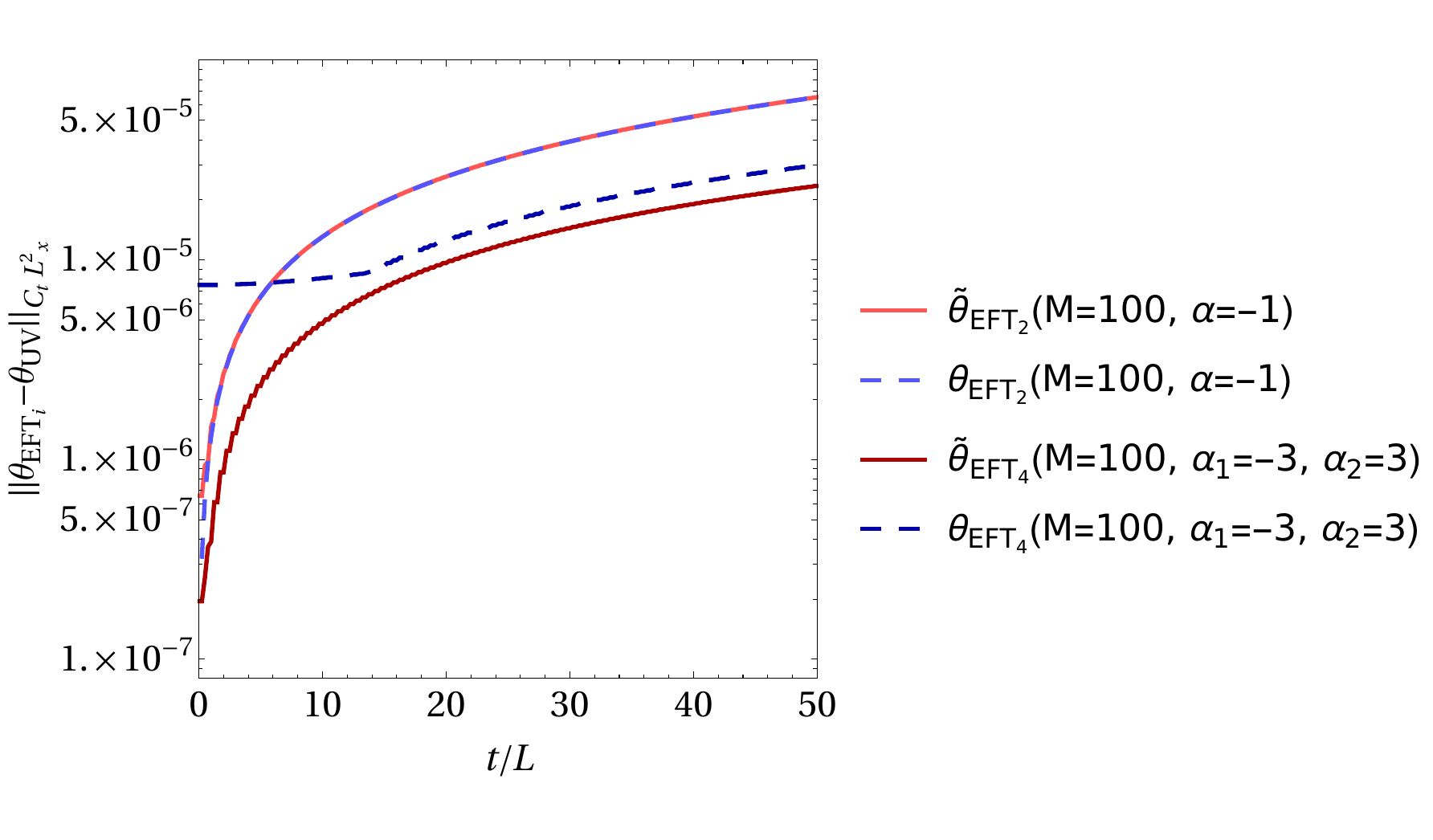}
    \captionsetup{width=.9\linewidth}
    \caption{Effect of field redefinitions on the solution for the regularised EFTs. For the EFT$_2$, the effect of the field definition \eqref{eq:field_redef_M2} is of higher order and does not affect the accuracy of the solution. On the other hand, for the EFT$_4$, the solution for $\tilde\theta$ (defined by \eqref{eq:field_redef_M4}) provides a noticeably better approximation to the UV solution than the bare solution $\theta$.}
    \label{fig:field_redef_eft4}
\end{figure}

For the EFT$_2$, the field redefinition \eqref{eq:field_redef_M2} adds a term $\frac{\alpha}{M^2}\Box\theta$ to the new field $\theta$; since in the regime of validity of the EFT one has that $\Box\theta\sim O(M^{-2})$, then the difference between the original field and the new one should be $O(M^{-4})$, and hence beyond the accuracy of this EFT. Therefore, for the EFT$_2$, one should expect that the error between the EFT solution and the full UV solution is not affected by the field redefinition. On the other hand, for the EFT$_4$, the field redefinitions \eqref{eq:field_redef_M4} modify the original $\theta$ field at $O(M^{-4})$, which is the same order as the accuracy of this EFT. Therefore, in this case we should expect that the error is sensitive to the field redefinition. In Fig. \ref{fig:field_redef_eft4} we compare the UV solution with the solution from these two EFTs, with and without the field redefinition. As Fig. \ref{fig:field_redef_eft4} shows, for the EFT$_2$ the error with respect to the UV theory is insensitive to the field redefinition, i.e. the differnce between $\theta$ and $\tilde \theta$ is negligible. On the other hand, for the EFT$_4$ we see that including the field redefinition provides a more accurate solution, as expected. Furthermore, the bare EFT$_4$ solution has a significant initial error compared to the UV solution, suggesting that in EFT$_4$ one has to take the field redefinition into account in order to get a solution accurate up to the desired order in the $1/M$ expansion.

\begin{figure}[t!]
    \centering
    \includegraphics[width=0.6\linewidth]{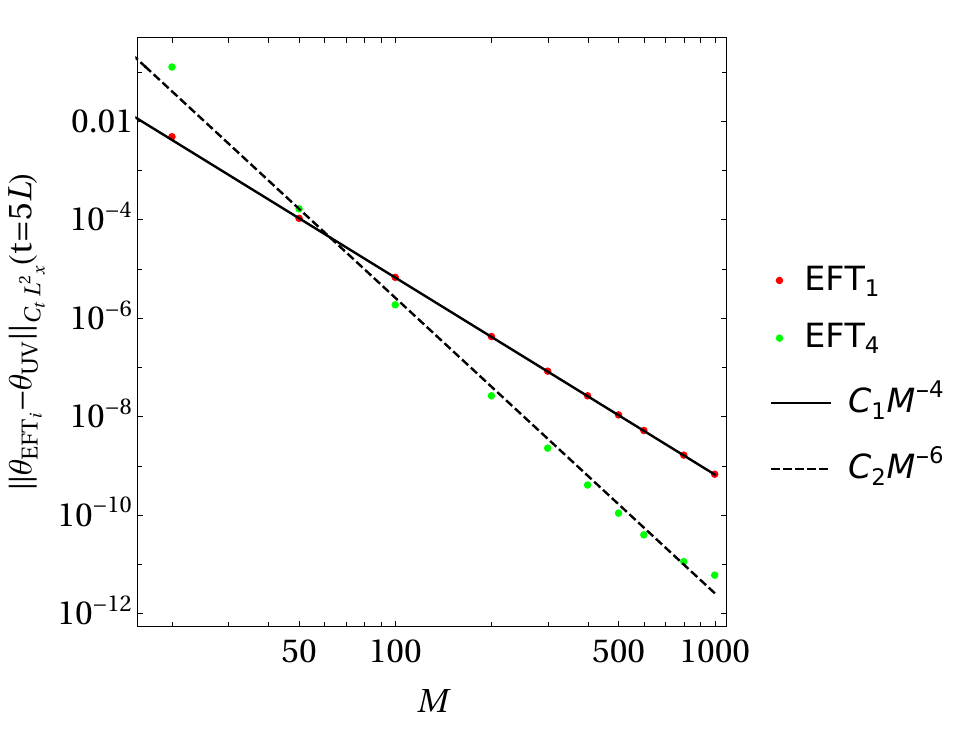}
    \captionsetup{width=.9\linewidth}
    \caption{Scaling of the error with $M$ for the EFT$_1$ and EFT$_4$. In the former case the error scales as $1/M^4$ while in the latter case it scales as $1/M^6$, as expected for such EFTs. The data points corresponding to EFT$_2$ are indistinguishable from those of EFT$_1$ for this range of masses and hence we do not show them.}
    \label{fig:mass_scale}
\end{figure}

The EFT$_1$ and EFT$_2$ are obtained from truncating the EFT expansion at $O(M^{-2})$. In the particular model that we are considering, the expansion parameter is $1/M^2$, and we expect that the error of solutions of these EFTs (measured from the UV solution) should scale as $1/M^4$. Similarly, the error of solutions to EFT$_4$ is expected to scale as $1/M^6$. In Fig. \ref{fig:mass_scale} we check that this is indeed the case for the choice of initial conditions \eqref{eq:init_dat_1}. Note that in this plot we do not show the results for EFT$_2$ since they are indistinguishable from those of the EFT$_1$ (for the range of masses considered). The fact that the error for the EFT$_2$ and the EFT$_4$ scales with the expected power of $M$ is a non-trivial check that the regularisation scheme provides a consistent way of constructing non-linear EFT solutions at the desired order in the $1/M$ expansion. This is particularly relevant at $O(M^{-4})$, where the original equations contain up to four-derivatives of the light field $\theta$ and, apart from the linearization scheme, no other method for constructing non-linear solutions is available.

\subsection{Long time behaviour of the errors}

\begin{figure}[t!]
    \centering
    \begin{subfigure}[C]{0.42\linewidth}
    \includegraphics[width=1\linewidth]{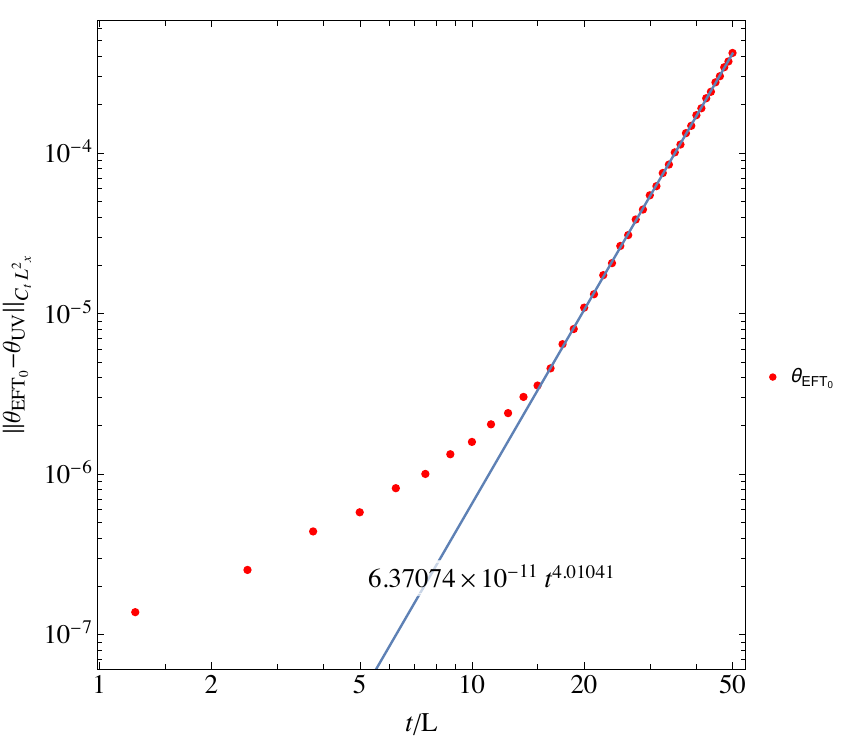}
    \end{subfigure}
    \begin{subfigure}[C]{0.52\linewidth}    
    \includegraphics[width=1\linewidth]{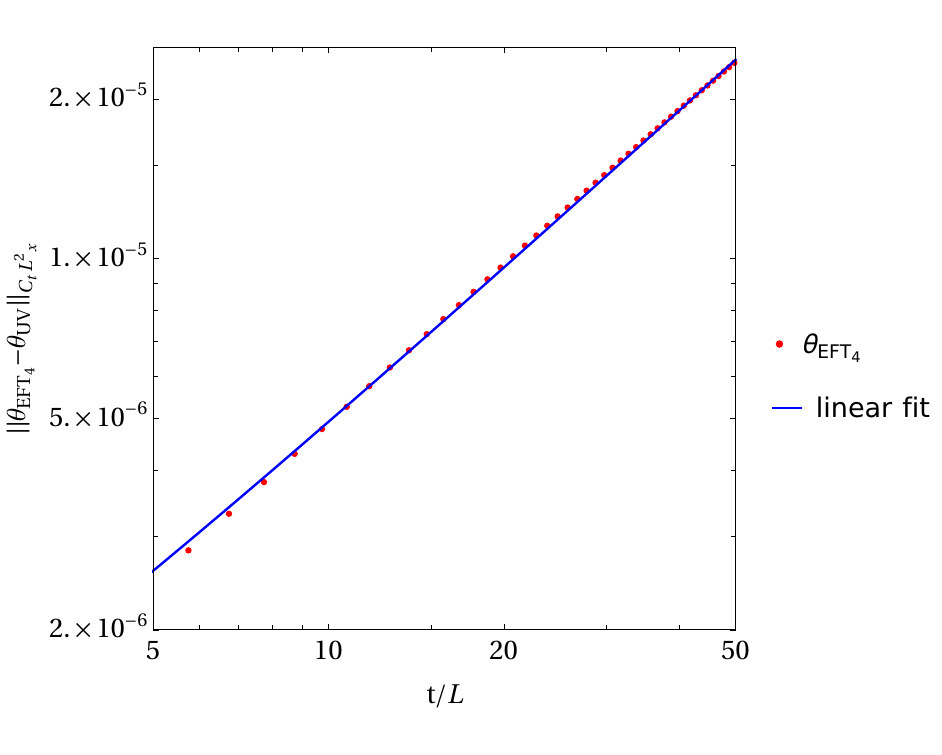}
    \end{subfigure}
    \captionsetup{width=.9\linewidth}
    \caption{The $C^0_tL^2_x$ error of $\text{EFT}_0$ (\textit{left}) and of the $\text{EFT}_4$ (\textit{right)} as a function of time. The former scales as $t^4$, in accordance with \cite{Reall:2021ebq}, whilst the latter scales linearly with time.}
    \label{fig:lin_plot}
\end{figure}

Any solution to a truncated EFT will inevitably introduce some error with respect to the solution of the full (UV complete) theory. Furthermore, the error of an EFT solution measured from the ``true'' (UV) solution will generically grow over time since they evolve according to different equations. This secular growth of the error has been pointed out in several references, e.g., \cite{Reall:2021ebq,Flanagan:1996gw}. 

In Section \ref{sec:linearisation} we quoted the estimates of \cite{Reall:2021ebq} for the growth in time of the error for the linearised EFT (i.e., EFT$_0$). Intuitively this growth in time arises as follows. The zeroth order solution $\theta^{(0)}$ does not experience secular growth. On the other hand, $\theta^{(2)}$ is a solution to an equation whose r.h.s. is at best bounded in time, so $\theta^{(2)}$ is expected to grow linearly in time. In turn, $\theta^{(4)}$ solves an equation whose r.h.s. grows linearly in time, and therefore $\theta^{(4)}$ is expected to grow quadratically in time, and so on. Iterating this reasoning gives the estimate \eqref{eq:EFT0_secular_growth} for the growth in time of the error. 

On the other hand, the regularisation scheme provides a single non-linear equation corresponding to any EFT truncated at some given order in the derivative expansion.\footnote{In our numerical scheme we solve a system of second order wave equations rather than a single higher order nonlinear equation. However, in this formulation there is only one equation that contains a truncation error, the rest of the equations are simply the defining equations of the auxiliary fields and hence they have vanishing error.} Equation \eqref{eq:eom_reg} of EFT$_2$ contains a truncation error of $O(M^{-4})$ compared to the UV equations of motion and, similarly, equation \eqref{eq:eoms_M4_reg} of EFT$_4$ has an error $O(M^{-6})$. Assuming smooth initial data and that the parameters in the field redefinition are chosen such that the masses of the auxiliary fields are positive, we expect that (in the regime of validity of EFT) both the EFT and UV solutions and their derivatives will be uniformly bounded for sufficiently long times. Then the truncation errors in the regularised EFT equations of motion will be bounded. Based on this (and on the argument of the previous paragraph), it follows that at time $T$ the solutions of \eqref{eq:eom_reg} and \eqref{eq:eoms_M4_reg} will have errors $O(T/M^4)$ and $O(T/M^6)$, respectively. Therefore, the errors of the solutions to the regularised EFTs (measured from the corresponding UV solutions) are expected to grow linearly in time. 

In Fig. \ref{fig:lin_plot} we display the $C^0_tL^2_x$ error for both the $\text{EFT}_0$ (left) and of the $\text{EFT}_4$ (right). Our results confirm the expected scaling of the error with time for the two EFTs. For the $O(M^{-4})$-accurate iterative scheme (EFT$_0$), the estimate \eqref{eq:EFT0_secular_growth} indicates that the secular growth of the error should scale like $t^4$, in agreement with our numerical results. On the other hand, Fig. \ref{fig:lin_plot} (right) confirms that the secular growth of the error for the regularised EFT$_4$ is linear in time. Note that this is the slowest growth of the error that one can expect for a truncated EFT; we will come back to this point and its implications for practical applications in the Discussion. 

\subsection{Dependency on the choice of frame}
\label{sec:frame_choice}

\begin{figure}[t]
    \centering
    \includegraphics[width=0.8\linewidth]{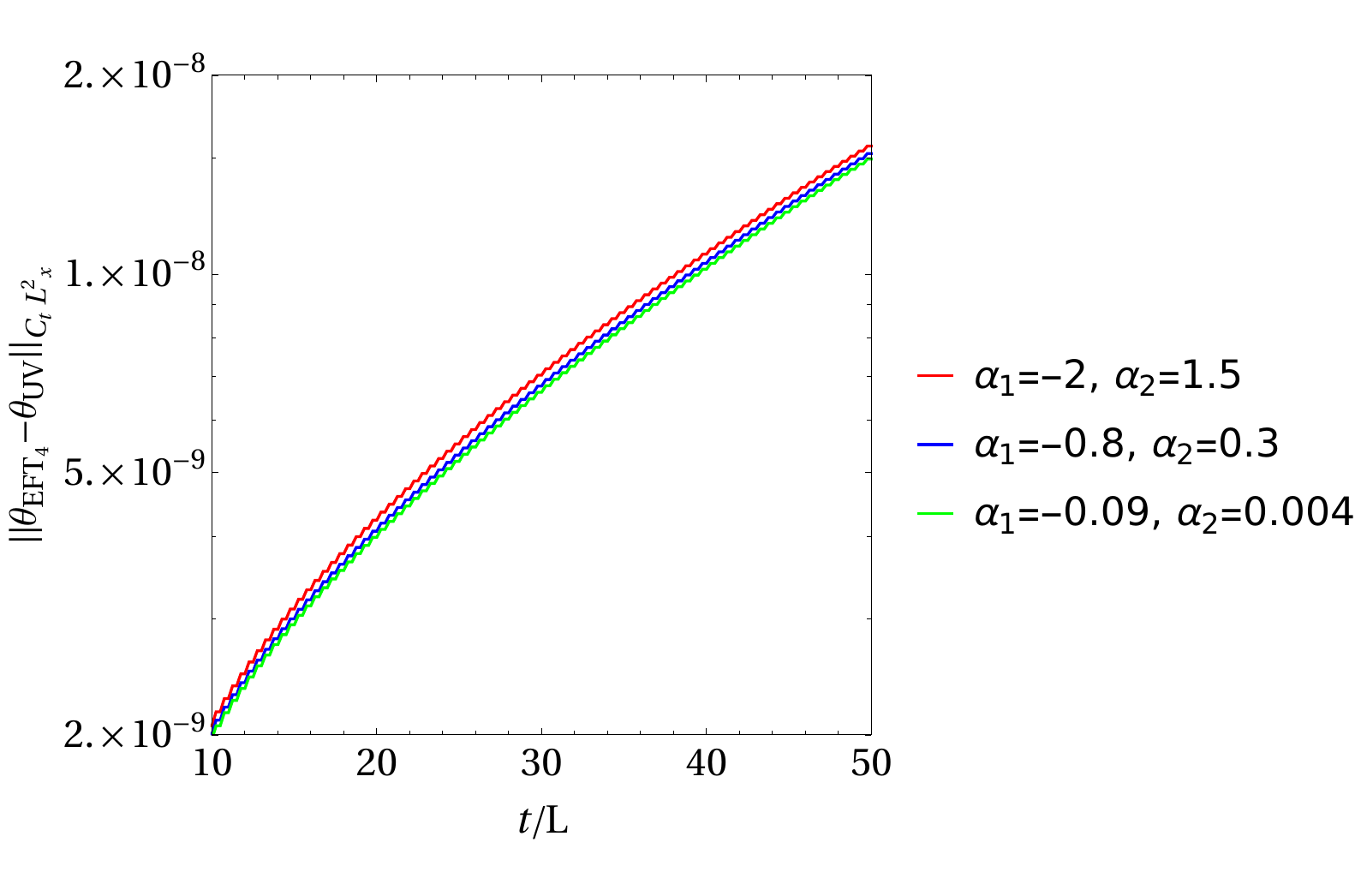}
    \captionsetup{width=.9\linewidth}
    \caption{The $C_t^0L^2_x$ error from the UV solution for three different choices of the $\alpha_1$ and $\alpha_2$ parameters in $\text{EFT}_4$ and $M=500$, starting from the initial data \eqref{eq:init_dat_1}. The different choices of these parameters, and hence masses of the auxiliary fields, have a negligible impact in the error of the EFT solution with respect to the UV solution.}
    \label{fig:diff_alphas}
\end{figure}

Recall that in the regularisation scheme, the field redefefinitions depend on some free parameters, e.g., \eqref{eq:field_redef_M2} in EFT$_2$ and \eqref{eq:field_redef_M4} in EFT$_4$. The only restriction on these parameters is that the corresponding masses (squared) of the new degrees of freedom should be positive. So, one may ask whether different choices of these parameters affect the obtained solutions. Note that the choice of the parameters in the field redefinitions is analogous to the choice of hydrodynamic frame in the context of the BDNK formulation of relativistic viscous hydrodynamics. For generic choices of these parameters, the masses of the new degrees of freedom will typically be $O(M)$, so we would expect that in the regime of validity of EFT, the solutions should not be sensitive to the choices of parameters since the corresponding massive modes should not be significantly ``excited''.  To check this explicitly, we solved the EFT$_4$ for different choices of $\alpha_1$ and $\alpha_2$ and compared the corresponding solutions. In Fig.\,\ref{fig:diff_alphas} we show the results for three different sets of parameters for solutions to EFT$_4$, starting from the initial data \eqref{eq:init_dat_1} and $M=500$. For these choices of the parameters, the values of the masses of the massive fields are as follows:
\begin{itemize}
    \item \textbf{Case (i)}: $\alpha_1=-2$, $\alpha_2=1.5$: $m_+=M$, $m_-=M/3$;
    \item \textbf{Case (ii)}: $\alpha_1=-0.8$, $\alpha_2=0.3$: $m_+=5M/3$, $m_-=M$;
    \item \textbf{Case (iii)}: $\alpha_1=-0.09$, $\alpha_2=0.004$: $m_+=12.5M$, $m_-=10M$.
\end{itemize}
Note that in Case (iii) the masses are an order of magnitude higher than in cases (i) and (ii). As we can see from Fig. \ref{fig:diff_alphas}, the difference between the solutions obtained in these three cases is negligible compared to the errors measured with respect to the UV solution, which confirms that, in the regime of validity of EFT, the specific choice of these masses does not affect the low energy dynamics of the EFT.

\subsection{Breakdown of the EFT}
\label{sec:breakdown}

In the regime of validity of the EFT, all derivatives of the field $\theta$ should be $O(1)$ (i.e., small compared to powers of $M$). However, even if we set up initial data satisfying this condition, the field $\theta$ may be driven, through its non-linear evolution, into the strongly coupled regime where $\partial^k\theta \sim O(M^k)$, $\forall k>0$. In such cases the EFT is expected to break down and it no longer provides an accurate description of the UV evolution. In this subsection we examine how this breakdown happens for the EFT$_1$ and the EFT$_2$; the breakdown of the EFT$_4$ should be qualitatively similar to the latter and therefore we shall not discuss it in detail. 

We consider the initial data \eqref{eq:init_dat_1} with a sufficiently small value of the mass parameter, $M=10$, so that on the initial data slice $\partial_x\theta$ and $\partial_t\theta$ are almost $O(M)$. As a result, both EFT$_1$ and EFT$_2$ break down after a fairly short time. We will show below that solutions to both of these EFTs break down when $(\partial\theta)^2\sim M^2/6$ but the nature of the breakdown is different in each case. On the other hand, the solution to the UV theory continues to exist beyond this point. Moreover, note that even prior to the breakdown of the EFT solutions, the errors of these solutions measured from the UV solution can be as large as $\sim 50 \%$. This already indicates that in this regime any type of effective field theory approach is not suitable. 

\begin{figure}[t!]
    \centering
    \includegraphics[width=0.48\linewidth]{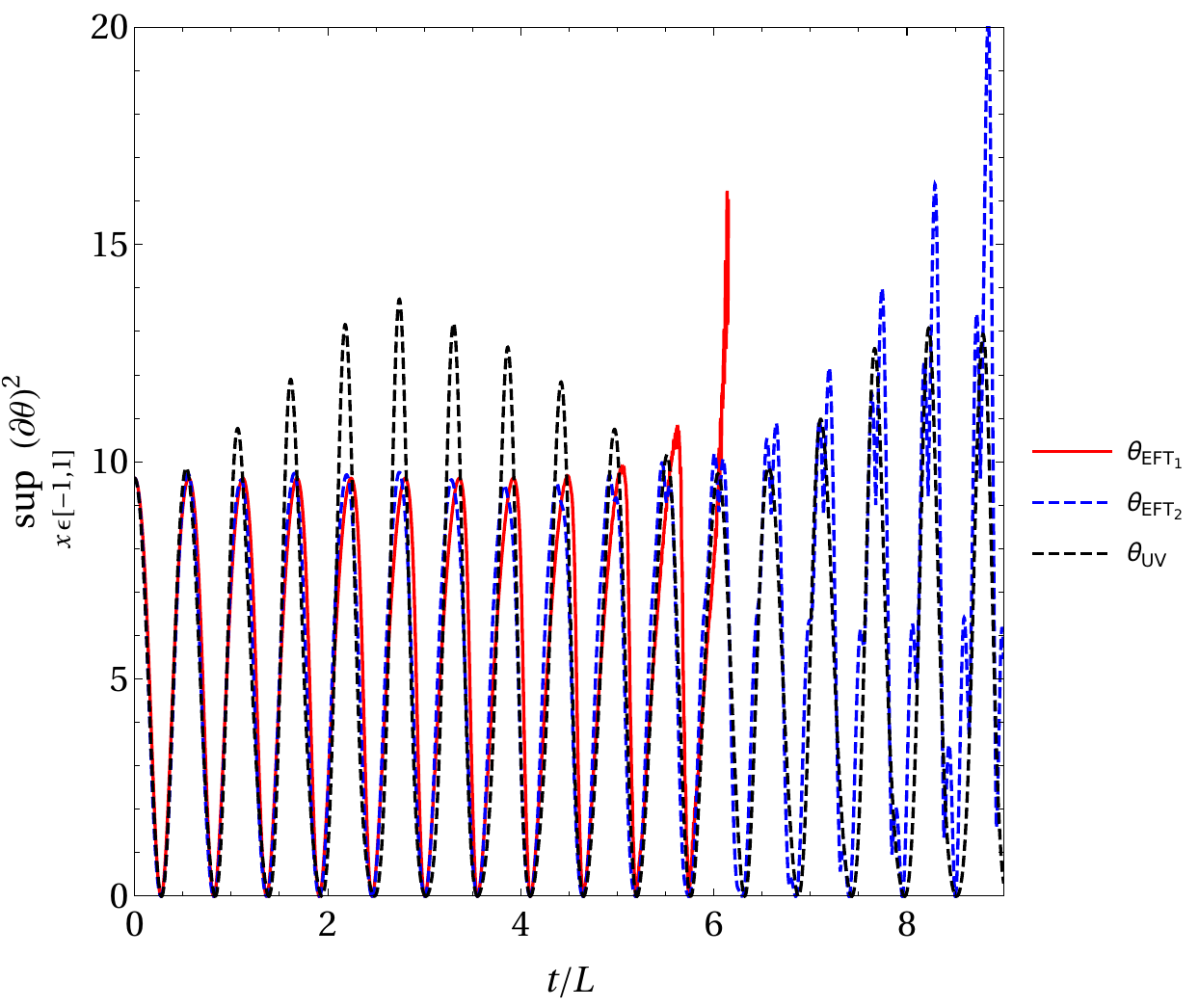}
    \includegraphics[width=0.4\linewidth]{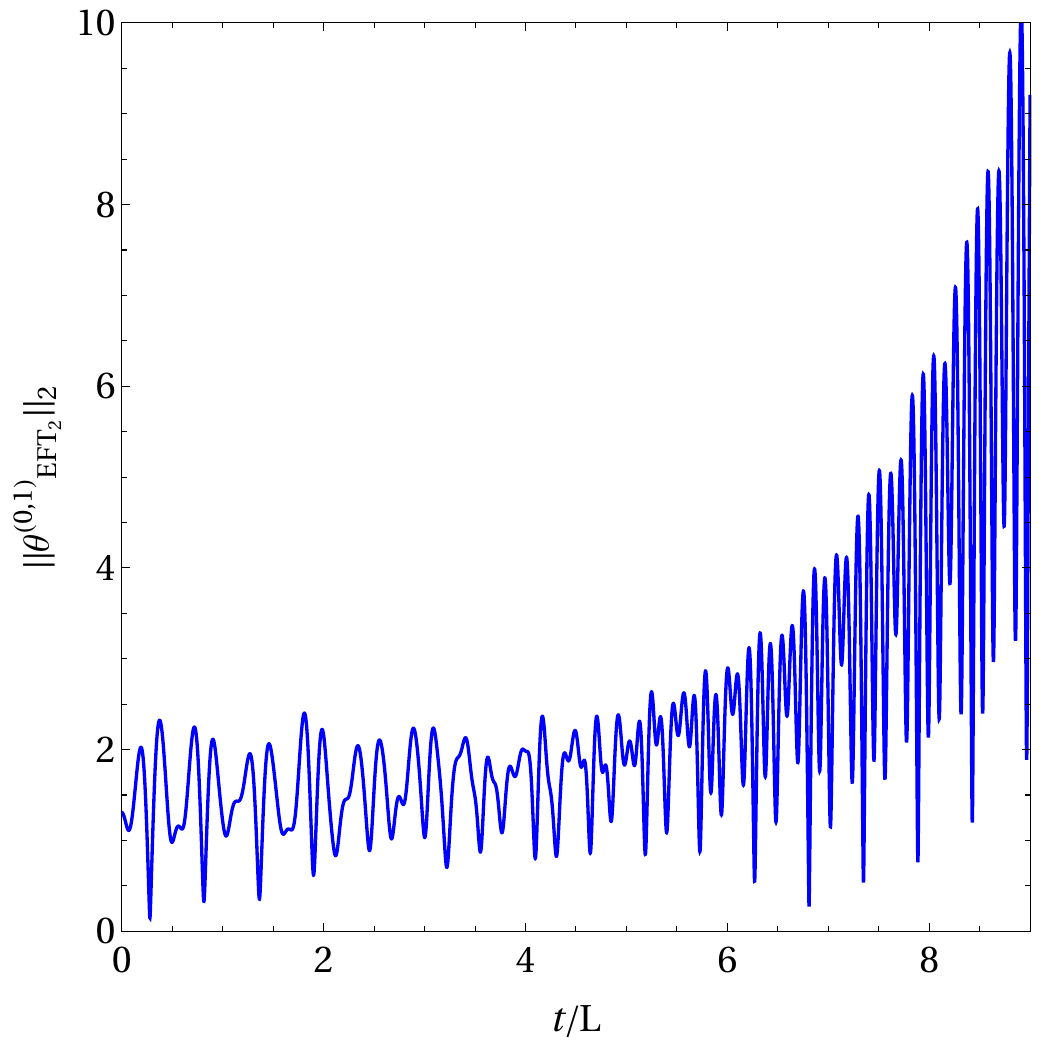}
    \captionsetup{width=.9\linewidth}
    \caption{\textit{Left}: Evolution of $\sup_{x\in[-1,1]}(\partial \theta)^2$ for the full UV theory, the EFT$_1$ and the EFT$_2$. Both the EFT$_1$ and the EFT$_2$ break down when $(\partial\theta)^2\sim M^2/6\approx 16.7$ for $M=10$; this value is attained at $t/L\sim 6$ for EFT$_1$ and at $t/L\sim 8$ for EFT$_2$. On the other hand, the UV solution persists for longer times. \textit{Right}: at $t/L\sim 5$ the field $\theta^{(0,1)}$ in the EFT$_2$ exhibits a ghostlike instability and starts to grow exponentially.}
    \label{fig:eft_breakdown}
\end{figure}

The breakdown of EFT$_1$ (and similar theories) at strong coupling has been studied in several previous works, see e.g., \cite{Allwright:2018rut,Ripley:2019hxt,Ripley:2019irj,Bernard:2019fjb,Figueras:2020dzx,Thaalba:2023fmq}. In this theory the breakdown occurs due to a change of character of the equation of motion: the effective metric \eqref{eq:eff_metric_EFT1} changes signature from Lorentzian to Euclidean. From \eqref{eq:eff_metric_EFT1} we see that this happens when $(\partial\theta)^2\sim M^2/6$. Fig. \ref{fig:eft_breakdown} (left) shows that for the initial data \eqref{eq:init_dat_1} and $M=10$, the quantity $(\partial\theta)^2$ reaches $M^2/6$ at a certain point at time $t/L\sim 6$, and the evolution of $\theta$ stops. Note that when this happens, the determinant of the inverse effective metric \eqref{eq:eff_metric_EFT1} goes to zero, implying that one of the characteristic speeds of the system goes to zero. This corresponds to a Tricomi-like transition. 

On the other hand, the evolution equations of EFT$_2$, eqs. \eqref{eq:eq_theta}--\eqref{eq:eq_box_theta}, are a collection of wave equations whose characteristics are null w.r.t. the (inverse) Minkowski metric $\eta^{ab}$. Therefore, the system is always hyperbolic, regardless of the size of $\partial\theta$. Nevertheless, when a solution of EFT$_2$ enters the strongly coupled regime, i.e., $\partial\theta\sim O(M)$, a different type of blow-up may occur due to the onset of a tachyonic instability. To see this, it is more convenient to work with the $4$th order equation \eqref{eq:eom_reg}\footnote{We remind the Reader that this is equivalent to \eqref{eq:eq_theta}--\eqref{eq:eq_box_theta}.} which we may also write as
\begin{equation}
    \left(\eta^{ab}\,\Box +\frac{M^2}{2\,\alpha} \, G^{ab} \right)\partial_a\partial_b\theta=0\,
    \label{eq:4order_eq_EFT2}
\end{equation}
where $G^{ab}$ is the effective metric \eqref{eq:eff_metric_EFT1}. Recall that we require $\alpha<0$ in order to avoid tachyonic instabilities in the EFT regime $(\partial\theta)^2\ll M^2$. When the determinant of $G^{ab}$ changes sign, which happens when $(\partial\theta)^2\sim M^2/6$, then the second term in \eqref{eq:4order_eq_EFT2} will behave as a tachyonic mass term. This suggests that the solution may start growing (at least) exponentially fast when the condition $(\partial\theta)^2\sim M^2/6$ is met. The subsequent non-linear evolution of $\theta$ is likely to lead to a blow-up in a very short time. It is interesting to note that the onset of this tachyonic instability in EFT$_2$ is independent of the specific choice of the parameter $\alpha$ in the field redefinition \eqref{eq:field_redef_M2}.

Next, we comment on the dynamics of EFT$_1$ and EFT$_2$ shortly before the breakdown. We observe in Fig. \ref{fig:eft_breakdown} that until $t\approx 5L$ solutions of EFT$_1$ and EFT$_2$ show fairly good agreement with each other. At around $t\approx 5L$, both the solution of EFT$_1$ and that of EFT$_2$ exhibit growth but the growth rate in EFT$_2$ is slower. Moreover, $\theta_{\text{EFT}_2}$ tracks the growth of $\theta_{\text{UV}}$ until around $t\approx 9L$. To understand this behaviour, we employ a heuristic argument in a simplified setting. Consider the following linear equation (playing the role of EFT$_1$):
\begin{equation}
    A^{ab}\partial_a\partial_b \theta+B^a\partial_a\theta+ C\,\theta=0 \label{eq:toy}
\end{equation}
where we assume for simplicity that $A$, $B$ and $C$ are constant tensors and $A$ is a Lorentzian metric  with $A^{00}<0$. We further assume that $B$ and $C$ are such that this equation admits a growing mode solution of the form $\theta\sim e^{\lambda t+\text{i}kx}$ (this can be arranged if e.g., $B^0<0$). In other words, we assume that the characteristic equation of \eqref{eq:toy} can be written as
\begin{equation}
    |A^{00}|(\lambda-\lambda_+(k))(\lambda-\lambda_-(k))=0
\end{equation}
such that $\text{Re}\,\lambda_+(k)>0$ for some choice of the wavenumber $k$. Now, let us add a ``regularising'' term to \eqref{eq:toy} leading to the modified equation 
\begin{equation}
    -\frac{1}{\mu^2}\Box(\Box \theta)+A^{ab}\partial_a\partial_b \theta+B^a\partial_a\theta+ C\,\theta=0\,, \label{eq:toy_v2}
\end{equation}
with $\mu^2>0$.  The corresponding characteristic equation is
\begin{equation}
   \frac{(\lambda^2+k^2)^2}{\mu^2}+ |A^{00}|(\lambda-\lambda_+(k))(\lambda-\lambda_-(k))=0\,. \label{eq:char_mod}
\end{equation}
For large enough $\mu$ this equation will admit a root that is close to $\lambda_+(k)$ and hence, by continuity, it has positive real part for some $k$. On the other hand, since the additional term (i.e., the first term in \eqref{eq:char_mod}) is manifestly positive, the real part of this root will be smaller than $\text{Re}\,\lambda_+(k)$. Therefore, we conclude that the growing mode of \eqref{eq:toy} (the analog of EFT$_1$) is inherited by \eqref{eq:toy_v2} (the analog of EFT$_2$) but the ``regularising'' term decreases the growth rate.

\begin{figure}[t!]
    \centering
    \includegraphics[width=0.7\linewidth]{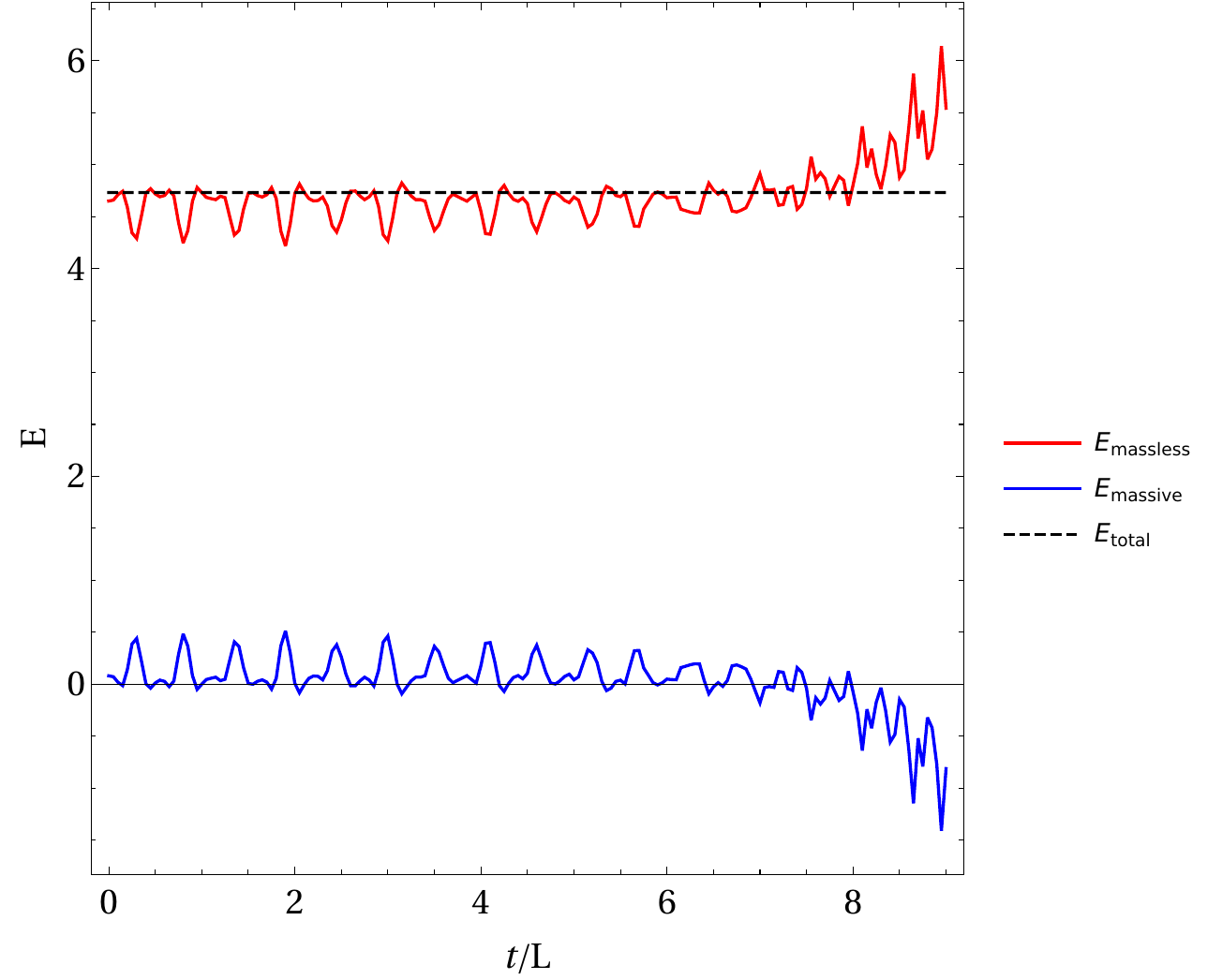}
    \captionsetup{width=.9\linewidth}
    \caption{The evolution of the energy associated with the ``massless'' and ``massive'' degrees of freedom in EFT$_2$ near the breakdown. The cascade of energy towards large negative values carried by the massive ghost field (in blue) and towards large positive values carried by the massless field (in red) is indicative of a ghost-like instability. The total energy is conserved.}
    \label{fig:eft2_energy_dof}
\end{figure}
As mentioned before, EFT$_2$ admits a conserved energy
\begin{align}
    \mathcal{E}_{\text{EFT}_2}[\theta](t)&=\int {\rm d}x \Biggl[\frac12 (\partial_t\theta)^2+\frac12 (\partial_x\theta)^2-\frac{1}{2M^2}\left(-(\partial_t\theta)^2+(\partial_x\theta)^2\right) \left(3\,(\partial_t\theta)^2+(\partial_x\theta)^2\right) \nonumber \\
    &\qquad \qquad +\frac{2\alpha}{M^2}\left((\partial_t\theta)\,\partial_t\theta^{(0,1)}+(\partial_x\theta)\,\partial_x\theta^{(0,1)}+\frac12 (\theta^{(0,1)})^2\right) \Biggr]\,.
   \label{eq:energy_EFT2} 
\end{align}
We may split the total energy into a sum of two contributions: the terms in the first line may be thought of as the energy carried by a ``massless'' degree of freedom, and the terms in the second line as the energy carried by the massive ghost field. However, we have to emphasise that this split is somewhat artificial since the non-linear interaction terms cannot be cleanly separated between a massless and a massive part.

For initial conditions that are consistent with the EFT expansion, the first two terms in the first line of \eqref{eq:energy_EFT2} (associated with a massless free scalar field) will be dominant, ensuring that the total energy is positive. On the other hand, the contributions of the ghost field do not have a definite sign. This means that EFT$_2$ may admit runaway solutions where the energy associated with the ``massless'' contributions grow to arbitrarily large positive values and the ``massive'' ghost contributions may acquire arbitrarily large negative values, whilst conserving the total energy. In particular, we show that the solution arising from the data \eqref{eq:init_dat_1} and $M=10$ is an example of such a runaway behaviour. In Fig. \ref{fig:eft2_energy_dof} we display the evolution of the energy carried by the ``massless'' (red curve) and ``massive'' ghost (blue curve) terms, as well as the total energy (black dashed line), which remains constant in time. This cascade of energy may be interpreted as a ghost-like instability, triggered by a low wavenumber growing mode described above. In practice the simulations crash shortly after this tachyonic instability kicks in.

\begin{figure}[t!]
    \centering
    \includegraphics[width=0.48\linewidth]{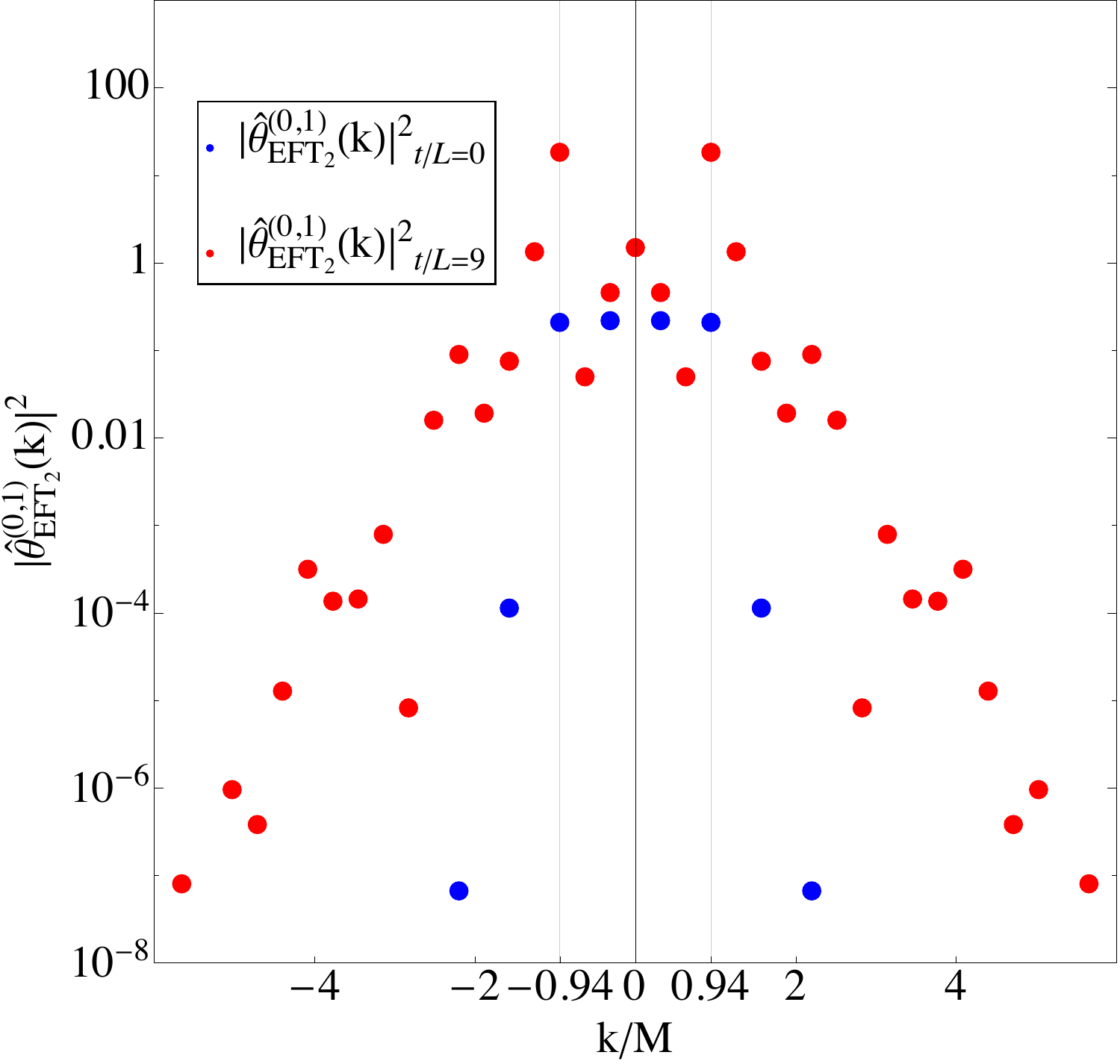}
    \includegraphics[width=0.46\linewidth]{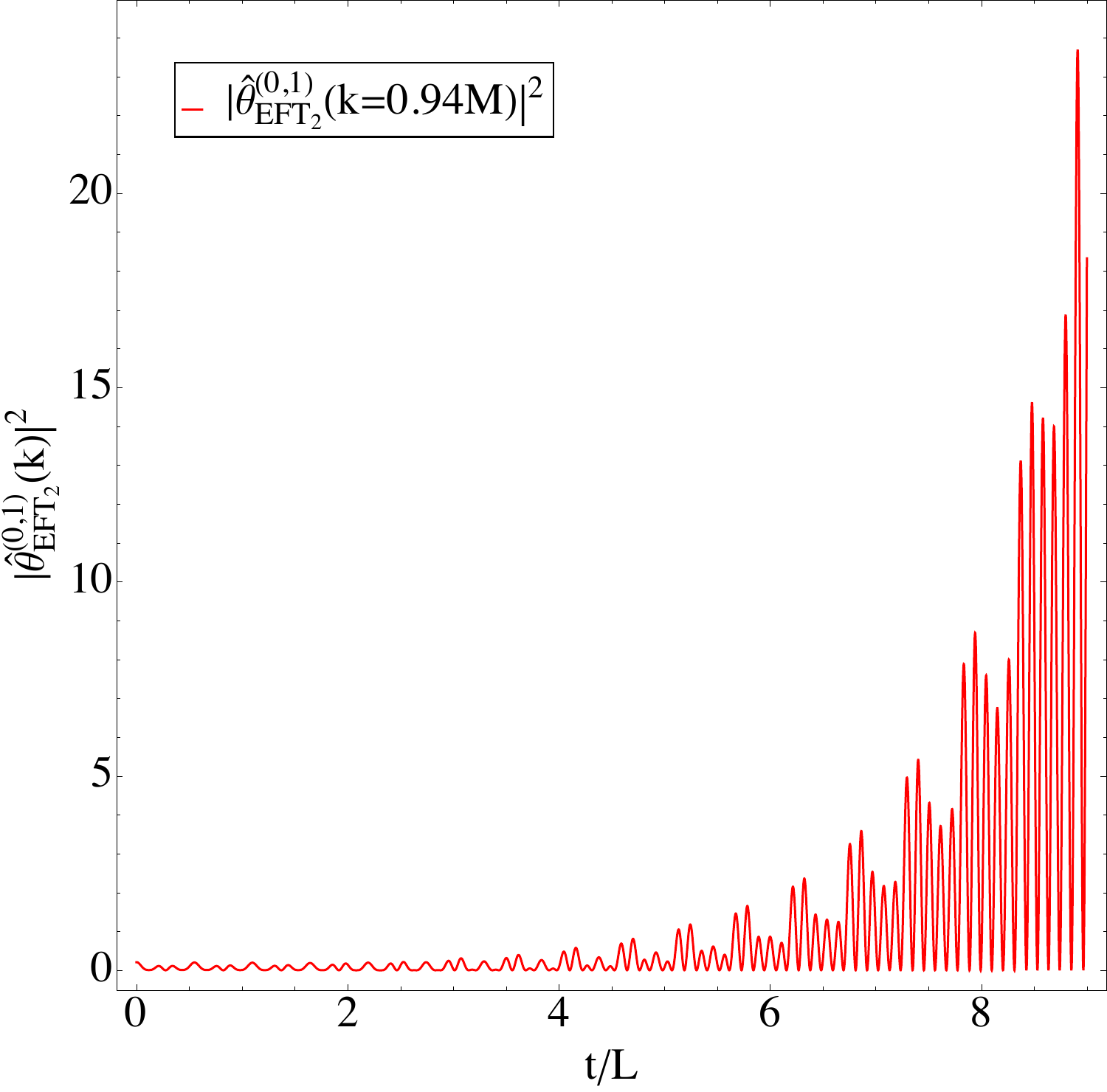}
    \captionsetup{width=.9\linewidth}
    \caption{\textit{Left}: The spatial Fourier transform of $\theta^{(0,1)}_{{\rm EFT}_2}$ computed before (at $t/L=0$) and after (at $t/L=9$) the breakdown of EFT$_2$, showing a growth of the amplitudes of high wavenumber modes. \textit{Right}: The evolution of the  amplitude of the $k=0.94 M$ spatial Fourier mode of $\theta^{(0,1)}_{{\rm EFT}_2}$. This Fourier mode has fairly small amplitude at the early stages of the evolution but exhibits rapid growth at late times.}
    \label{fig:spec_eft_breakdown}
\end{figure}

It is also interesting to analyse the breakdown of EFT$_2$ in Fourier space. We computed the spatial Fourier transform of $\theta^{(0,1)}_{{\rm EFT}_2}$ and followed the time evolution of the various Fourier modes. The results are shown on Fig. \ref{fig:spec_eft_breakdown}. On the left panel of Fig. \ref{fig:spec_eft_breakdown} we show the amplitudes of the Fourier modes on the initial data slice (blue dots) and after the breakdown at $t/L=9$ (red dots). The initial data already has a mode $k\sim M$ with amplitude $O(1)$ (i.e. the same amplitude as other low wavenumber modes) which already indicates that EFT may not be applicable. At later times we can see significant growth of the amplitudes of modes with $k\gtrsim M$, indicating a cascade towards wavenumbers beyond the EFT cutoff. On the right panel of Fig. \ref{fig:spec_eft_breakdown} we track the time evolution of the mode with wavenumber $k=0.94 M$. This mode is already present in the initial data and exhibits small amplitude oscillations until $t\sim 4L$. Afterwards, however, the amplitude of this mode grows rapidly and becomes much larger than the amplitudes of low wavenumber modes, which is a clear indication of the breakdown of the effective field theory description.

\subsection{Gaussian initial data}
In this subsection we analyse another class of initial conditions that include more Fourier modes along the compact spatial direction. More specifically, we choose
\begin{equation}
    \theta_0 = e^{-a\,x^2}\,,\quad \theta_1 = c(1-e^{-b\,x^2})\,.
    \label{eq:gaussian_data}
\end{equation}

We fix $M=200$ and choose $a=b=40$, $c\in\{0,\,1\}$ such that the initial data has $O(1)$ amplitude and $|\partial\theta|\sim M/5$ initially. This means that initially the EFTs are quite strongly coupled and the nonlinearites are significant. For this reason, we use an implicit time integrator to evolve this initial data. Choosing larger amplitudes or steeper initial gradients while keeping $M$ fixed would lead to the blow-up of EFT solutions after a fairly short amount of time, similarly to what we have seen in Section \ref{sec:breakdown}. 

\begin{figure}[t]
    \centering
    \includegraphics[width=\linewidth]{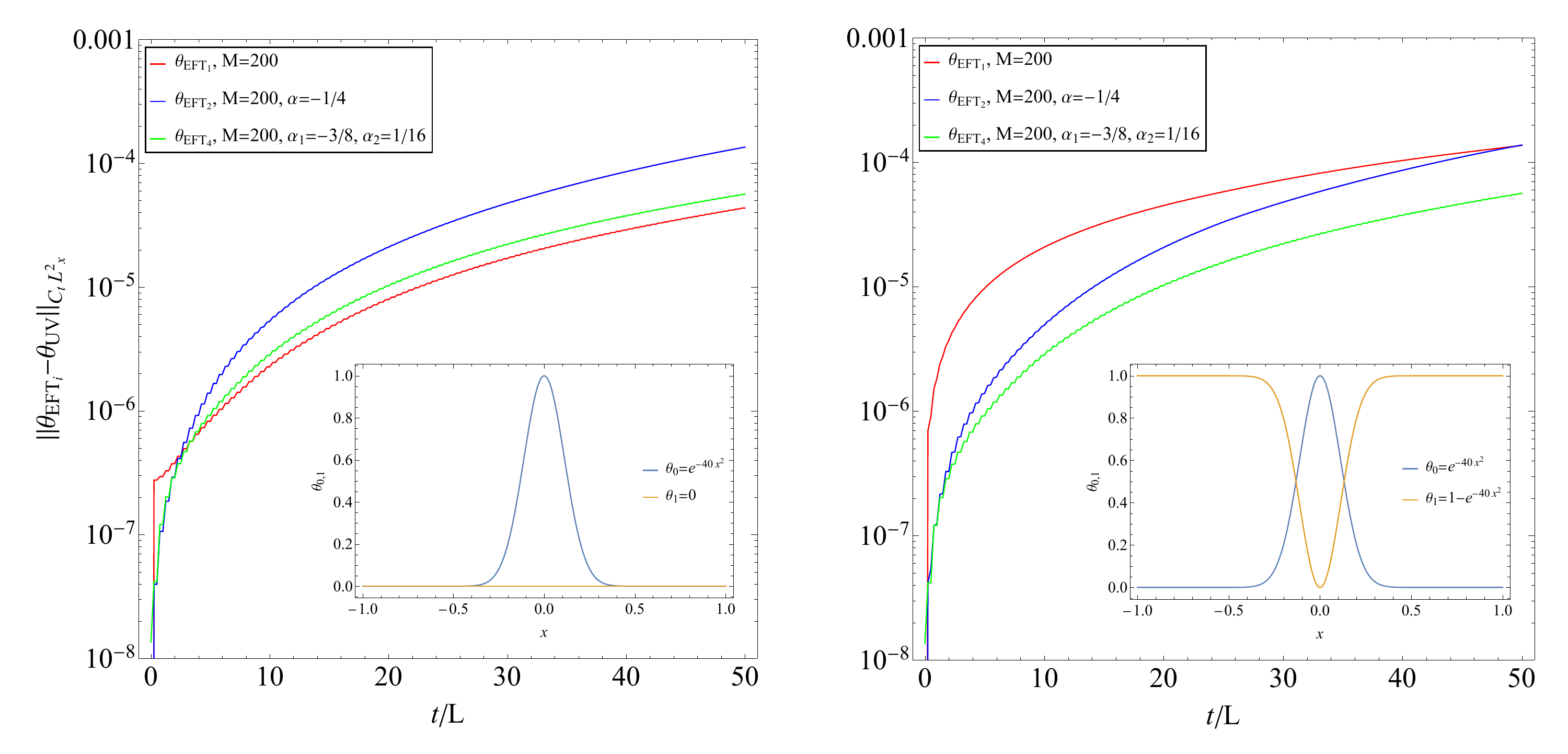}
    \caption{Errors for the different EFTs with initial data \eqref{eq:gaussian_data} and $M=200$. With these choices, the EFTs are at the boundary of their regime of applicability.}
    \label{fig:gaussian_data}
\end{figure}

With this choice of initial data, we see that while we can evolve the EFTs for long times, the hierarchy between the different EFTs is sensitive to the choice of initial conditions: Fig. \ref{fig:gaussian_data} (left) shows that with $\theta_1=0$, the error in the EFT$_4$ is larger than in the EFT$_1$ or EFT$_2$. On the other hand, Fig. \ref{fig:gaussian_data} (right) shows that with $\theta_1\neq 0$, EFT$_4$ provides a more accurate solution, as one would naively expect. These results indicate that with this choice of initial conditions and mass scale $M$, the EFT expansion is at the border of its regime of applicability. 

\section{Discussion}
\label{sec:discussion}

\subsubsection*{Summary of the main results}

In this paper we have studied the regularisation scheme proposed in \cite{Figueras:2024bba} to obtain a well-posed initial value formulation for EFTs with higher derivative equations of motion, and we have applied it to the Abelian-Higgs model \eqref{eq:UV_action}. This is a useful toy model because we can solve the UV theory and compare its solutions to solutions of the corresponding low energy EFT for the the massless field, whenever effective field theory is applicable. 

The EFT truncated at the leading order correction is the two-derivative theory \eqref{eq:action_M2} (we call it EFT$_1$), which is of the Horndeski type. As such, this theory only propagates a massless scalar and it has second order equations of motion, which can be straightforwardly solved in the weakly coupled regime. The EFT at the same truncation order can be reformulated using the regularisation method of  \cite{Figueras:2024bba}, which results in a different EFT that, in addition to the original massless degree of freedom, also propagates a massive ghost. We call this theory EFT$_2$. Truncating the EFT at the level of the next-to-leading order corrections in the derivative expansion \eqref{eq:action_M4} yields a theory with fourth order equations of motion for the light field, see eq. \eqref{eq:eoms_M4}. It is not clear how to formulate the initial value problem for PDEs of this type at the fully non-linear level using traditional methods. The regularisation scheme of \cite{Figueras:2024bba} provides a manifestly well-posed formulation of these equations of motion in terms of the original massless field  plus two massive ghost fields. We call this theory EFT$_4$. A key aspect of the regularisation method is that all the additional massive degrees of freedom introduced by the scheme have masses of the order of the UV mass scale $M$. Therefore, the expectation is that at sufficiently low energies, such additional massive modes should not be significantly excited and hence the EFTs obtained with the regularisation method should still be able to provide a consistent description of the low energy physics. The goal of this paper was to analyse this in detail. For comparison, we have also considered solving the EFT perturbatively in a $1/M$ expansion, which amounts to iteratively solving linear wave equations sourced by the lower order solutions. We call this approach EFT$_0$. 

We carried out simple numerical experiments using the initial data \eqref{eq:init_dat_1} for the massless scalar field $\theta$. Consistent initial data for the massive UV field $\rho$ can obtained from the EFT expansion and using the reduction of order procedure. Similarly, initial data for the auxiliary fields (higher derivatives of $\theta$) in the EFT$_2$ and EFT$_4$ is obtained by order reduction. Notably, the initial data \eqref{eq:init_dat_1} has an amplitude of $O(1)$ hence the non-linearities in the equations play an important role on the initial slice already. Varying the mass scale $M$ while keeping the initial data fixed allows us to interpolate between the regime of validity of EFT and a regime where the EFT expansion breaks down. In the regime of validity of EFT, we see a clear improvement in the accuracy of the solutions provided by EFT$_4$ compared to the solutions provided by EFT$_1$ and EFT$_2$. The latter two EFTs provide equivalent descriptions in agreement with the expectations. We have also checked that the errors in the solutions constructed with the EFTs scale with the expected powers of the UV mass scale $M$. Furthermore, since we solve the evolution equations of these EFTs fully non-linearly, the late time error of these EFT solutions grows linearly in time, which is optimal. This should be contrasted with the errors of the solutions constructed with the linearisation scheme, which grow polynomially in time, with a power that depends on the truncation order of the EFT \eqref{eq:EFT0_secular_growth}.

In the regularised EFTs (i.e., in EFT$_2$ and EFT$_4$), one has the freedom to choose the parameters appearing in the field redefinitions, \eqref{eq:field_redef_M2} and \eqref{eq:field_redef_M4} respectively. However, there is a restriction on these parameters imposed by the requirement that the theories linearised around the trivial solution $\theta=0$\footnote{Using the shift symmetry, there is no loss of generality in choosing this particular constant solution.} have no tachyonic modes. Such a choice is always possible and does not require fine-tuning. Once a choice of these parameters is fixed (together with initial data compatible with the EFT regime), the solutions obtained with the regularisation scheme remain bounded within the computational domain. Even though the regularisation scheme introduces high frequency oscillations $\omega\gtrsim M$ into the solution, we find that the amplitudes of the high frequency modes are highly suppressed. This suggests that the regularised EFT solutions satisfy the uniform boundedness requirements of \cite{Reall:2021ebq} and the $M\to\infty$ limit may be well-defined. Furthermore, we have confirmed that the solution for the light field $\theta$ provided by the regularised EFTs is not sensitive to the specific choices of the free parameters (and hence masses) in the field redefinitions. 

Both the UV theory and the EFTs that we have considered derive from an action and admit a conserved energy. The conserved energy of the UV theory is positive definite, reflecting the fact that it is a ``healthy'' theory. On the other hand, the conserved energy of higher derivative theories is unbounded from below: the higher derivative contributions to the energy do not have a definite sign.\footnote{This is the case even for EFT$_1$, which is a Horndeski theory and does not propagate ghosts} Choosing initial data in the regime of validity of EFT ensures that these non-positive terms are subleading to the manifestly positive energy contribution of the massless scalar. This ensures that the conserved energy associated with the initial data is positive. Furthermore, for such initial data, we find that the energy carried by the ghost fields remains bounded and small compared to the positive definite terms throughout the evolution. However, for strongly coupled initial data we find that the energy associated to the ghost fields may acquire large negative values in a finite time, leading to the breakdown of EFT.

It is interesting to compare the mechanism of the breakdown in EFT$_1$ and EFT$_2$ since these two theories are equivalent up to perturbative field redefinitions. The breakdown of EFT$_1$ occurs when $(\partial\theta)^2\sim M^2/6$ and the equation of motion changes character from hyperbolic to elliptic through a Tricomi-like transition. On the other hand, the cause of the breakdown of EFT$_2$ is the transition of the massive degree of freedom from ghost-like to tachyonic when $(\partial\theta)^2\sim M^2/6$ (i.e., the same condition as in EFT$_1$). We also show that the ultimate breakdown of both of these two EFTs is preceded by a growth due to a long wavelength unstable mode. The growth rate in EFT$_2$, however, is always slightly smaller than in EFT$_1$ and hence the solution of EFT$_2$ persists slightly longer.

\subsubsection*{On the global nonlinear stability of the vacuum in EFT}

In our 1+1-dimensional numerical experiments we found long-lived solutions for the regularised EFTs that exhibit stable evolution throughout the computational domain. These solutions arise from initial data with a characteristic length scale significantly larger than $M^{-1}$. One might still question whether these EFT solutions would exhibit blow-up at later times. It is entirely possible that this would happen in our (1+1)-dimensional setting (especially with a compact spatial dimension). The reason for this is that the vacuum is not stable even in the UV theory: with no decay mechanism present in 1+1 dimensions, perturbations can grow (linearly in time). Thus it is possible that, for some initial data, solutions of the UV theory may dynamically evolve to a regime where EFT is no longer applicable. This would lead to the eventual breakdown of the regularised EFT solutions.

However, we expect the situation to be different in 3+1 (and higher dimensions) for the following two reasons. Firstly, the vacuum is stable in the UV theory \cite{Dong:2019ttn} and therefore we expect that for small enough perturbations of the vacuum, the UV solution will always remain in a regime where EFT is applicable. It is then also reasonable to expect that the regularised EFT solutions will capture this behaviour. Secondly, the ghost fields in the regularised EFTs are governed by equations of motion that have a similar structure as the model PDE
\begin{equation}
    (\Box-M^2)\,\Phi=-4\lambda\, \Phi^3. \label{eq:ghost_toy}
\end{equation}
It is a well-known result from PDE theory \cite{Sogge} that for $M=0$ and in lower than $3+1$ dimensions, \eqref{eq:ghost_toy} does not admit global solutions for generic small initial data. On the other hand, in $3+1$ dimensions there exist global solutions to \eqref{eq:ghost_toy} for arbitrary small initial data, regardless of the sign of $\lambda$! This is a very surprising result since for positive $\lambda$ the solution $\Phi=0$ sits at the top of a potential $V(\Phi)=-\lambda\Phi^4$, which is unbounded from below; therefore,  naively, one might think that arbitrarily small perturbations of this solution would trigger a cascade (i.e., the field $\Phi$ would roll down the potential). The reason why this does not happen in $3+1$ (and higher) dimensions is that one can establish sufficiently fast decay for the linear wave equation. Then a continuity argument reveals that for small enough initial perturbations of the $\Phi=0$ solution, the linear behaviour will dominate and the perturbations decay before the cascade has time to set in. This example suggests that one might hope for global existence of solutions even in the presence of ghost fields, at least for a restricted (but still open) set of initial data.

\begin{figure}[t]
    \centering
    \includegraphics[width=0.6\linewidth]{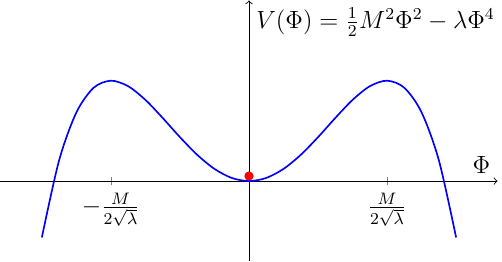}
    \captionsetup{width=.9\linewidth}
    \caption{The shape of the potential for the toy equation \eqref{eq:ghost_toy} that mimicks the behaviour of the massive ghost fields in the regularised EFTs.}
    \label{fig:potential}
\end{figure}

It is interesting to note that the global behaviour of solutions is expected to be better when the ghost fields are massive (which is the case in the regularised EFTs discussed in this paper), this was shown in rigorous mathematical works \cite{Kla1985,Kat2012} and also recently observed by \cite{Salvio:2019ewf,Deffayet:2025lnj}. This is because massive fields generally exhibit faster decay than massless fields. The improved stability can also be understood heuristically by noticing that the mass term creates a well in the potential, see Fig. \ref{fig:potential}. The field $\Phi$ (which here is a representative of a ghost field, i.e., the analogue of a higher derivative of the light field $\theta$ in the EFTs discussed in this paper) has to travel a distance of $O(M)$ in field space to overcome the potential barrier and exhibit runaway behaviour.

Based on this picture we conjecture that in at least $3+1$ dimensions, any choice of initial data with a characteristic length scale significantly larger than $M^{-1}$ will give rise to global EFT-compatible solutions. We plan to rigorously investigate this problem in future works.

\subsubsection*{Future directions}

\begin{figure}[t]
    \centering
    \includegraphics[width=0.6\linewidth]{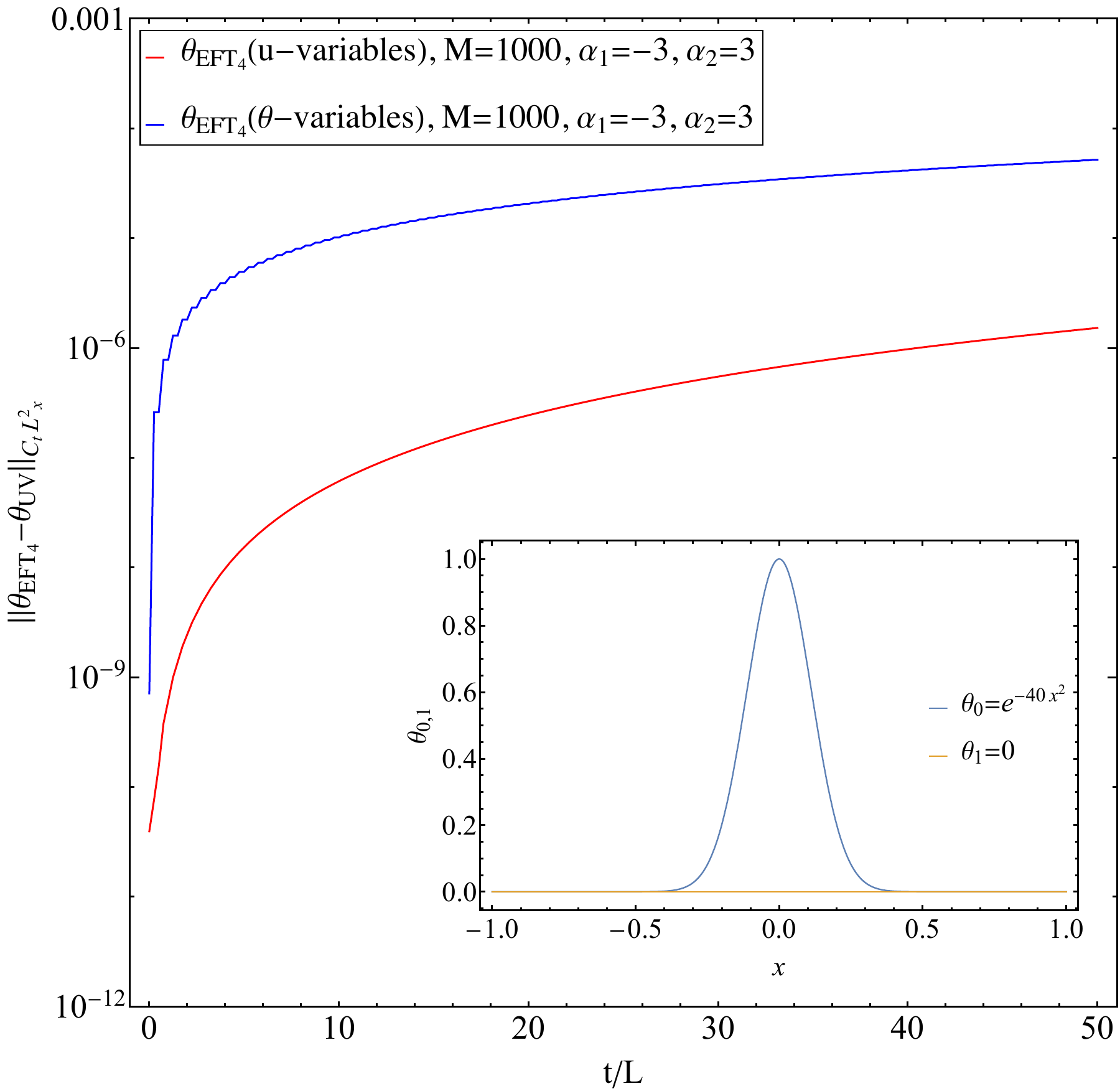}
    \captionsetup{width=.9\linewidth}
    \caption{Comparison of the errors for the EFT$_4$ obtained with the original variables (blue) and the rescaled variables (red) for a large mass $M=1000$. The errors are visibly smaller with the rescaled variables. }
    \label{fig:rescaled}
\end{figure}

In practical applications, the UV mass scale, $M$ in our case, must be suitably large so that the EFT provides an accurate description of the UV physics. Examining the equations of motion of the regularised EFTs, e.g., \eqref{eq:eq_box_theta} and \eqref{eq:eq_theta02_M4}, we observe that the high powers of $M$ appearing in these equations may cause a loss of numerical accuracy if $M$ is very large. This problem may be exacerbated in the regularised formulation of higher derivative theories of gravity \cite{Figueras:2024bba}, because the equations will only be more non-linear and this loss of accuracy may lead to numerical instabilities. To overcome this issue, it is useful to define dimensionless variables to absorb the powers of $M$:
\begin{equation}
    \theta^{(m,n)} \equiv \partial_{a_1}\ldots\partial_{a_m}\Box^n\theta = M^{m+2n}\,u^{(m,n)}\,.
    \label{eq:theta_dimensionless}
\end{equation}
In terms of these new dimensionless variables, the evolution equations for the EFT$_2$ become
\begin{align}
\Box u = &~M^2\,u^{(0,1)}\,,\\
\Box u^{(1,0)}_a = &~M\,\partial_a u^{(0,1)} \,, \\
\left(\Box +\frac{M^2}{2\,\alpha} \right) u^{(0,1)} =&~  \frac{M^2}{\alpha} \left( u^{(1,0)}_a\,u^{(1,0)a}\,u^{(0,1)} + \frac{2}{M}\,u^{(1,0)a}\,u^{(1,0)b}\,\partial_a u^{(1,0)}_b\right)\,,
\end{align}
and for the EFT$_4$ we get
\begin{align}
    \Box u =&~M^2\,u^{(0,1)}\,, \label{eq:eq_u_M4_rescaled}\\
    \Box u^{(1,0)}_a=&~M\,\partial_a u^{(0,1)}\,,\\
    \Box u^{(0,1)} = &~M^2\,u^{(0,2)}\,, \\
    \Box u^{(2,0)}_{ab} =&~M\,\partial_{a} u^{(1,1)}_{b}\,, \\
    \Box u^{(1,1)}_a =&~M\,\partial_a u^{(0,2)}\,,\\
    \left(\Box+\frac{\alpha_1}{\alpha_2}\,M^2\right)u^{(0,2)} =&-\frac{M^2}{2\,\alpha_2}\,u^{(0,1)} \label{eq:eq_u02_M4_rescaled}\\
    &+\frac{M^2}{\alpha_2}\big[ u^{(0,1)}\eta^{ab} + 2\,u^{(2,0)ab}\big]u^{(1,0)}_a\,u^{(1,0)}_b \nonumber\\
    &+\frac{2\,M^2}{\alpha_2}\,\Big[\frac{1}{M}\,u^{(1,0)a}u^{(1,0)b}\partial_a u^{(1,1)}_b+u^{(0,1)}u^{(1,0)a}u^{(1,1)}_a \nonumber\\
&\hspace{1.75cm}+u^{(1,0)a}u^{(1,1)b}u^{(2,0)}_{ab}+u^{(0,1)}u^{(2,0)ab}u^{(2,0)}_{ab} \nonumber\\
&\hspace{1.75cm}+\frac{2}{M}\,u^{(1,0)a}u^{(2,0)bc}\partial_{(a}u^{(2,0)}_{bc)}\Big]\,.\nonumber
\end{align}

In Fig. \ref{fig:rescaled} we compare the results for the EFT$_4$ obtained by solving the system with and without rescaling for a large mass scale $M=1000$. This plot clearly shows that for such large masses, the errors obtained with the rescaled formulation are significantly smaller. It is also worth noting that the code based on the rescaled dimensionless variables exhibits much better convergence properties than the code based on the original variables, even if both codes use the same differencing and time integration schemes.

The insights into the regularisation scheme that we have obtained in this paper should prove useful in the implementation of this scheme in higher derivative theories of gravity, as originally envisioned in \cite{Figueras:2024bba}. An attractive feature of this method is that it is fully covariant, unlike other currently known approaches such as the MIS method (also known as the `fixing-of-the-equations' approach). In this paper, we have shown that the regularisation scheme is robust and can provide accurate and long-lived EFT solutions in realistic situations. Another key aspect of the regularisation method is that it allows one to construct solutions of the fully non-linear EFT evolution equations, thus avoiding a faster than linear secular growth of error with respect to the UV theory. This appears to saturate the optimal scaling of the errors for solutions that remain bounded. This feature of the solutions constructed with the regularisation scheme could be useful in order to identify deviations from GR introduced by higher derivative corrections: even if the Wilson coefficients in the action of these theories are very small (due to the high powers of the UV length scale), small deviations can accumulate over the long inspiral phase of a black hole binary and might be detectable or constrained with graviational wave observations. The implementation of the regularisation scheme in a higher derivative theory of gravity is in progress.

Higher derivative theories arise very naturally in physics and their study has a long history. The regularisation scheme introduced in \cite{Figueras:2024bba} and further explored here, sheds new light on this old problem and provides a new way of constructing physically acceptable solutions to large classes of higher derivative theories.

\section*{Acknowledgements}

We would like to thank Ramiro Cayuso, Katy Clough, Aaron Held, Luis Lehner, Eugene Lim, Harvey Reall and Bob Wald for discussions. PF and ADK are supported by the STFC Consolidated Grant ST/X000931/1. SY acknowledges the support from the China Scholarship Council.

\appendix

\section{Consistency checks}
\label{sec:consistency_chks}

\begin{figure}[t]
    \centering
    \includegraphics[width=0.48\linewidth]{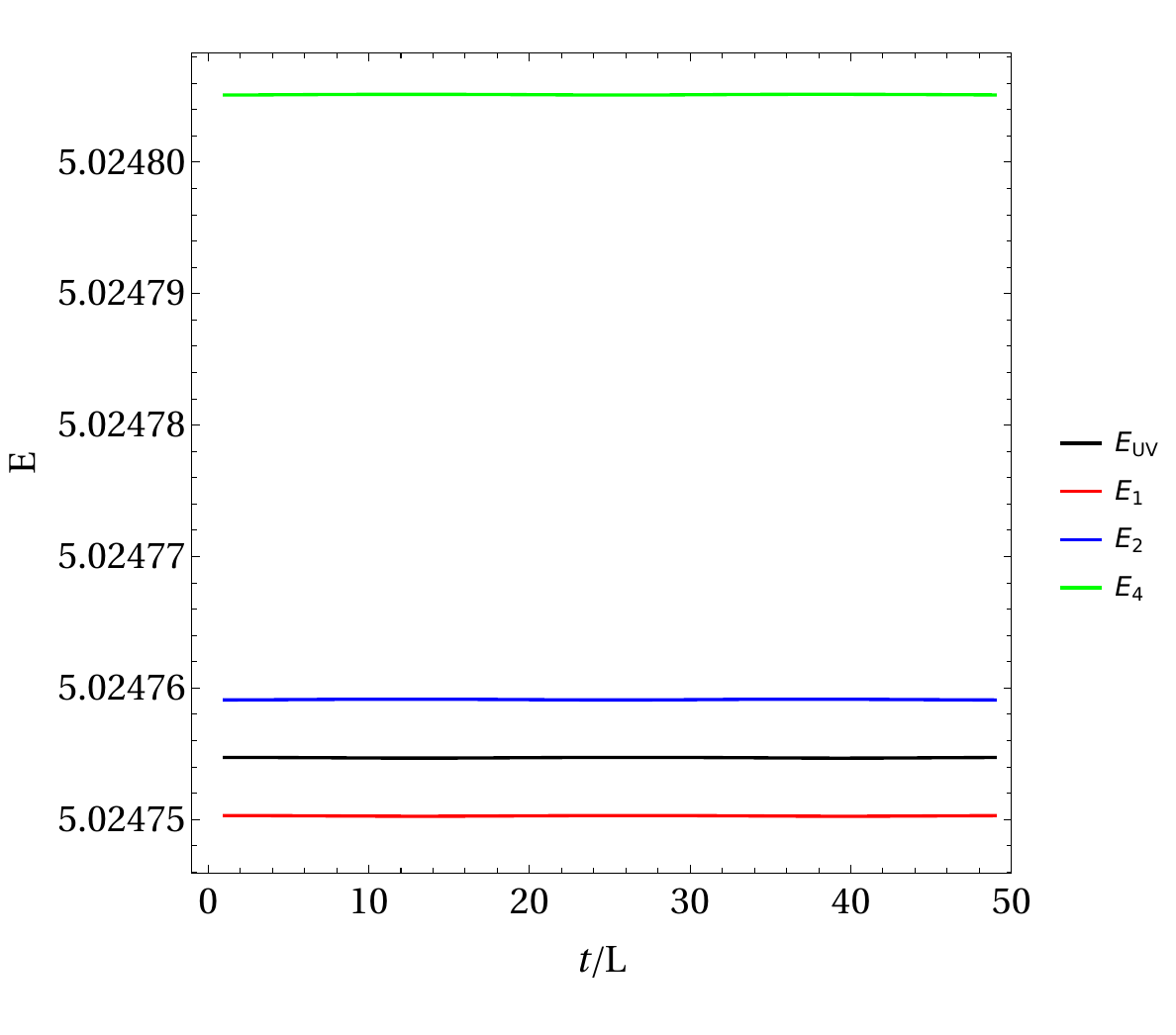}
    \includegraphics[width=0.48\linewidth]{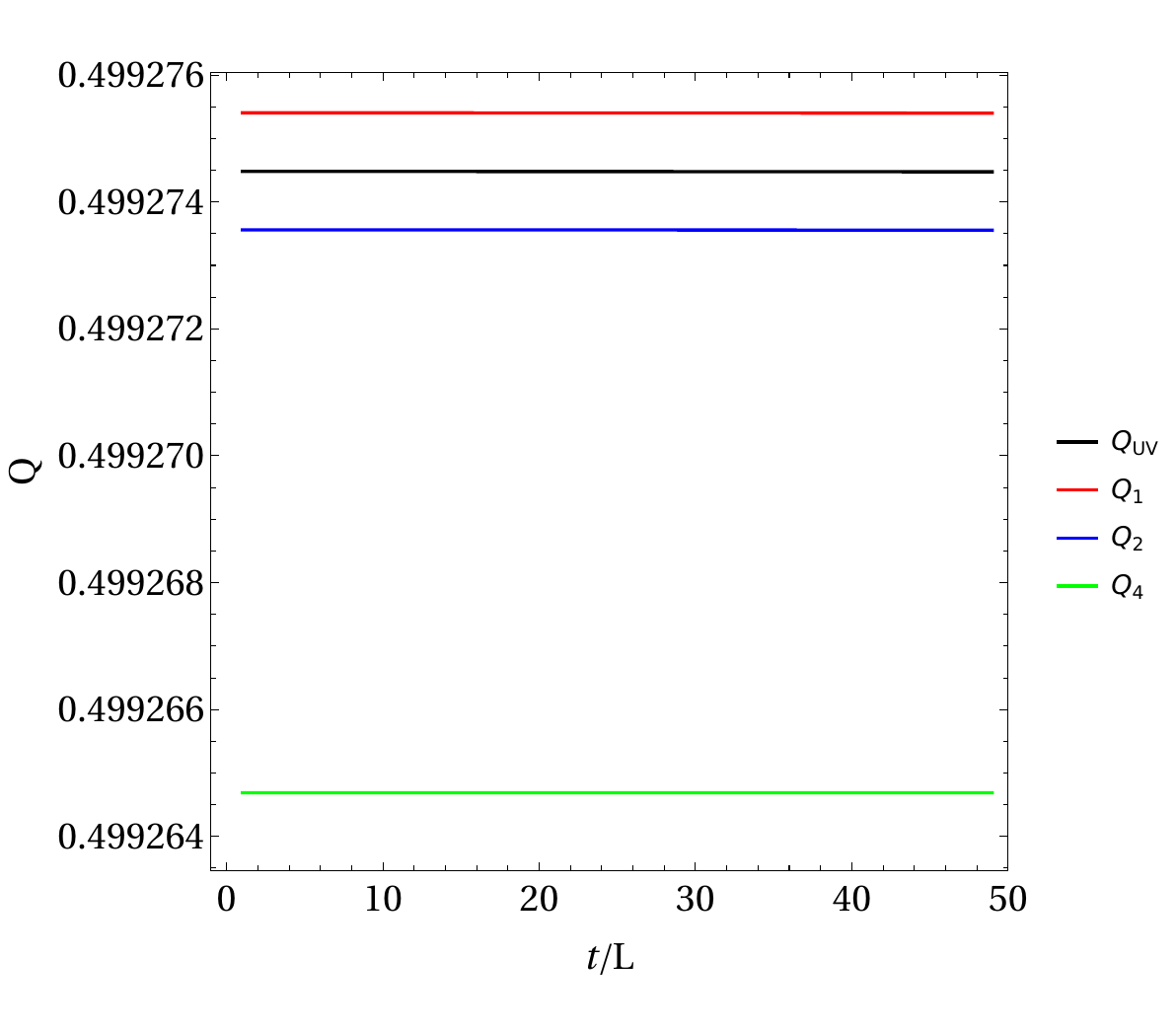}
    \captionsetup{width=.9\linewidth}
    \caption{Conservation of energy and the charge ${\cal Q}$ in the UV theory and the different EFTs for $M=100$ and the initial data \eqref{eq:init_dat_1}.}
    \label{fig:charge_plot}
\end{figure}

In this section we carry out some consistency checks of our numerical simulations. As discussed in Section \ref{sec:norms}, the UV theory and the various EFTs admit conserved charges. To check the accuracy of our numerical simulations, we monitored the conservation of the energy and the charge associated with the shift symmetry throughout the evolution. Here we present results for the initial data \eqref{eq:init_dat_1} and $M=100$. In this setup, both the energy ${\cal E}$ and the charge ${\cal Q}$ are positive in each of the theories considered. In Fig. \ref{fig:charge_plot} we demonstrate that ${\cal E}$ and ${\cal Q}$ remain constant (within numerical accuracy) in time throughout our simulations. Note that for each theory, the expressions for ${\cal E}$ and ${\cal Q}$ are different, and hence these plots merely serve the purpose of checking the consistency of our simulations for each theory.

\begin{figure}[t]
    \centering
    \includegraphics[width=0.6\linewidth]{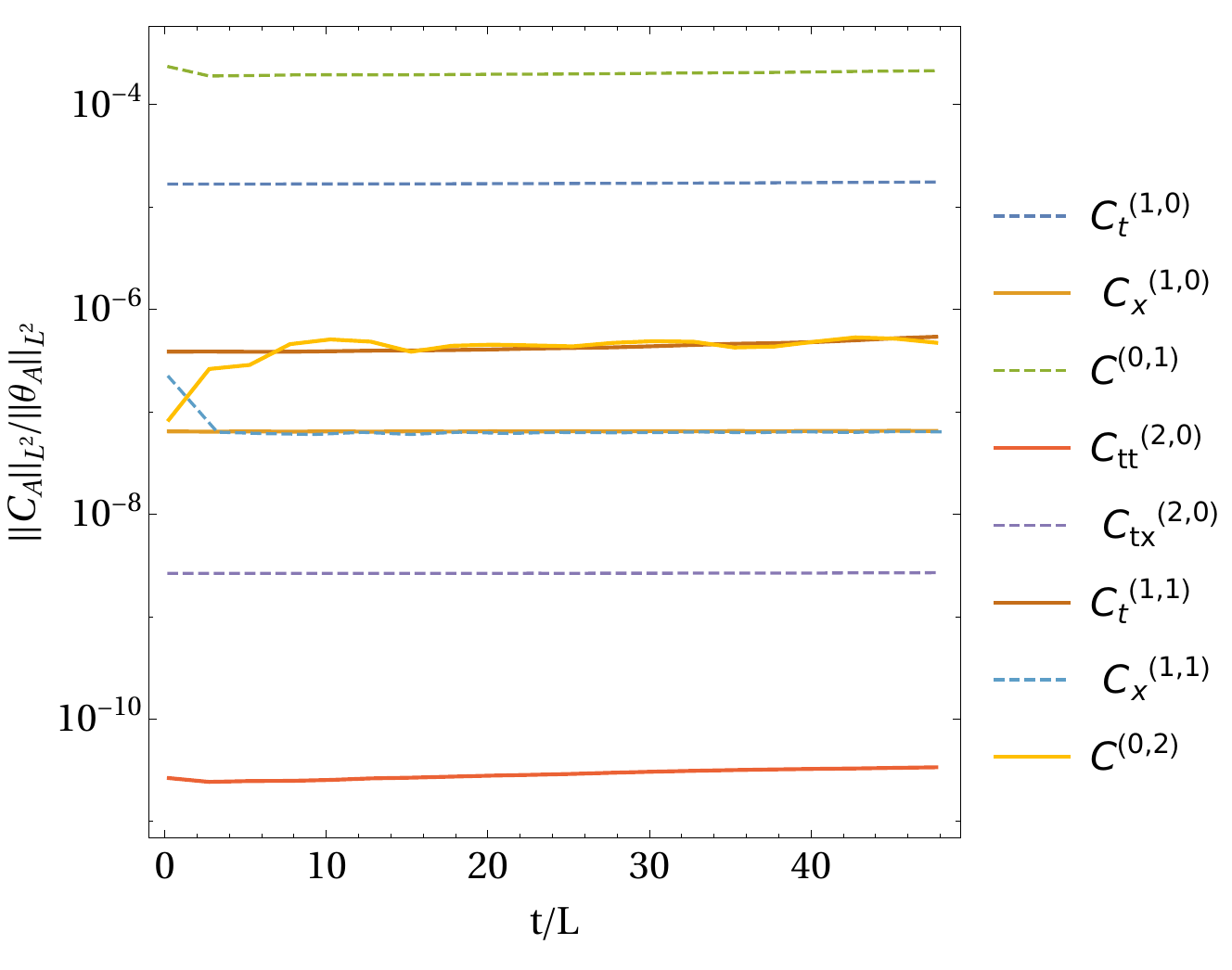}
    \captionsetup{width=.9\linewidth}
    \caption{The relative constraint violations associated with the various auxiliary variables in $\text{EFT}_4$ with $M=100$, $\alpha_1=-3$, $\alpha_2=3$, starting from the initial data \eqref{eq:init_dat_1}.}
    \label{fig:constr_plot}
\end{figure}

Recall that we solve the higher derivative equations of motion of EFT$_2$ and EFT$_4$ by integrating a system of second order wave equations for the auxiliary variables $\{\theta,\theta_a^{(1,0)},\theta^{(0,1)}\}$ and $\{\theta,\theta_a^{(1,0)},\theta^{(0,1)},\theta_a^{(1,1)},\theta_{ab}^{(2,0)},\theta^{(0,2)}\}$, respectively. The second order systems are equivalent to the original higher derivative equations as long as the constraints
\begin{equation}
    {\cal C}_a^{(1,0)}\equiv \theta^{(1,0)}_a-\partial_a \theta, \qquad \qquad {\cal C}^{(0,1)}\equiv \theta^{(0,1)}-\Box\theta
\end{equation}
in EFT$_2$ and
\begin{align}
    &{\cal C}_a^{(1,0)}\equiv \theta^{(1,0)}_a-\partial_a \theta, \qquad \qquad {\cal C}^{(0,1)}\equiv \theta^{(0,1)}-\Box\theta, \nonumber \\
   & {\cal C}_a^{(1,1)}\equiv \theta^{(1,1)}_a-\partial_a \Box\theta, \qquad \qquad {\cal C}_{ab}^{(2,0)}\equiv \theta_{ab}^{(2,0)}-\partial_a\partial_b \theta, \qquad \qquad {\cal C}^{(0,2)}\equiv \theta^{(0,2)}-\Box(\Box\theta)
\end{align}
in EFT$_4$ are satisfied. Hence, for each auxiliary variable $\theta_A$, there is an associated constraint relating $\theta_A$ to derivatives of $\theta$. Assuming that the equations of motion hold, these constraints propagate according to a system of wave equations that are linear and homogeneous in the constraint variables. In EFT$_4$ the system governing the propagation of constraints is
\begin{align}
\Box {\cal C}^{(0,1)}&={\cal C}^{(0,2)}, \\
\Box {\cal C}_a^{(1,0)}&=\partial_a{\cal C}^{(0,1)}, \\
\Box {\cal C}_a^{(1,1)}&=\partial_a {\cal C}^{(0,2)}, \\
\Box {\cal C}_{ab}^{(2,0)}&=\partial_a {\cal C}_b^{(1,1)}, \nonumber\\
    \left(\Box+\frac{\alpha_1}{\alpha_2}\,M^2\right){\cal C}^{(0,2)} =&-\frac{M^4}{2\,\alpha_2}\,{\cal C}^{(0,1)}+\frac{M^2}{\alpha_2}\big[ {\cal C}^{(0,1)}\eta^{ab} + 2\,{\cal C}^{(2,0)ab}\big]\theta^{(1,0)}_a\theta^{(1,0)}_b \nonumber\\
    &+\frac{2M^2}{\alpha_2}\big[ \theta^{(0,1)}\eta^{ab} + 2\,\theta^{(2,0)ab}\big]\theta^{(1,0)}_a{\cal C}^{(1,0)}_b \nonumber\\
    &+\frac{2}{\alpha_2}\,\big[~2{\cal C}^{(1,0)a}\theta^{(1,0)b}\partial_a\theta^{(1,1)}_b+\theta^{(1,0)a}\theta^{(1,0)b}\partial_a{\cal C}^{(1,1)}_b+{\cal C}^{(0,1)}\theta^{(1,0)a}\theta^{(1,1)}_a \nonumber\\
   &\hspace{1.1cm} +\theta^{(0,1)}{\cal C}^{(1,0)a}{\theta}^{(1,1)}_a+\theta^{(0,1)}\theta^{(1,0)a}{\cal C}^{(1,1)}_a \nonumber\\
&\hspace{1.1cm}+{\cal C}^{(1,0)a}\theta^{(1,1)b}\theta^{(2,0)}_{ab}+\theta^{(1,0)a}{\cal C}^{(1,1)b}\theta^{(2,0)}_{ab}+\theta^{(1,0)a}\theta^{(1,1)b}{\cal C}^{(2,0)}_{ab} \nonumber \\
&\hspace{1.1cm}+{\cal C}^{(0,1)}\theta^{(2,0)ab}\theta^{(2,0)}_{ab}+2\theta^{(0,1)}\theta^{(2,0)ab}{\cal C}^{(2,0)}_{ab}+2\,{\cal C}^{(1,0)a}\theta^{(2,0)bc}\partial_{(a}\theta^{(2,0)}_{bc)} \nonumber \\
&\hspace{1.1cm} +2\,\theta^{(1,0)a}{\cal C}^{(2,0)bc}\partial_{(a}{\theta}^{(2,0)}_{bc)}+2\,\theta^{(1,0)a}\theta^{(2,0)bc}\partial_{(a}{\cal C}^{(2,0)}_{bc)}\big]\,.
\end{align}
Therefore, if the constraints are initially satisfied then they will continue to hold at all times. Of course, in numerical simulations the finite resolution will inevitably introduce some constraint violations. To quantify the accuracy to which we solve the original higher derivative equations, we compute the {\it relative} constraint violation $||{\cal C}_A||_{L^2}/||\theta_A||_{L^2}$ for our auxiliary variables. In Fig. \ref{fig:constr_plot} we show the relative errors for these variables in $\text{EFT}_4$ with $M=100$ and initial data \eqref{eq:init_dat_1}, confirming that the constraint violations are negligibly small.

Note that not all these constraints are independent. For example, for EFT$_4$ in $1+1$ dimensions an independent set of constraints ($7$ in total) is given by ${\cal C}^{(0,1)}$, ${\cal C}_t^{(1,0)}$, ${\cal C}_t^{(1,1)}$, ${\cal C}_{ab}^{(2,0)}$, ${\cal C}^{(0,2)}$, since e.g., $\partial^a {\cal C}_a^{(1,0)}$ and $\partial^a {\cal C}_a^{(1,1)}$ can be expressed in terms of the other constraints. This is in agreement with the fact that EFT$_4$ propagates $3$ degrees of freedom: we have $10$ auxiliary fields and $7$ independent constraints. Likewise, for EFT$_2$ we have an independent set of 2 constraints (which can be chosen to be ${\cal C}^{(0,1)}$, ${\cal C}_t^{(1,0)}$ since $\partial^a {\cal C}_a^{(1,0)}$ is expressible with the other constraints) for the 4 dynamical fields, giving rise to 2 degrees of freedom.

\section{Alternative formulation of EFT$_4$}
\label{sec:alt_formulation}

As mentioned in the main body of the paper, the EFT$_4$ propagates the original massless mode $\theta$ and two additional massive ghosts. We can formulate the equations of motion for this EFT by writing explicit massive wave equations for these new modes. The formulation presented here has the advantage that the highest power of the UV mass scale that explicitly appears is $M^2$, which could be an advantage in terms of providing numerically accurate solutions in situations where $M$ is chosen to be very large. 

As before, define
\begin{equation}
    \theta_a^{(1,0)} \equiv \partial_a\theta\,,\quad \theta^{(2,0)}_{ab} \equiv\partial_a\partial_b\theta\,,\quad \theta^{(1,1)}_a\equiv\partial_a\theta^{(0,1)}\,.
\end{equation}
Then, equation \eqref{eq:eoms_M4_reg} can be written as
\begin{align}
    \Box\theta =&~ \theta^{(0,1)}\,,\label{eq:eq_theta_M4_new}\\
    \Box\theta^{(1,0)}_a =& ~\theta_a^{(1,1)}\,,\\
    (\Box-m_-^2)\theta^{(0,1)} =& ~M^2\theta^{(0,2)}\,,\\
    (\Box-m_-^2)\theta^{(1,1)}_a =& ~ M^2 \partial_a\theta^{(0,2)}\,,\\
    \Box\theta^{(2,0)}_{ab} =&~\partial_a\theta_b^{(1,1)}\,,\nonumber\\
    (\Box-m_+^2)\theta^{(0,2)} =&~\frac{1}{\alpha_2}\big[ \theta^{(0,1)}\eta^{ab} + 2\,\theta^{(2,0)ab}\big]\theta^{(1,0)}_a\theta^{(1,0)}_b \nonumber\\
    &+\frac{2}{\alpha_2 M^2}\,\big[~\theta^{(1,0)a}\theta^{(1,0)b}\partial_a\theta^{(1,1)}_b+\theta^{(0,1)}\theta^{(1,0)a}\theta^{(1,1)}_a \nonumber\\
&\hspace{1.1cm}+\theta^{(1,0)a}\theta^{(1,1)b}\theta^{(2,0)}_{ab}+\theta^{(0,1)}\theta^{(2,0)ab}\theta^{(2,0)}_{ab} \nonumber\\&+2\,\theta^{(1,0)a}\theta^{(2,0)bc}\partial_{(a}\theta^{(2,0)}_{bc)}\big]\,,
\label{eq:eq_theta02_M4_new}
\end{align}
with
\begin{equation}
    m_\pm^2 = -\frac{M^2}{2\,\alpha_2}\Big(\alpha_1\pm\sqrt{\alpha_1^2-2\,\alpha_2} \Big)\,,
\end{equation}
and $\alpha_1<0$ and $0<\alpha_2\leq\frac{1}{2}\alpha_1^2$. Note that one can freely arrange the masses  $m_+$ and $m_+$ in the equations above between the two modes; above we have chosen the masses such that the largest one is associated to the field with the highest derivatives. In practice, this particular choice should not make a significant difference, since both $m_+$ and $m_-$ are of the same order, typically $O(M)$. 

We have performed some preliminary numerical experiments with this formulation using the same initial data and parameters as in the example of Fig. \ref{fig:rescaled}, but we have not observed any significant improvement with respect to the results obtained with the original (unrescaled) variables. Likewise, we can rewrite \eqref{eq:eq_theta_M4_new}--\eqref{eq:eq_theta02_M4_new} in terms of dimensionless variables, eq. \eqref{eq:theta_dimensionless}, and the new system behaves in a similar way, qualitatively and quantitatively, to the rescaled system \eqref{eq:eq_u_M4_rescaled}--\eqref{eq:eq_u02_M4_rescaled}. 

\bibliographystyle{JHEP}
\bibliography{AH_EFT}

\end{document}